\documentclass[11pt, onesid, letter]{article}
\usepackage[margin = 20mm, paperwidth = 176mm, paperheight = 250mm]{geometry}
\usepackage{setspace}
\usepackage{tabto}
\usepackage{amsmath,amscd}

\usepackage{bbold}
\usepackage{makecell}
\usepackage{mathrsfs}
\usepackage{amsfonts}
\usepackage{amssymb}
\usepackage{bm}
\usepackage{graphicx}
\usepackage{epstopdf}
\usepackage{subcaption}
\usepackage{verbatim}
\usepackage{rotating}
\usepackage{indentfirst}
\usepackage{booktabs}
\usepackage{float}
\usepackage{eurosym}
\usepackage{natbib}
\usepackage{todonotes}
\usepackage{booktabs}
\usepackage{array}
\usepackage{paralist}
\usepackage{verbatim}
\usepackage{color}
\usepackage[colorlinks=true,linkcolor=red,urlcolor=blue,citecolor=blue]{hyperref}
\usepackage[font=footnotesize,labelfont=bf]{caption}
\usepackage{braket}
\usepackage[graphicx]{realboxes}
\usepackage{authblk}
\usepackage[super]{nth}
\usepackage{tikz}
\usetikzlibrary{arrows.meta}
\usepackage{changepage}

\usepackage[normalem]{ulem}
\usepackage[ruled,vlined]{algorithm2e}

\providecommand{\keywords}[1]
{
  \small	
  \textbf{\textit{Keywords:}} #1
}

\definecolor{persimmon}{rgb}{0.93, 0.35, 0.0}

\newcommand{\coma}{\text{,}}
\newcommand{\rev}{``}

\title{Italian Business-to-Business Invoicing Data: \\ A Network Analysis}

\author[]{Valerio Astuti}
\author[]{Daniele Piras}

\affil[]{Bank of Italy\thanks{The views expressed in the paper are solely those of the authors and do not necessarily represent the views of the Bank of Italy.

\vspace{0.1cm}

\noindent For correspondence: valerio.astuti@bancaditalia.it, daniele.piras@bancaditalia.it.
}}

\date{}

\begin{document}

\maketitle

\begin{abstract}
  \noindent We present a comprehensive {description} of the network of Italian Business-to-Business commercial relationships, based on the universe of electronic invoices collected by the Italian Tax Office.
  The firm-to-firm detail of the data is exploited to describe the distribution of the numbers of buyers {(customers)} and sellers {(suppliers)} per firm, the centrality of each firm in the production network, and the average distance between firms. 
  We characterize for the first time the firm-to-firm network describing the Italian production system, and present its geographic and sectoral breakdown.
  The analysis reveals heavy tailed distributions for the numbers of buyers and sellers per firm, implying a scale-free structure of the network. 
  The distributions of centrality values display heavy tails as well, indicating a strong concentration of importance in a relatively small number of firms. 
  We estimate the tail exponents for all these distributions, finding in all cases lower exponents for downstream (buyer-side) distributions than for the upstream (seller-side) ones.
\end{abstract}

\keywords{Production networks, input-output analysis, firm-level data, complex networks, power-law distributions.}
%\JEL J{C61, C63, C83, E42, E58}

\clearpage
%\tableofcontents
%\clearpage

\section{Introduction}

The structure of the production network of a country is among the fundamental elements characterizing its economy, and many insights can be derived from its study.
{For example it determines how shocks propagate in the production system (\cite{acemoglu2010cascades}), it helps identify the most important firms in the economy (\cite{acemoglu2012network}), and it influences long-term growth (\cite{mcnerney2022production}).}
While traditionally most of the economic literature focuses on the study of aggregate quantities like gross domestic product, inflation or business cycle fluctuations, it has been relatively recently pointed out that some results cannot be {obtained without considering the micro structure of production networks}.
For this reason, a growing amount of attention is dedicated to the \emph{relationships} between firms, which can provide a microfoundation for a significant part of the aggregate fluctuations (see \cite{gabaix2011granular, acemoglu2012network, carvalho2014micro, baqaee2019macroeconomic}).
In this paper we contribute to the growing literature on production networks by {describing} for the first time the microstructure of the network of commercial exchanges between Italian firms, using a comprehensive dataset of firm-to-firm transactions. 
Our analysis reveals several key structural features of this network. 
We focus in particular on the distribution of the number of connections and various measures of centrality, offering a characterization of the production network which can be leveraged in future applications. 

The network structure of the Italian productive system is derived from the 2019 Italian electronic invoicing data. 
The introduction of electronic invoicing and its distribution through a centralized system, known as Sistema di Interscambio (managed by Agenzia delle Entrate, the Italian Tax Office), marks a transformative milestone in the digitization of Italy's fiscal and administrative framework.
Initially introduced in 2014, from the 1st of January 2019 electronic invoicing is mandatory, up to minor exceptions,\footnote{Firms with legal personality that are exempt from electronic invoicing, VAT associated documentary and accounting obligations are only agricultural producers with an annual turnover below 7,000 euro — of which at least two thirds derive from the sale of agricultural and fishery products (listed in Table A, Part I of Article 34, paragraph 6 of Presidential Decree 633/1972).} for both business-to-business and business-to-consumer transactions.
This extension not only reinforces Italy's commitment to fiscal transparency and the fight against tax evasion, but also offers invaluable opportunities for statistical analysis and informed policy-making.

The dataset we are provided with by Agenzia delle Entrate contains information on the gross amount of invoices issued by Italian entities to both firms and final consumers on a quarterly basis. 
For firms with legal personality we add information on the main sector of economic activity and the province in which the head office is located.
The dataset records invoices as they are issued, so that it maps, for each issuer, its customer base (buyers), namely its downstream connections in the network of commercial exchanges. 
By focusing on each customer, it is then straightforward to map all the issued invoices, so to also reconstruct its supplier (seller) base, namely its upstream connections.

Similar datasets have been used in several studies for other countries, but often with limitations on the data collection procedure. In particular, it is common to have a reporting threshold, below which the transactions are not reported. 
In our dataset no such limitation is present, and we have a complete coverage of commercial transactions.
To represent this dataset of commercial exchanges as a network, firms are represented as nodes, and each link represents the presence (and optionally the magnitude) of the relationship between two firms when one has issued invoice(s) to the other. 
We study both the existence of these connectivity relations - taking into consideration only whether two firms are connected - and their strength, measured by weighting relations by the yearly total amount exchanged between the firms.
The resulting network is among the largest studied in the literature, with more than 1.6 million nodes and almost 60 million links.

We observe power-law distributions for both the number of buyers (customers) and sellers (suppliers) of each firm, suggesting the underlying network exhibits scale-free properties. 
Even though we find a strong positive correlation between these two numbers, the right tail of the distribution of the number of buyers is much heavier than the corresponding one for the number of sellers.
{Although our analysis is static, this aligns with} the economic hypothesis that a firm grows by {preferentially} increasing the number of buyers (extensive margin) and strengthening the already existing connections with its suppliers (intensive margin) (see \cite{mizuno2014structure, oberfield2018theory, carvalho2019production, afrouzi2020growing, bernard2022origins}).
Scale-free networks are widely studied in the literature, and examples cover scientific citations (\cite{redner1998popular}), protein interactions (\cite{jeong2001lethality}), and the World Wide Web (\cite{albert1999diameter}). 
{See also \cite{eikmeier2017revisiting} for a review of power-law distribution in real-world networks.}
{Like many other scale-free networks, the Italian production network also presents} the \emph{small-world} property: even though the overwhelming majority of nodes has a {small} number of neighbors, from almost every node it is possible to explore the whole network in few steps (on average less than 4, and almost all pairs of nodes can be connected in less than 6 steps).
This effect is caused by the presence of few nodes with a large number of connections, which can act as connecting hubs between less connected nodes.
In turn this phenomenon implies that the network connectivity structure is strongly dependent on few firms, and is critically vulnerable to targeted shocks to these firms, although more resilient to diffused  shocks, since the probability of randomly hitting these nodes is very low.
The network is negatively assortative, that is, firms with above-average number of connections tend to be linked to firms with below-average number of connections, in agreement with previous findings in the literature (\cite{newman2003assortativity, fujiwara2010large, bacilieri2026firm}).

Finally, for firms within the network, we compute and describe the properties of five measures of centrality.
These measures are commonly used to assess the {relevance} of a firm within the value chain, and to estimate the effects of various types of shocks on the network. 
For these measures we give a sectoral and geographical breakdown, revealing that buyer firms in the service sector exhibit higher centrality when considering the number of connections, whereas utility companies emerge as the most influential nodes in terms of the monetary volume exchanged.
From a geographical perspective, we find a pronounced difference between the provinces in the central and northern regions of Italy and the rest of the country, both in terms of number of connections and network centrality. 
More specifically, alongside major provinces - which unsurprisingly retain considerable significance - the analysis identifies additional smaller provinces within the central and northern {regions} distinguished by the presence of relevant firms operating in the construction, manufacturing, and real estate sectors.

\subsection{Related literature}
The study of production systems from a network perspective was initiated in the economics literature by Leontief (see \cite{leontief1991economy, leontief1941structure}).
In this branch of literature the structure of the production network is depicted as consisting of a representative firm for each sector composing the economy, and the focus of the study is the input-output network between these representative firms.
More recent applications include an analysis of the importance of wealthy families in Renaissance Florence (\cite{padgett1993robust}), the role of the network structure in the diffusion of information (\cite{banerjee2013diffusion, banerjee2014gossip}), and the transmission of microeconomic shocks to the aggregate economy (\cite{acemoglu2010cascades}).
The importance of the microstructure of production networks was pointed out also in \cite{carvalho2014micro}, in which the author shows that idiosyncratic firm-level shocks can generate aggregate fluctuations when propagated through input-output linkages. 
These findings challenged the conventional belief that microeconomic shocks would cancel out in the aggregate, highlighting instead how network structures can amplify initially localized perturbations.

Recent innovations in data collection processes have expanded the range of possibilities for characterizing the underlying network structure of many systems.
In particular, the theory of networks has been applied to the World Wide Web (see \cite{albert1999diameter}), to the analysis of power grids (see \cite{pagani2013power} for a review), and to the study of human mobility (\cite{gonzalez2008understanding}).
The increasing availability of micro-data allowed at the same time to characterize real-world production networks with unprecedented detail, often at level of firm's transaction.
In \cite{bernard2019production} the geographic dimension of production networks and its role in the marginal costs of production was studied using Japanese transaction data.
\cite{dhyne2021trade} used Belgian data on firm-to-firm sales to assess the dependence on foreign markets and the sensitivity to trade shocks.
Micro-data are used also in \cite{criscuolo2024estonia} to characterize the network structure of the Estonian production system, and a recent review of similar works can be found in \cite{bacilieri2026firm}.

The relation between a node centrality and its economic importance is explored in \cite{acemoglu2012network}, while in \cite{banerjee2013diffusion, banerjee2014gossip} the authors analyzed the role of network centrality in information diffusion and financial participation.
The most direct application of the concept of centrality is in the analysis of shock propagation: in \cite{acemoglu2010cascades, acemoglu2013network} the importance of the network structure and in particular of the most central nodes in the amplification of economic shocks is analyzed.
\cite{haldane2011systemic} and \cite{cont2013network} investigate the central role of network structures in the propagation of shocks in financial systems.
Finally, see \cite{watts1998collective}, \cite{albert2000error} and \cite{motter2002cascade} for the analysis of shock propagation and the role of central nodes in more general networks. 
Perhaps the most famous application of node centrality has been the introduction by Google of the PageRank algorithm to rank web pages (\cite{brin1998anatomy, page1999pagerank}).
This measure is equivalent to the probability for a user randomly clicking on hyperlinks to land on a given page, and it proved extremely efficient in the classification of web pages.
There are many centrality measures available in the literature (see for example \cite{das2018study} for a review), and each of them captures a different notion of importance of the node considered.  

The rest of the paper is organized as follows: in Section \ref{sec:FromDataToNetwork} the basic features of the dataset and the associated network are described. Section \ref{sec:degree_distribution} focuses on the distribution of the connections in the network, and Section \ref{sec:global_centrality} contains a comprehensive study of the centrality measures of nodes in the network.

\section{From Transaction Data to Network}
\label{sec:FromDataToNetwork}

The dataset concerns the electronic invoices issued in Italy in 2019.
The records of the dataset are constructed as quarterly aggregation of invoices exchanged between a seller and a buyer, reporting both the number of invoices exchanged and the aggregated gross transaction value. 
The dataset also includes additional information, but we introduce only those variables that are directly relevant to our analyses.
No threshold is imposed to register a transaction before the quarterly aggregation.
The invoices {in our dataset} refer to transactions between Italian entities, which can be of one of following types:
\begin{itemize}
    \setlength\itemsep{0px}
    \item[(\textbf{B})] Business: refers to firms with legal personality;
    \item[(\textbf{G})] Government: refers to public sector organizations;
    \item[(\textbf{D})] Natural Person: refers to firms that, by their legal nature, must be treated as natural persons, e.g. sole proprietorship (this category appears as both seller and buyer, in aggregated form); 
    \item[(\textbf{C})] Consumer: represents the final consumers (this category appears only as buyer, in aggregated form).
\end{itemize}

\begin{figure}[H]
\centering
  \includegraphics[width=1\linewidth]{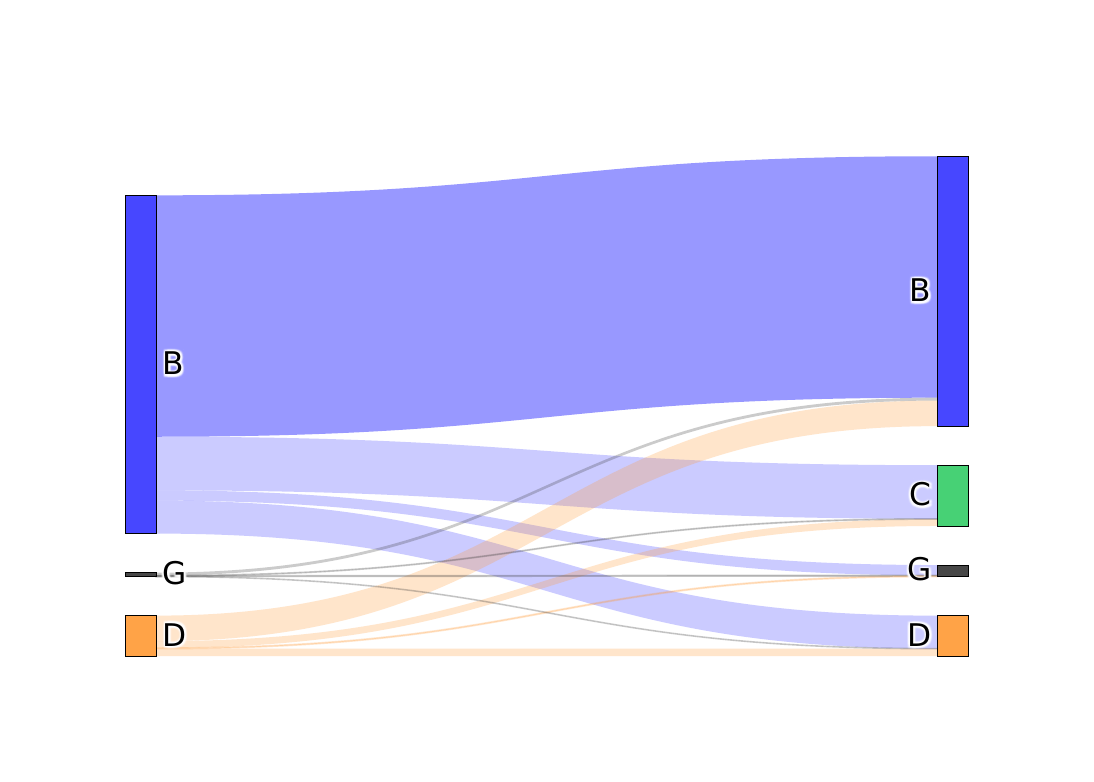}
  \caption{Flows between categories in terms of aggregated gross transaction amount. Bars' height represents the total gross transaction amount of invoices by type of entity. Our analysis will focus on the transactions where both the seller and buyer are firms of type B, the flow in darker blue.}
  \label{fig:categories_flows}
\end{figure}

No individual transactions for natural persons (type D) or consumers (type C) appear in the dataset: for type C, data are aggregated - in each quarter - across all consumers; for type D, data are aggregated by 2-digit sector of economic activity (both for invoices received and invoices issued). For firms of type B and G, the dataset also reports the economic sector of activity (up to 4 digits when available) and the province where the {head} 
office of the firm is located.
The transformation of transactional data into a network involves identifying the entities, which will be the nodes of the network, and the relationships between them, that will be used to define the links connecting related nodes. In our case, this 
is relatively straightforward. 
Firms will be the nodes, and a link will exist between two nodes if one firm has issued at least one invoice to the other. {Since the roles of the issuer (seller) and the receiver (buyer) of an invoice are not symmetrical, the links have a specific direction, making the resulting network oriented (or directed)}.
We will only consider transactions where both firms are of type B (corresponding to the darker blue flow in Figure \ref{fig:categories_flows}), which we further aggregate on an annual basis, so that it will be possible to assign a weight to each link, equal to the total yearly amount of the invoices issued by the seller to the buyer.

We also remove self-loops, corresponding to records where the same firm is both seller and buyer. 
These documents are issued primarily for fiscal purposes and do not represent an actual production relation or a transactional linkage between buyer and seller.
For a comparison between the original dataset and the one restricted to transactions between type B firms, {which covers 62.44\% of the overall amount of gross transactions in 2019}, see Table \ref{tab:stats_dataset}. 
\begin{table}[H]
    \centering
    \small
    \begin{tabular}{lccc}
    %\toprule
     & Complete Dataset & B $\rightarrow$ B & \% \\
    \midrule
    Total gross amount (euro billion) &  2,705.61 & 1,689.36 &  62.44 \% \\
    Number of entities (million)      & 1.74      & 1.64     &  94.18 \% \\
    Number of rows (million)          & 163.15    & 59.07     &  36.21 \% \\
    \bottomrule
    \end{tabular}
    \caption{Comparison of the complete dataset with the business-to-business subset of transactions.}
    \label{tab:stats_dataset}
\end{table}

The resulting network consists of 1,642,731 nodes, sharing 59,071,160 {oriented} edges, corresponding to a total amount of gross invoices of more than 1,689 {billion} euro.
The network can be thus identified by a pair of sets $(\mathcal N, \mathcal E) $, where $\mathcal N $ is the set of nodes $\mathcal N = \{ u \,:\,u$ is a firm of type B and is a seller or buyer in a transaction$\}$, and $\mathcal E$ is the set of links, namely the set of ordered pairs of nodes connected by a link, $\mathcal E = \{(u,v)\,:\,u,v\in \mathcal N$ and $u \text{ issued an invoice to } v \}$. We also define the set of weights of the links $\mathcal W = \{w_{uv} \,:\,(u,v) \in \mathcal E \}$, and $w_{uv} $ is the total yearly amount of invoices that $u$ issued to $v$. 
We will refer to the number of links as $E$ and to the number of nodes as $N$, so that $|\mathcal E | = E$ and $|\mathcal N | = N$. As already mentioned $N = 1\coma 642\coma 731$ and $E=59\coma 071\coma 160$.
 
%\subsection{A Global Picture of the Network}
\subsection{A Summary of the Network Structure}
\label{sec:density_and_mean_degree} 
A first insight on the structure of the network is given by {its} global properties, such as network density and mean degree. 
%Entarmbe queste misure sono indicatori che permettono di avere una prima idea globale della vulnerabilità strutturale e la resilienza del sistema agli shock.
%La densità, dà una misura del grado di quanto sono connesse tra loro le imprese: maggiore è l'interconnessione, maggiore è la facilità con cui si possono propagare shocks across the economy, amplifying systemic risk.
Both measures serve as indicators that provide a preliminary understanding of the structural vulnerability of the system and its resilience to shocks.
Specifically, network density quantifies the extent of interconnections among firms: the higher the density the greater the inter-firm connectivity.
On one hand this implies a greater level of diversification in the network, on the other it can facilitate the propagation of shocks throughout the economy, thereby amplifying systemic risk.
%The density of a network gives a measure of how interconnected its nodes are\mi{cercare di spiegare perché può essere utile la densità}.
%{
%The higher the density value the more inter-firm connections are present, which can facilitate the transmission of both positive and negative shocks throughout the network. Conversely, a lower density suggests a more sparsely connected structure, which may present, for instance, clusters of firms with many internal links, with relatively few connections with the rest of the network.
%}
The definition of density compares the number of links in the actual network with the number of links in a complete network with the same number of nodes.
% \begin{figure}[H]
% \centering
%   \includegraphics[width=0.5\linewidth]{plots/density_example.png}
% \caption{density example}
%   \label{fig:density_example}
% \end{figure}
A complete directed network (a network where all nodes are pairwise connected) with $n$ nodes has exactly $n\cdot(n-1)$ links, since each of the $n$ nodes is connected to the remaining $n-1$ and we do not consider self-loops.
The density of a network $\rho$, representing the share of actual connections compared to this theoretical maximum, is thus defined as:
\begin{equation}
\label{eq:def_density}
    \rho = \frac{E}{N \cdot(N -1)}.    
\end{equation}
Since $N$ is approximately 1.6 million, the theoretical maximum number of possible links is close to 2,700 billion. The actual number of observed links is approximately 60 million, resulting in a network density of $0.0022 \%$. 
%This value aligns with findings reported in the literature for production networks of other countries (e.g., see \cite{dhyne2021trade} for Belgium and  \cite{criscuolo2024estonia} for Estonia) \al{Anticipare risultato introduzione. Mettere richiamo all'appendice in cui discutiamo effetti threshold.}.

To gain further insight into the network density, one may consider the average number of links per node, also called mean degree of the network.
It is defined as:
\begin{equation}\label{eq:mean_deg_lafond}
    \overline{\text{k}} = \frac{1}{N} \sum_{i = 1}^{N} \text{k}_{in}^{(i)} =
    \frac{1}{N} \sum_{j = 1}^{N} \text{k}_{out}^{(j)}
\end{equation}
where $\text{k}_{in}^{(i)}$ (resp. $\text{k}_{out}^{(j)}$) is the number of in-going (resp. out-going) links of node $i$ (resp. $j$). Since we are considering a closed system (we do not consider links going outside the Italian production network), we notice that the sum of the incoming degrees is exactly equal to the sum of the out-going degrees, and both are equal to the total number of links. 
Equation \eqref{eq:mean_deg_lafond} can thus be written as:
\begin{equation}\label{eq:mean_deg_lafond_E_N}
\overline{\text{k}} = \frac{E}{N}.
\end{equation}
For our network we find $\overline{\text{k}} = 35.96${, meaning that on average each firm in the network is connected to around other 36 firms (via either downstream or upstream links).}
Definition \eqref{eq:mean_deg_lafond} (and \eqref{eq:mean_deg_lafond_E_N}) treats on the same ground both in- and out-going connections. As we will see in Section \ref{sec:degree_distribution}, the two directions can also be studied independently.
To provide a basis for comparison with other countries, we present in Table \ref{tab:density_average_degree} a selection of datasets identified in the literature (see \cite{bacilieri2026firm} for Ecuador and Hungary, \cite{dhyne2016three} for Belgium and  \cite{criscuolo2024estonia} for Estonia) whose data collection methodologies and structural constraints are closely aligned with ours. 
We also selected these datasets based on the availability of information on both the number of firms and the number of links, allowing the calculation of both network density and average node degree.
\begin{table}[h]
    \centering
    \small
    \begin{tabular}{lcrrrrr}
    \multicolumn{1}{c}{Country} & Year&\makecell{Transaction \\  threshold (euro)} & \multicolumn{1}{c}{$N$} & \multicolumn{1}{c}{$E$} & \multicolumn{1}{c}{$\overline{\text{k}}$} & \multicolumn{1}{c}{
    \makecell{$\rho$ \\ (percentage)}
    }\\
    \midrule
        \textbf{Italy}	  & \textbf{2019}&\textbf{0.00}  &\textbf{1,642,731}  &\textbf{59,071,160}  &\textbf{35.96} &	\textbf{0.0022}\\
        Ecuador & 2007 & 0.00 & 56,058 & 1,873,023 & 33.41 & 0.0596\\
         & 2011 & 0.00 & 72,200 & 2,774,900 & 38.43 & 0.0532\\
         & 2015 & 0.00 & 86,345 & 3,372,929 & 39.06 & 0.0452\\
        %Hungary & 2015 & 9.260,41 & 119.469 & 356.788 & 2,99 & 0,0025\\
        % & 2019 & 926,04 & 313.117 & 2.116.912 & 6,76 & 0,0022\\
        Hungary & 2021 & 2.50 & 493,616 & 18,710,235 &37.90 & 0.0077\\
        Estonia & 2019 & 0.00 & 103,742 & 856,508 & 8.26 & 0.0080\\
        %Belgium & 2002 & 250,00 & 88.301 & 4.187.000 & 47,42 & 0,0537\\
        % & 2007 & 250,00 & 95.941 & 4.848.000 & 50,53 & 0,0527\\
        % & 2012 & 250,00 & 98.745 & 5.026.000 & 50,90 & 0,0515\\
        % & 2012 & 250,00 & 79.788 & 3.505.207 & 43,93 & 0,0551\\
        % & 2012 & 250,00 & 250.000 & 8.700.000 & 34,80 &	0,0139\\
        Belgium & 2014 & 250.00 & 321,824 & 8,900,000 & 27.65 & 0.0086\\
        % Belgium &  & 250,00 &  &  &  & \\
    \bottomrule
    \end{tabular}
    \caption{Comparison of network density and average number of links per node between Italy and other similar datasets.}
    \label{tab:density_average_degree}
\end{table}
% \sout{A final remark concerning the comparison of these global quantities across networks
% from different countries pertains to data collection methods, that may vary and be subject to some constraints. A discussion on this matter is provided in Appendix \ref{sec:appendix_rep_thres}.}\mi{Aggiungere frase che anticipa appendice B.}
A final remark concerning the comparison of these quantities across networks from different countries pertains to data collection methods, that may vary among datasets and be subject to different constraints. 
The presence of such methodological variations makes direct comparisons not straightforward: similar values might result from offsetting effects of contrasting factors rather than genuine structural analogy between the systems represented by the respective datasets. 
We provide a discussion on this matter in Appendix \ref{sec:appendix_rep_thres}, analyzing the dataset behavior through the systematic application of synthetic limitations.

%\subsection{Density and Mean Degree: Sectoral and Geographical Breakdown}
%\lb{Sectoral and Geographical Breakdown of global network quantities - per far capire in maniera meno tecnica cosa contiene il paragrafo. Stiamo parlando di lettori distratti!}
\subsection{Sectoral and Geographical Breakdown of Global Network Quantities}
\label{sec:density_and_mean_degree_breakdown} 
%\subsubsection{Network Sectoral Statistics}
%\subsubsection{Network Geographical Statistics}
%The nodes of the network represent Italian companies that have issued or received an electronic invoice. It is thus of interest to examine how the values represented by the network fluctuate when grouped by geographic area or sector.
%Specifically, the focus is on the distribution of the number of enterprises, cumulated turnover, as well as the average and median number of incoming and out-going connections of nodes, when grouping nodes by sector (represented by the 2-digit ATECO code) and by territory (represented by the province of the registered office).
%The results are reported below in two subsections, one for the ATECO and another for the province.
Since we included information on the main economic sector of activity and the province of the head office, it is of interest to examine how network density and mean degree behave when these two additional dimensions are taken into account and to compare sectoral and geographical results with the values observed for the entire network. 
We assigned each  {firm} to its corresponding {\rev one-letter" sector} of activity, according to the NACE classification\footnote{\rev Nomenclature statistique des Activités économiques dans la Communauté Européenne", the European industry standard classification system used for statistical analysis and comparison of economic activities.} (see Appendix \ref{sec:app_tabella_settori} for the full mapping), and employed this classification to improve the readability of the plots.

\begin{figure}[H]
    \centering
    \includegraphics[width=1\linewidth]{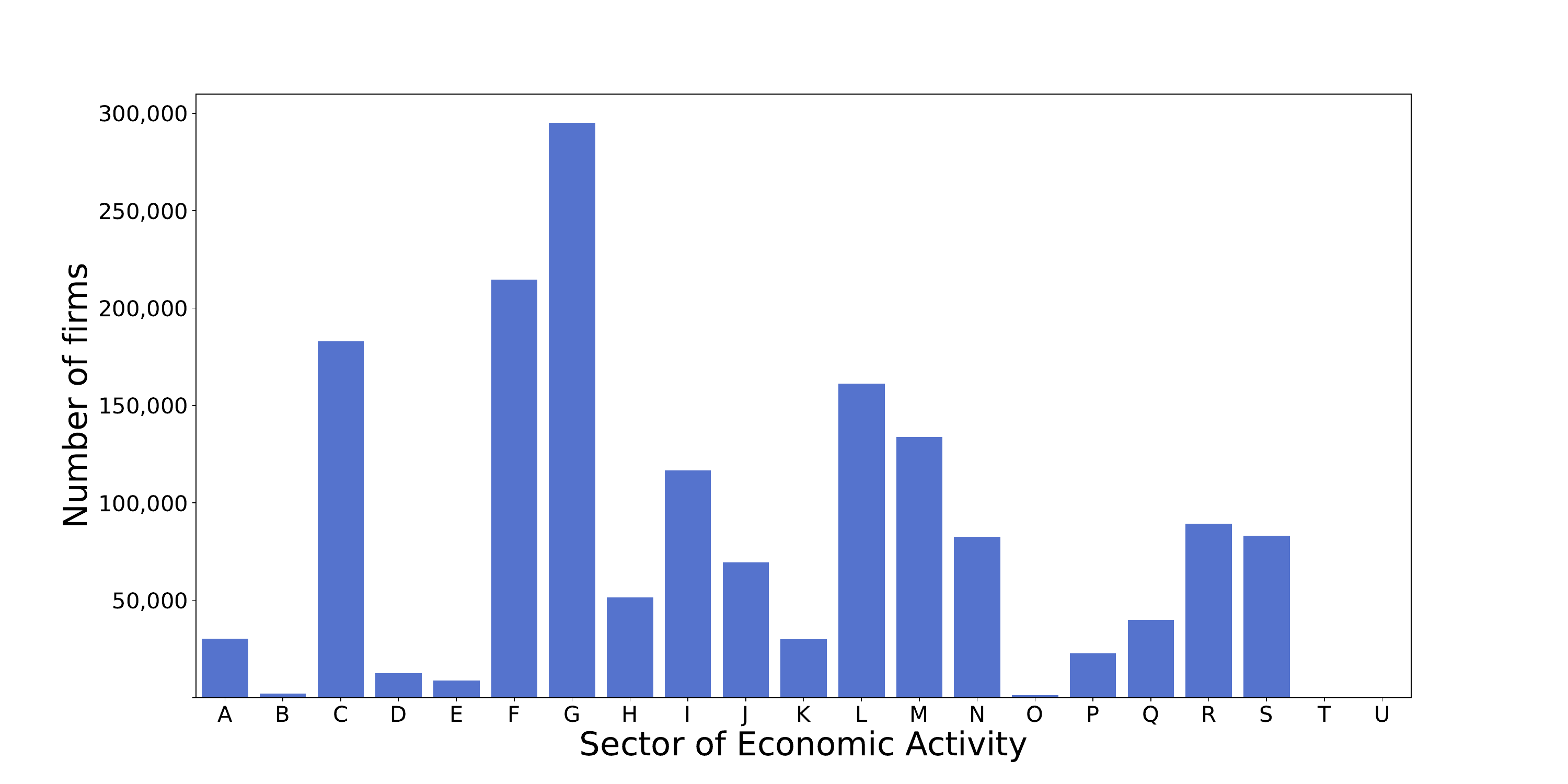}
    \caption{Number of firms per {sector} of economic activity. See Appendix \ref{sec:app_tabella_settori} for the full mapping between one-letter and 2-digit sectors. Sector U is the least represented, with fewer than 100 firms in the dataset.}
    \label{fig:firms_per_category_power}
\end{figure}

In terms of sectoral distribution (see Figure \ref{fig:firms_per_category_power}), wholesale and retail trade (sector G) accounts for the largest number of firms, with almost 40\% more firms than the second-largest sector, construction (F). The third position is held by firms operating in the manufacturing sector (sector C).
%al{inserire tavola con traduzione lettere settori} \{già messa in appendice e riportato in nota 2}.

\begin{figure}[H]
    \centering
    \includegraphics[width=1\linewidth]{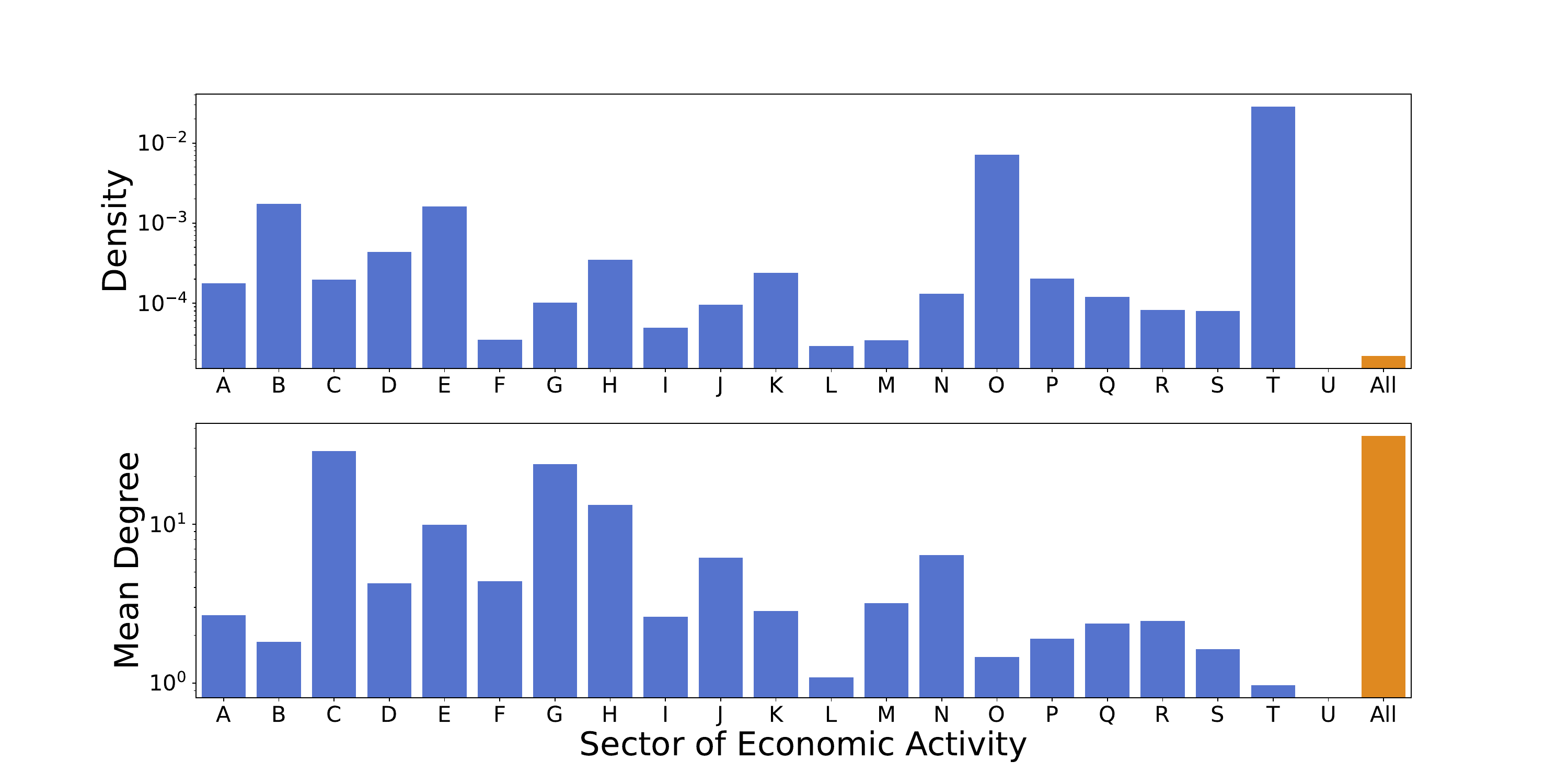}
    \caption{Density (upper plot) and mean degree (lower plot) of the networks of transactions within the same sector of economic activity, compared to the value for the whole network (reported in orange).}
    \label{fig:density_mean_degree_within_category}
\end{figure}

\begin{figure}[H]
\centering
  \includegraphics[width=1\linewidth]{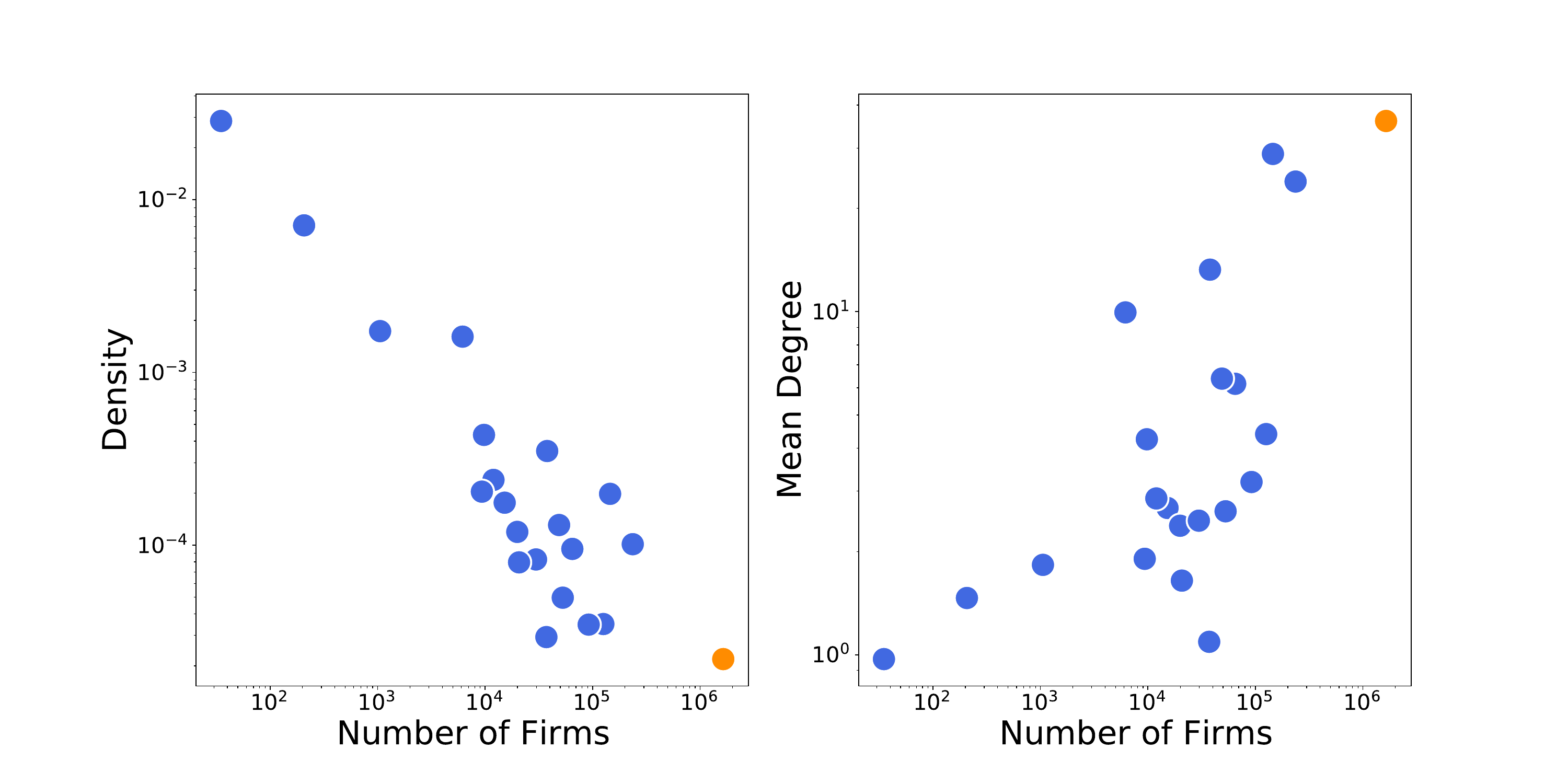}
  \caption{Plot of density (left hand side) and mean degree (right hand side) as a function of the number of nodes in the sub-networks of transactions within the same sector of economic activity.}
  \label{fig:density_mean_degree_wrt_Ncategory}
\end{figure}

To provide a concrete illustration of the concepts of density and mean degree, we compare the behavior observed within the sub-networks defined by transactions between firms in the same sector of economic activity to those observed in the complete network. In the sub-networks we observe a higher density and a lower mean degree (see Figure \ref{fig:density_mean_degree_within_category}). 
% \sout{Category U (extraterritorial organizations and entities) consistently shows no within-category links.}
This behavior becomes even more intriguing when considering the dependency on the number of firms in each sub-network (see Figure \ref{fig:density_mean_degree_wrt_Ncategory}).
From the definition of density (Equation \eqref{eq:def_density}) and mean degree (Equation \eqref{eq:mean_deg_lafond_E_N}), we can see how both these quantities are related to the number of firms and the number of links: the mean degree is the ratio between the number of links $E$ and the number of firms $N$, while the density is the ratio between the number of links $E$ and (approximately) the number of firms squared $N^2$.
%$$
%\text{Mean Degree } = \frac{E}{N} \qquad \text{Density } = \frac{E}{N\cdot(N-1)},
%$$
The fact that the density becomes smaller and the mean degree becomes larger as the number of firms grows suggests that  $E$ is proportional to $N^\gamma$ with $1<\gamma<2$, that is the number of links grows with respect to the number of firms as $N^\gamma$. This expectation is confirmed by a regression of the number of links in each sector sub-network with respect to the number of firms. The estimated coefficient\footnote{Here and in the rest of the paper we indicate in parentheses the standard error of the estimated coefficients.} is $\gamma = 1.26\, (0.07)$.

\begin{figure}[H]
\centering
\begin{minipage}{.5\textwidth}
  \centering
  \includegraphics[width=1\linewidth]{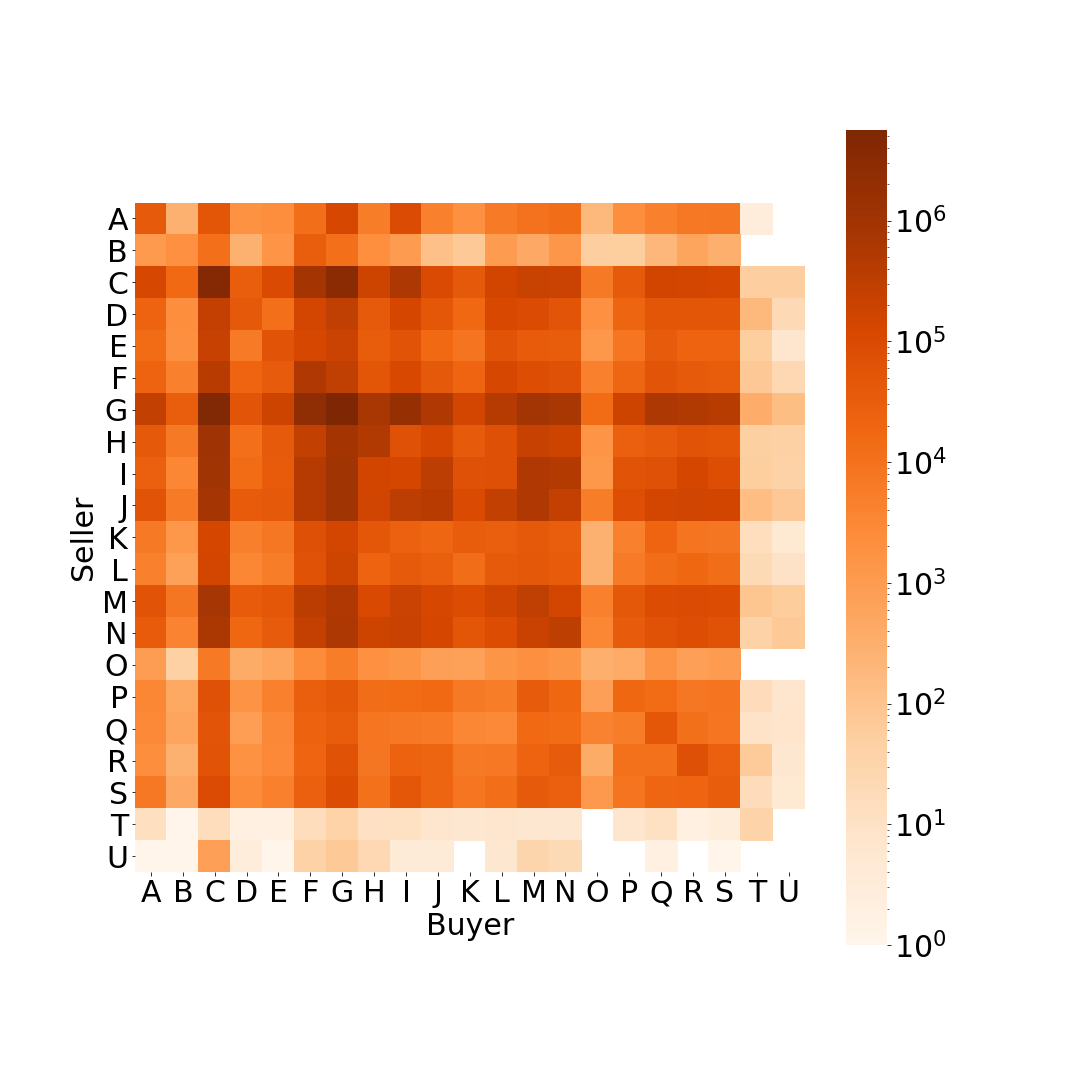}
\end{minipage}%
\begin{minipage}{.5\textwidth}
  \centering
 \includegraphics[width=1\linewidth]{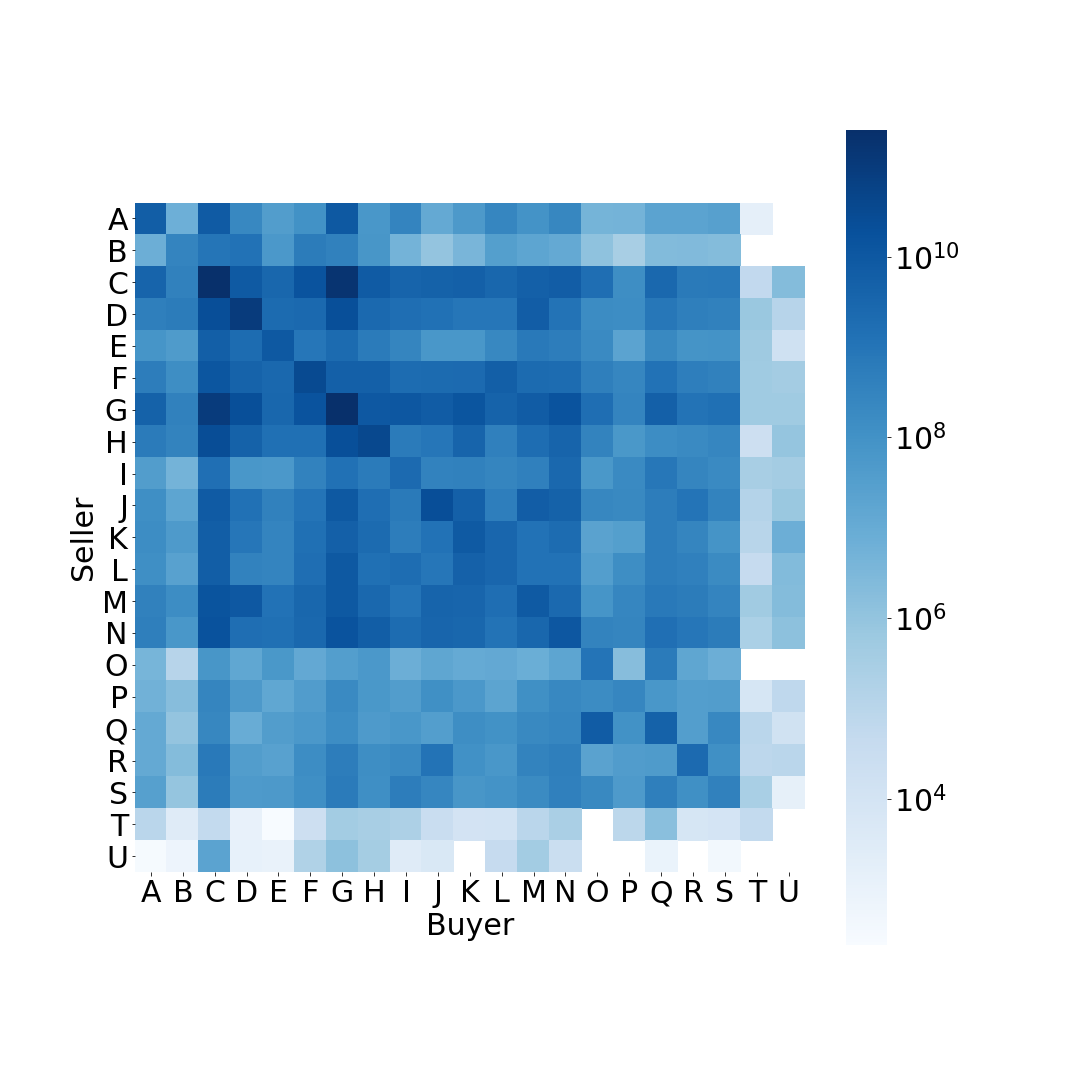}
\end{minipage}
\caption{Distribution of number of connections (left hand side) and total gross amount of the invoices (right hand side) between sectors.}
  \label{fig:category_heatmap_links_imponibile_totale}
\end{figure}

{We now go back to the full network to analyze} the invoice exchanges between {sectors} of economic  activity, visualized in the heatmaps of Figure \ref{fig:category_heatmap_links_imponibile_totale} with the economic activity of the seller in the rows and that of the buyer in the columns.

%The heatmaps of Figure  show the trade exchanges between the different Categories of Economic activity, with the Category of Economic activity of the Seller category in the rows and the one of the buyer in the columns. 

\begin{figure}[H]
\centering
\begin{minipage}{.5\textwidth}
  \centering
  \includegraphics[width=1\linewidth]{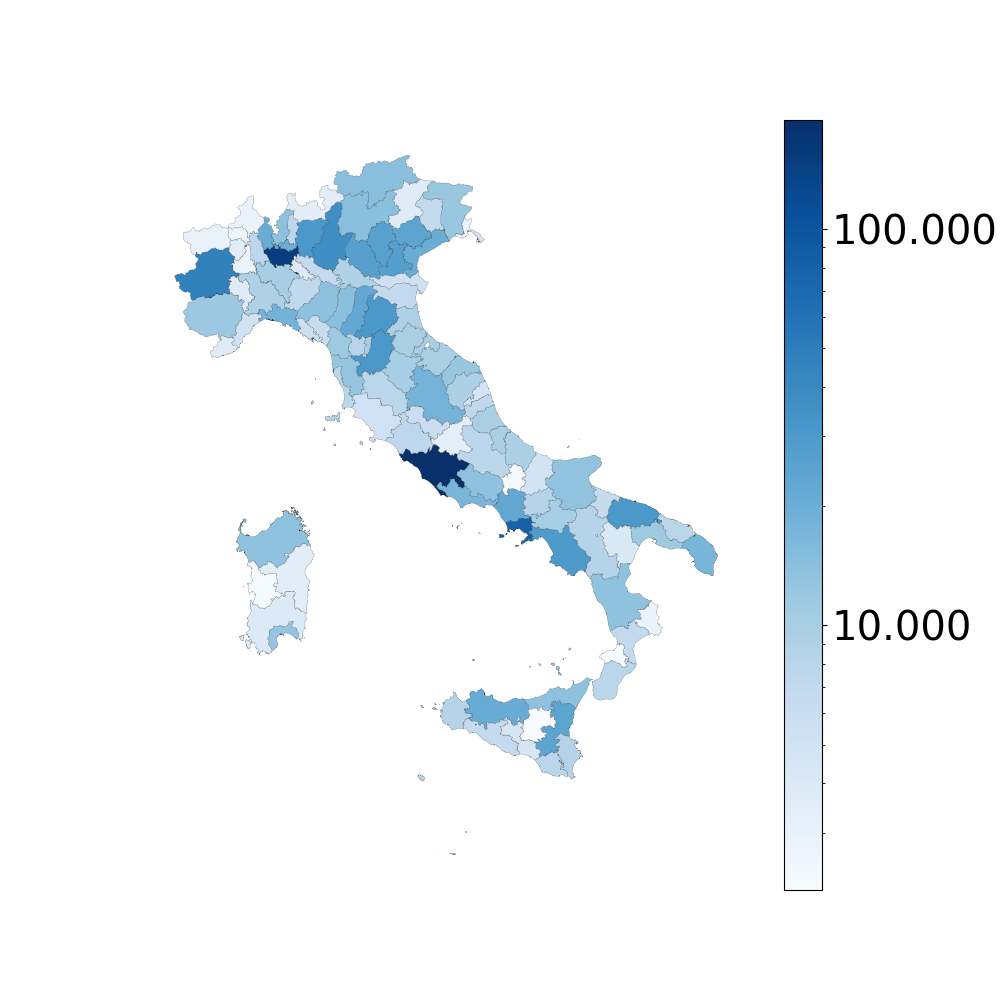}
\end{minipage}%
\begin{minipage}{.5\textwidth}
  \centering
  \includegraphics[width=1\linewidth]{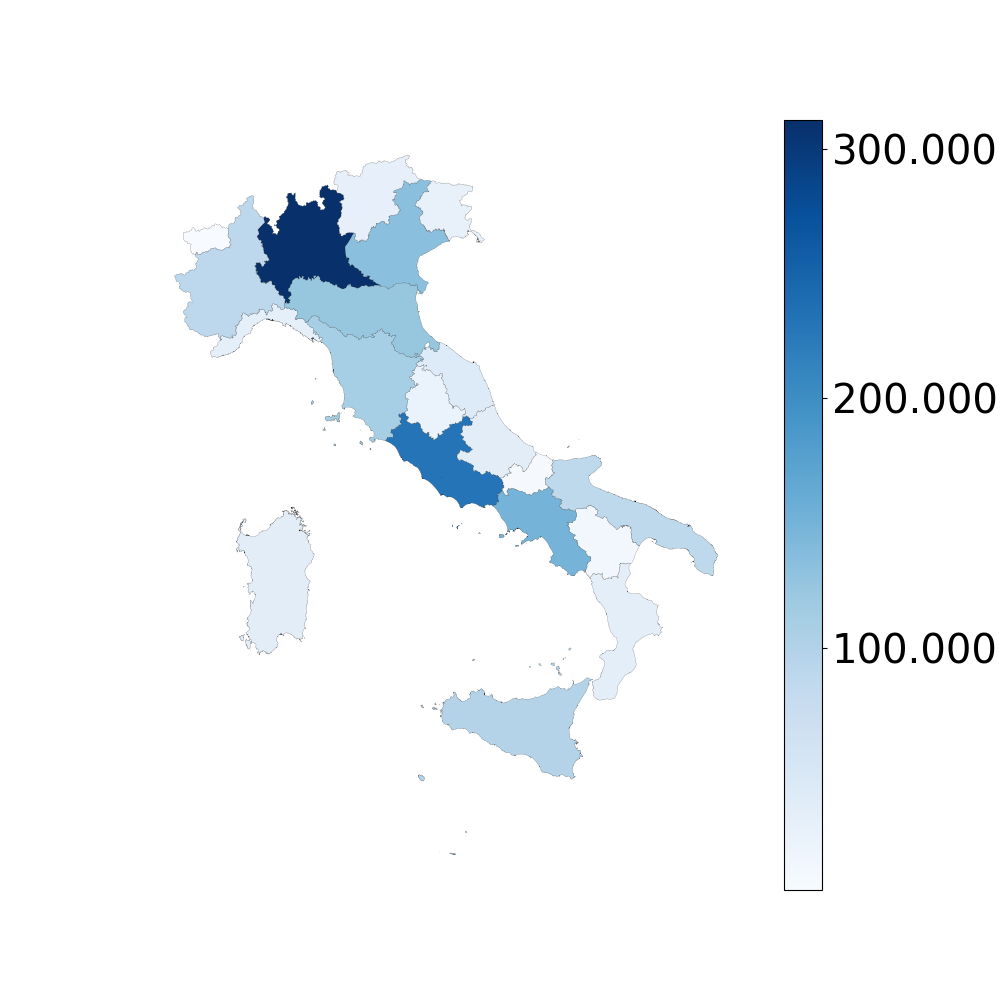}
\end{minipage}
\caption{Distribution of Italian firms by province (left hand side) and by region (right hand side) of registered office.}
  \label{fig:firms_per_region_provincie}
\end{figure}

\begin{figure}[H]
\centering
\begin{minipage}{.5\textwidth}
  \centering
  \includegraphics[width=1\linewidth]{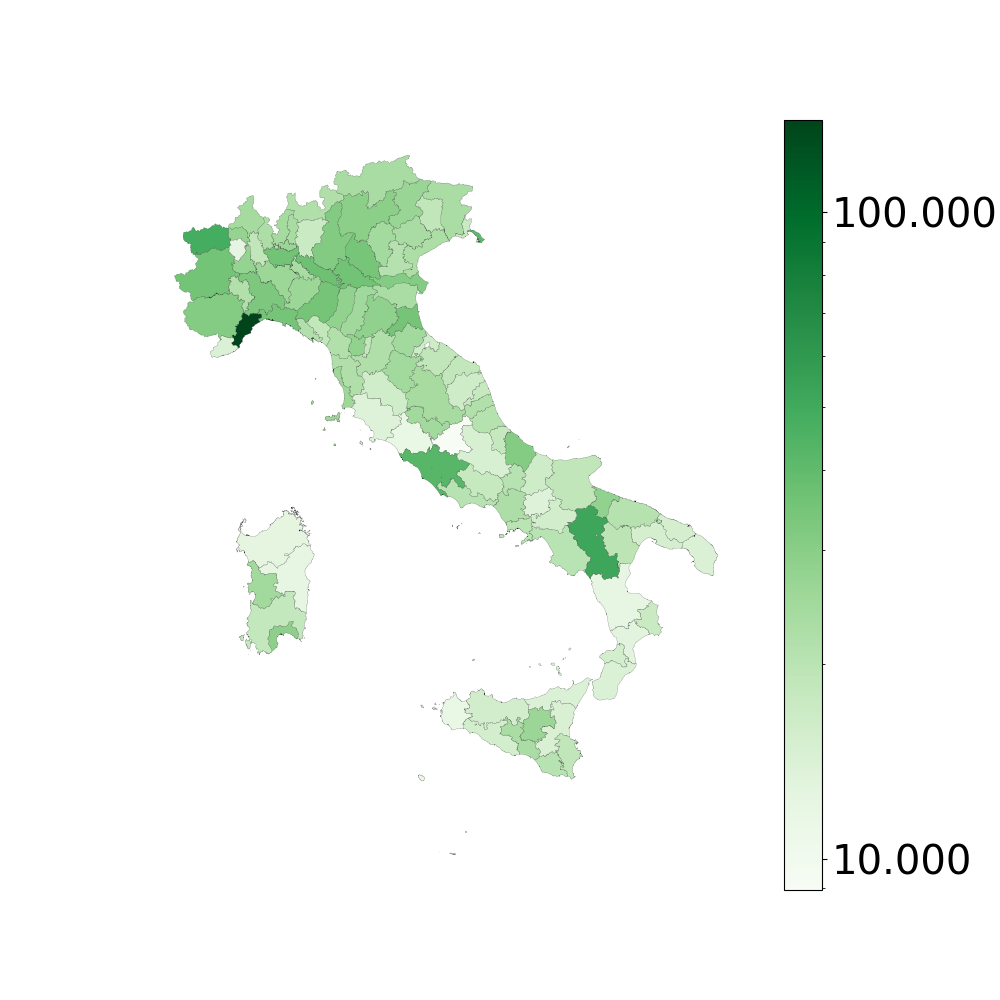}
\end{minipage}%
\begin{minipage}{.5\textwidth}
  \centering
  \includegraphics[width=1\linewidth]{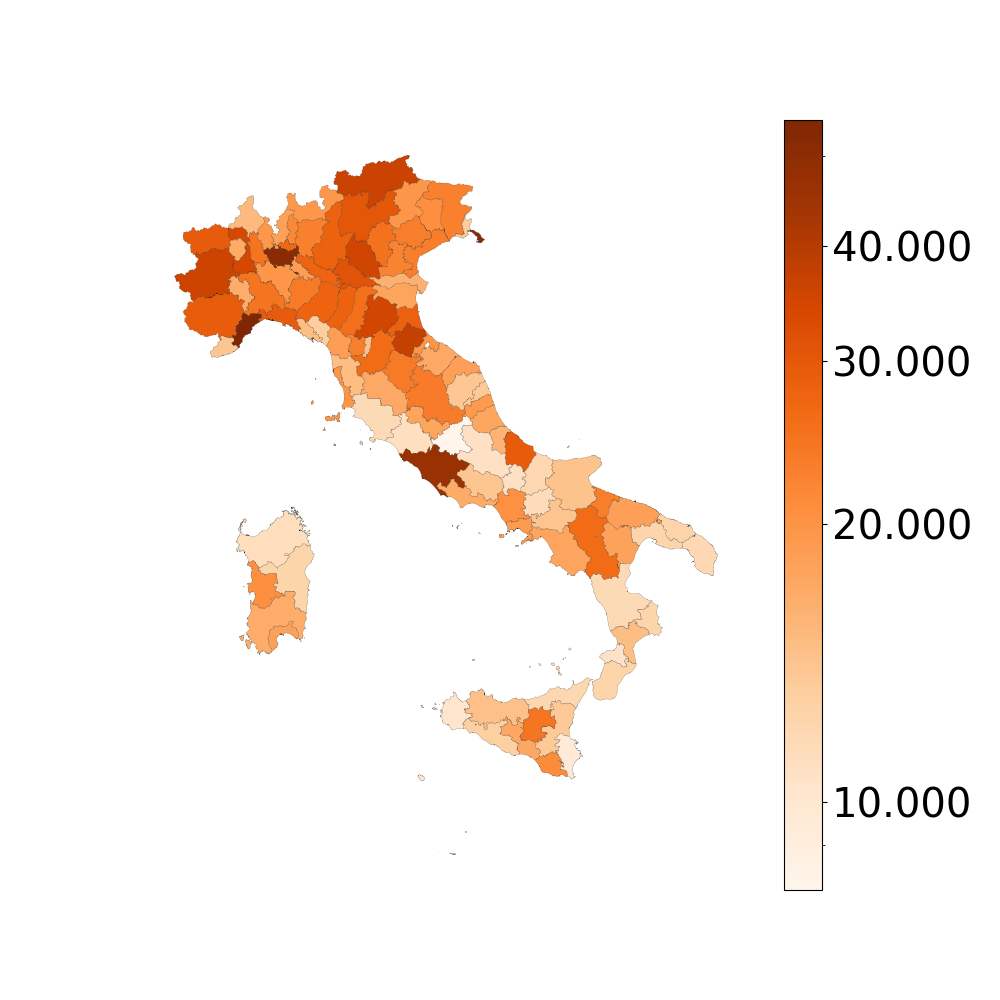}
\end{minipage}
\caption{Average gross invoice amount by province of the seller (left hand side) and buyer (right hand side).} %\va{Cambiare scala to plain numbers? (ved. figura 3)}}{FATTO}
  \label{fig:imponibile_medio_cedente_cessionario}
\end{figure}

Both plots show how the sectors with the most interactions are G (wholesale and retail trade; repair of motor vehicles and motorcycles) and C (manufacturing): together they account for more than 30\% of the links and more than 45\% of the total gross amount of the invoices. Looking at intra-sector relations, it is interesting to note how the total gross amount is the largest in invoices within sector C, while it is not surprising that the largest number of connections are within sector G, the sector with the highest number of firms. 
%\{CHECK, il quadratino più scuro sembra quello C-C}
% \begin{table}[H]
%     \centering
%     \small
%     \begin{tabular}{ccrcrc}
%     \multicolumn{1}{c}{Supplier} & \multicolumn{1}{c}{Customer}& \multicolumn{1}{c}{Links} & \multicolumn{1}{c}{Rank} & \multicolumn{1}{c}{Invoices} &\multicolumn{1}{c}{Rank}  \\
%     \midrule
%         C & C &4.193.815&$3^{rd}$&254.885.034.504,89&$1^{st}$\\
%         G & G &5.660.987&$1^{st}$&235.583.992.574,39&$2^{nd}$\\
%         G & C &3.263.865&$4^{th}$&182.332.668.510,82&$3^{rd}$\\
%         C & G &4.861.881&$2^{nd}$&97.975.772.130,30 &$4^{nd}$\\
%     \bottomrule
%     \end{tabular}
%     \caption{\{questa tabella può anche cadere}}
%     \label{tab:rank_one_letter_sector}
% \end{table}
% C 	G 	97975772130.30 	
% G 	C 	182332668510.82 
% G 	G 	235583992574.39
% C 	C 	254885034504.89
% G 	C 	3263865
% C 	C 	4193815
% C 	G 	 	4861881
% G 	G 		5660987
A geographical analysis of the distribution of Italian firms (see Figure \ref{fig:firms_per_region_provincie}) shows that they are predominantly concentrated in the northern regions of the country, although the province with the highest number of firms is Rome (RM). 
In terms of the number of registered companies, Lombardy holds the leading position, largely driven by the provinces of Milan (MI) - ranked second among the provinces - together with Bergamo (BG) and Brescia (BS). 
Lazio ranks second among the regions, primarily due to the presence of Rome (RM). 
Campania, located in the southern part of Italy, ranks third, with its position mainly attributed to the presence of Naples (NA), the third province in terms of company numbers.
Finally, we analyzed the invoices issued and received by province. {Since the} maps of total gross invoice amount by province of seller and buyer %(Figure \ref{fig:imponibile_totale_cedente_cessionario})
largely reflect the geographic distribution of firms (see Figure \ref{fig:firms_per_region_provincie}), we also computed for each province the average gross invoice amount by seller and buyer firm (see Figure \ref{fig:imponibile_medio_cedente_cessionario}). 
 Notably, in the map of average amount of invoice per seller, the province of Savona (SV) exhibits the highest value, despite ranking 84th in terms of firm count. This can be attributed to the presence of companies operating in the manufacturing of oil refinery products, chemical products, and ships (activities included in sector C), which issue invoices for large amounts. 
 When this activity is weighted against the relatively small number of firms with registered office located in the province, it results in the highest average invoice amount per seller.
From the buyer's perspective, while the province of Savona continues to hold significance, other provinces characterized by a high concentration of firms, such as Rome and Milan, also stand out. In addition, despite ranking 88th in terms of number of firms, the province of Trieste (TS) emerges as noteworthy due to the average purchase volume by firms operating in sector D (Electricity, Gas, Steam, and Air Conditioning Supply).

\section{Number of Links per Firm: Degree Distribution}
\label{sec:degree_distribution}
% An alternative to deal with the \va{perché sto Mean Degree con la lettera maiuscola?} Mean Degree is to compute the mean value of the distribution of the in-going (resp. out-going) degree of a node, considered as the average of the entire sequence of values taken by the variable \rev in-going degree" (resp. \rev out-going degree").
%\va{toglierei questa frase, lascerei solo una cosa tipo "It is instructive to consider the mean in- and out-degree excluding nodes which have zero connection of one of the two types. This change in the definition of the average eliminates the symmetry of equation \eqref{eq:mean_deg_lafond}, and points out the difference in shape of the in- and out-going degree distributions."}
%\sout{In the definition of mean degree (see equations \eqref{eq:mean_deg_lafond} and \eqref{eq:mean_deg_lafond_E_N}) presented in Section \ref{sec:density_and_mean_degree}, the denominator corresponds to the total number of firms in the network. Thus, the quantity inherently incorporates}
{There are} firms having exclusively one type of link, either out-going or in-going\footnote{Corresponding to firms solely engaged in sales activities (i.e., exclusively issuing invoices), or firms solely engaged in purchasing activities (i.e., exclusively receiving invoices) respectively.} (respectively 10,944 and 624,772 in 2019).
It is therefore instructive to compute the average in-degree and out-degree separately for nodes which have at least one incoming and at least one out-going connection respectively, pointing out the differences between the two distributions.
%Although this definition may seem to coincide with \eqref{eq:mean_deg_lafond} (or \eqref{eq:mean_deg_lafond_E_N}), the two definitions differ on one specific category of nodes: those representing firms that only sell or firms that only buy.
Following this intuition, we can define the mean in-degree $ \overline{\text{k}}_{in}$ as 
\begin{equation}\label{eq:mean_in_deg_distr}
    \overline{\text{k}}_{in} = \frac{1}{N_{in}} \sum_{ i = 1}^{N_{in}} \text{k}_{in}^{(i)} = \frac{E}{N_{in}}
\end{equation}
and out-degree  $\overline{\text{k}}_{out}$ as
\begin{equation}\label{eq:mean_out_deg_distr}
    \overline{\text{k}}_{out} = \frac{1}{N_{out}} \sum_{j = 1}^{N_{out}} \text{k}_{out}^{(j)} = \frac{E}{N_{out}},
\end{equation}
where $N_{in}$ is the number of elements of $\mathcal N _{in} = \{ u_i \in \mathcal N \,:\,\text{k}_{in}^{(i)} > 0\}$, the set of nodes that possess at least one in-going link, and $N_{out}$ is the number of elements in $ \mathcal N_{out}= \{ u_j \in \mathcal N \,:\, \text{k}_{out}^{(j)} > 0\}$, the set of nodes that possess at least one out-going link. %\va{Reintrodurre notazione matematica Daniele}
Following this approach we get $\overline{\text{k}}_{in} = 36.20$ and $\overline{\text{k}}_{out} = 58.03$.
As expected,\footnote{By definition $\mathcal N_{in} \subseteq \mathcal N$ and $\mathcal N_{out} \subseteq \mathcal N$, hence it follows that $N_{in} \leq N$ and $N_{out} \leq N$, and since the number of links $E$ does not change (see Equations \eqref{eq:mean_deg_lafond_E_N}, \eqref{eq:mean_in_deg_distr} and \eqref{eq:mean_out_deg_distr}), $\overline{\text{k}}_{in} \geq \overline{\text{k}}$ and $\overline{\text{k}}_{out} \geq \overline{\text{k}}$.} we find that $\overline{\text{k}}_{in} > \overline{\text{k}}$ and $\overline{\text{k}}_{out} > \overline{\text{k}}$, where $ \overline{\text{k}} = 35.96$ (see Section \ref{sec:density_and_mean_degree}).
To summarize, 99.33\% of our firms have in-going links (i.e. received at least one invoice) and for this subset the average number of links is 36.20, which is very close to the overall average ($\overline{\text{k}}=35.96$). 
More interestingly, only 61.98\% of our firms have out-going links (i.e. issued at least an invoice) and for this subset the average number of links is much larger (58.03).
In network theory nodes that only have out-going links are referred to as sources
(seller-only), while nodes that only have in-going links are referred to as sinks (buyer-only).
We refer to Appendix \ref{subsection:SinkSources} for a more in-depth analysis on buyer-only and seller-only firms, that represent the $38.02\%$ and $0.67\%$ of the firms respectively.
%{Inseriamo confronto con gli altri dataset sulle percentuali, nel paper di Lafond ci sta, tipo tabella nella sezione density/mean degree, ma la metterei nell'appendice dedicata +++ Daniele} FATTO, ma non lo metterei, i numeri fanno schifo 
%\sout{The in- and out-degree distributions can also be interpreted as a measure of the importance of a node. Namely, t}

The more connections a node has, the more it can be considered important in the network.
In a more general framework, one of the key defining properties of a node within a network is its \emph{centrality}, a concept that can capture multiple underlying properties and serves as a measure of a node’s structural importance.
We will expand on the concept of centrality and its applications in Section \ref{sec:global_centrality}. 
In this specific case, where centrality is defined in terms of the number of direct connections of the nodes, the measure is referred to as \emph{degree centrality}.
{In practice, the measure counts the number of downstream links (out-degree) associated with firms that issue invoices and the number of upstream links (in-degree) associated with firms that receive them.}
In more formal terms, this metric assigns importance to a node which is proportional to the number of its connections, hence it is one of the easiest measures to evaluate, yet it can be very useful to estimate the relevance of a node. 
The measure assigns weight 1 to each link, hence it is a function of the adjacency matrix of the network,\footnote{For a basic example of matrix representation of a network, see Appendix \ref{app:matrix_representation}.} defined as:
\begin{equation} 
\label{eq:adjacency_matrix}
A = (A_{ij})_{u_i,u_j \in \mathcal{N}} = \left\{
\begin{array}{cc}
      1 &  (u_i,u_j) \in \mathcal E \\
       & \\
      0 & \text{otherwise}
      \end{array}
\right.
\end{equation}
where $\mathcal N$ and $\mathcal E$ are the sets of nodes and links defined in Section \ref{sec:FromDataToNetwork}.
For each seller (row) and buyer (column), the degree centrality is obtained by summing the elements of the adjacency matrix $A$.
Denoting by $d_i$ the out-degree centrality of node $i$, it can be evaluated as:
\begin{equation}
\label{eq: def_degree_centrality}
    d_i \propto \sum_{j = 1} ^{N} A_{ij} =  \sum_{ j = 1} ^{N_i} 1
\end{equation}
where $N$ is the number of nodes in the network and $N_i$ is the number of nodes in the neighborhood of node $i$ (the set of nodes reachable from $i$ with a single step). The proportionality in \eqref{eq: def_degree_centrality} is given by the fact that the final centrality vector is considered up to a normalization.

\begin{figure}[H]
\centering
  \includegraphics[width=1\linewidth]{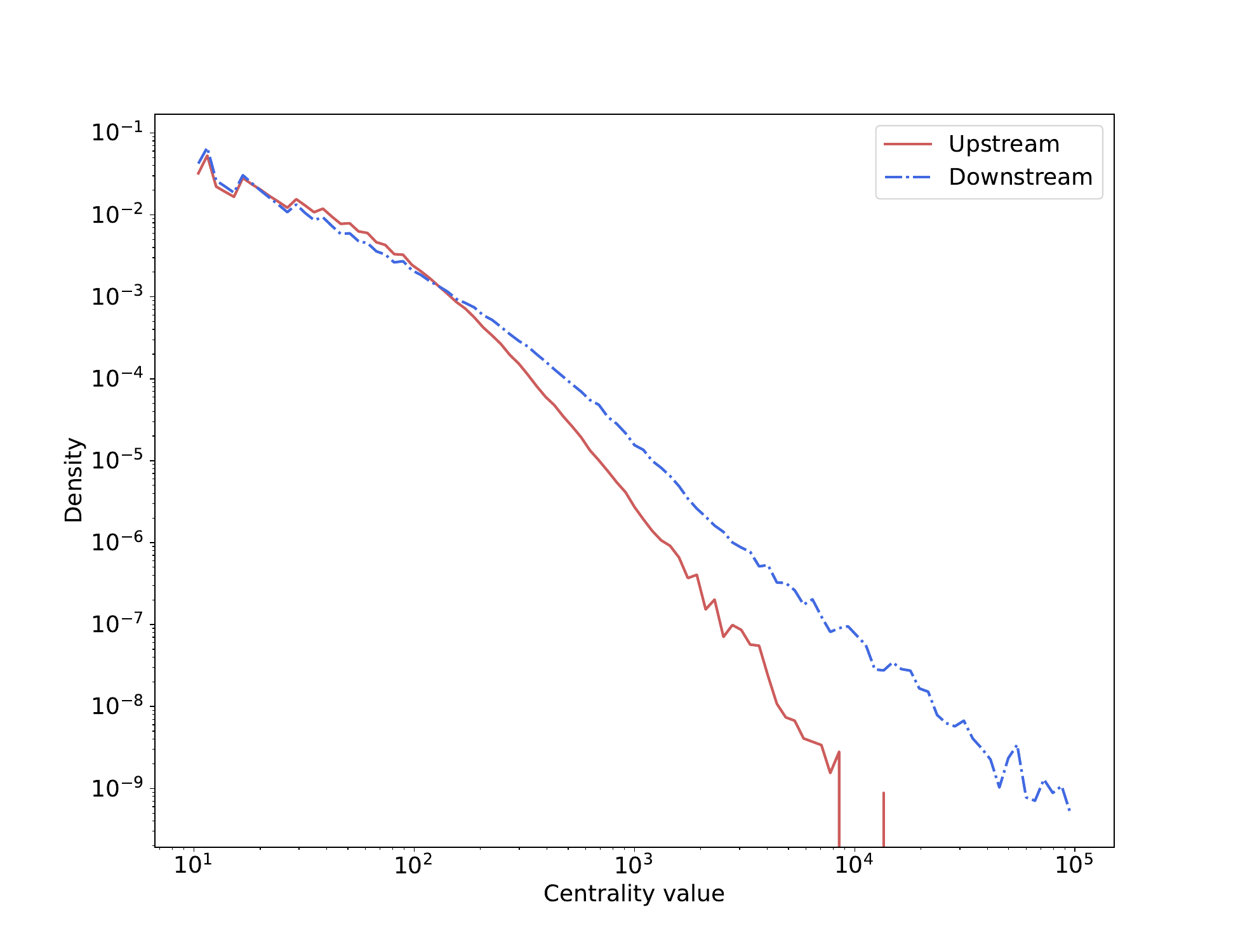}
  \caption{Density histograms of in-degree and out-degree centrality. The x-axis is partitioned into 100 logarithmically spaced bins, while the y-axis reports the density of firms falling within each bin on a logarithmic scale.}
  \label{fig:degree_centrality}
\end{figure}

\begin{figure}[H]
\centering
  \includegraphics[width=1\linewidth]{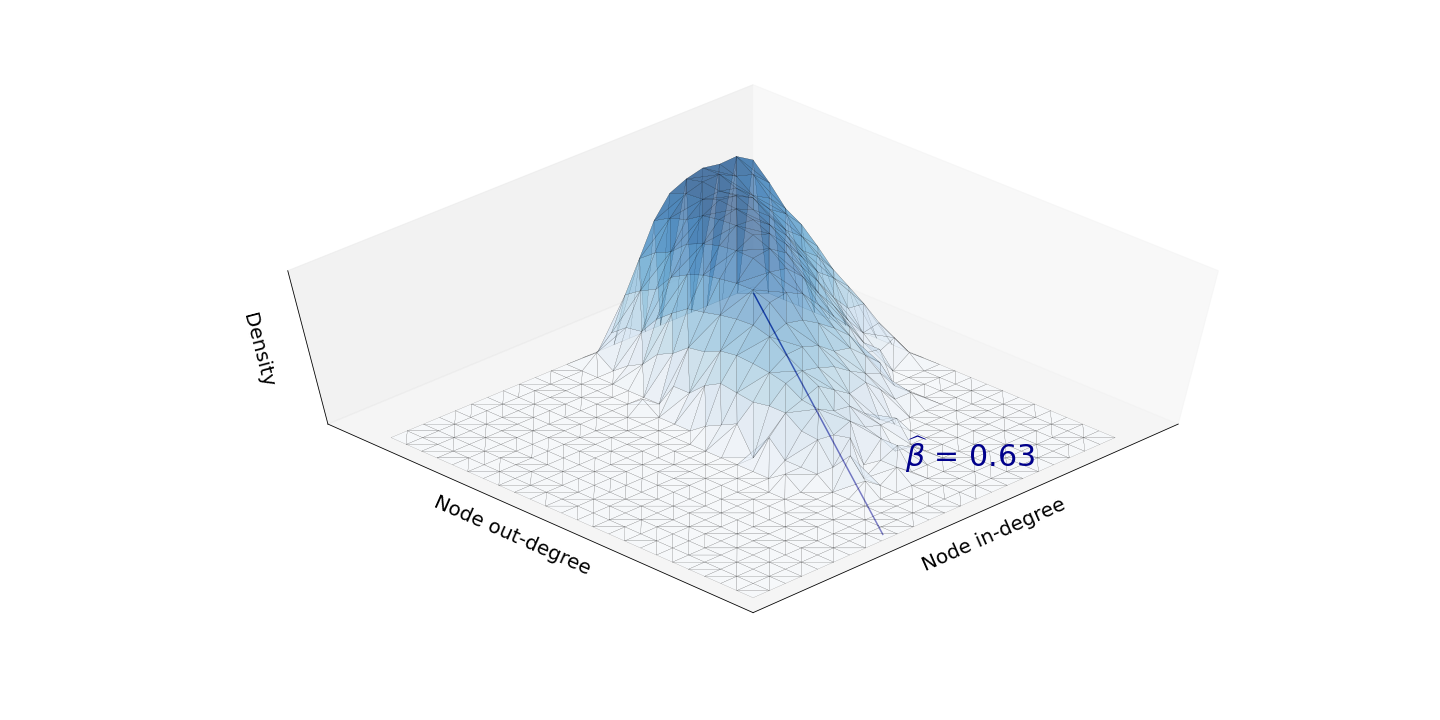}
  \caption{3D surface plot of the joint distribution of out-degree and in-degree. We divide the joint distribution domain in 625 squares in a logarithmic scale and the surface reports the number of nodes falling in each square. The blue line report the Total Least Squares estimated coefficient between the two variables.}
  \label{fig:hist2d_joint_2019}
\end{figure}

Both the in-degree and out-degree distributions are right skewed: the mean is larger than the median and the standard deviation is larger than both (see Table \ref{tab:degree_stats}).
\begin{table}[H]
    \centering
    \small
    % \begin{tabular}{lrrrrrrr}
    % %\toprule 
    % {} & \multicolumn{1}{c}{Min} &\multicolumn{1}{c}{p25} &\multicolumn{1}{c}{p50} &\multicolumn{1}{c}{p75} &\multicolumn{1}{c}{Max} &\multicolumn{1}{c}{Average} & \multicolumn{1}{c}{Standard Deviation}\\
    % \midrule
    % In-degree & 1 & 3 & 13 & 41 & 13,422 & 36.20 & 79.40   \\
    % Out-degree & 1 & 2 & 7 & 30 & 729,181 & 58.03 & 1,320.39\\
    % \bottomrule
    % \end{tabular}
    \begin{tabular}{lrrrrrrrrr}
    \toprule 
    {} & \multicolumn{1}{c}{p01} & \multicolumn{1}{c}{p25} &\multicolumn{1}{c}{p50} &\multicolumn{1}{c}{p75}& \multicolumn{1}{c}{p99} &\multicolumn{1}{c}{Average} & \multicolumn{1}{c}{Standard Deviation}\\
    \midrule
    In-degree  & 1 & 3 & 13 & 41 & 312 & 36.20 & 79.40   \\
    Out-degree & 1 & 2 & 7  & 30 & 633 & 58.03 & 1,320.39\\
    \bottomrule
    \end{tabular}
    %[  1.,   3.,  13.,  41., 312.])
    %[  1.,   2.,   7.,  30., 633.]

    \caption{Degree summary statistics. {Percentiles are evaluated separately for the in- and out-degree distributions, meaning that we are not considering here the correlation between the numbers of suppliers and customers for each firm. The population standard deviation for the out-degree distribution is formally divergent.}}
    \label{tab:degree_stats}
\end{table}
%{DA VEDERE CORREZIONI AF (larger dispersion)}
Comparing the two distributions and focusing on the right tails (Figure \ref{fig:degree_centrality}), we notice that they are described by power-laws.
% \footnote{All distributions are plotted using a log-log scale (meaning that the scales on the axes are logarithmically spaced). This is often convenient since heavy-tailed distributions are typically characterized by a high frequency of very small values and a low frequency of extremely large ones. As a result, representing such distributions on a linear scale would significantly compromise the readability and interpretability of the graphs. Given that the values on the x-axis are continuous, on the y-axis we plot the probability density, defined as the probability to be in a given bin divided by the bin width (not to be confused with the network density $\rho$ defined in Section \ref{sec:density_and_mean_degree}).\va{Eliminare-ridurre questa nota}}
In addition, the out-degree distribution has a significantly larger dispersion than the in-degree distribution.
%\sout{This finding is replicated also for other centrality measures in section \ref{sec:global_centrality}}
% Analogous findings hold also for the other centrality measures described in Section \ref{sec:global_centrality}.
%, and can be linked to the expansion pattern of a firm.
%{elimina da qui}a natural hypothesis is that in order to grow a firm tends to increase its customer base - hence expanding along the extensive margin - reinforcing the strength of already existing connections on the supply side - expanding along the intensive margin (for additional details about this hypothesis you can see for example \cite{mizuno2014structure} and \cite{carvalho2019production}, while quantitative models are introduced in \cite{oberfield2018theory}, \cite{afrouzi2020growing} and \cite{bernard2022origins}).
%{a qui}
%\va{Se sto paper va fuori dalla banca, il paragrafo precedente è da cancellare (appendice se necessario?). In ogni caso ci sono dei problemi nelle formule.}
% The density of nodes with a given degree is described by a distribution with power-law tails. 
% {Vogliamo scrivere una frasetta tipo lafond che dice che questa cosa è legata al grafico lineare nel plot log-log scale?}. 
The estimated values of the power-law exponents are $\alpha_{out} = 2.16$ $(0.03)$ for the out-degree distribution and $\alpha_{in} = 3.42$ $(0.03)$ for the in-degree distribution (see Figure \ref{fig:in_out_degree_log_scale_tail_fit} in Appendix \ref{app:additional_figures} for a detailed plot of the tails of the distributions). 
For a primer on the estimation of power-law exponents see \cite{clauset2009power}.
%Since the tails of the distributions appear linear in the log-log plots of the density of nodes of a given degree (see right hand side of Figure \ref{fig:in_and_out_degree_log_scale_tail} and Figure \ref{fig:in_out_degree_log_scale_tail_fit} ). Therefore, since the $ %logarithm of the number of nodes of degree k is proportional to the logarithm of k, the number of nodes with degree k is proportional to k raised to the power of alpha.Therefore, since the $\log (\text{Density of nodes of degree } k) = \beta + \alpha \log (k)$, it follows that $\text{Density of nodes of degree } k = \beta^\prime + k ^ \alpha$. %logarithm of the number of nodes of degree k is proportional to the logarithm of k, the number of nodes with degree k is proportional to k raised to the power of alpha. A lower exponent corresponds to a distribution with fewer finite moments, thereby indicating higher variability. From an economic perspective, this suggests that firms with the highest number of connections (in the right tail of the distributions) engage in sales transactions with a significantly larger number of firms compared to those from which they make purchases.
%In order to \sout{be well defined} \va{have a well defined distribution}, \va{the} power-law exponent must be greater than 1 \va{stesso commento che per il paragrafo di sopra, secondo me la frase precedenti si può togliere}. 
%\sout{The closer the exponent is to 1 the fewer finite moments the distribution has, implying greater variability.}
Since $\alpha_{out} < \alpha_{in}$, there is a higher probability to find a firm with a high number of buyers than one with a high number of sellers.
A natural hypothesis is that in order to grow a firm tends to increase its customer base - hence expanding along the extensive margin - while reinforcing the strength of already existing connections on the supply side - expanding along the intensive margin (for additional details about this hypothesis you can see for example \cite{mizuno2014structure} and \cite{carvalho2019production}, while quantitative models are introduced in \cite{oberfield2018theory}, \cite{afrouzi2020growing} and \cite{bernard2022origins}).
To provide a practical illustration, for each firm we compute the ratio between out-degree and in-degree and use the number of out-going connections (customers) as a proxy for its size. 
Considering only firms with at least one out-going link, we find a median value for this ratio of $0.43$. 
This value increases up to $1.87$ for firms with an out-degree of at least 100 - thus exceeding one - and continues to increase, reaching $5.59$ for firms with out-degree of at least 1,000 and $23.80$ for firms with out-degree of at least 10,000.
We present a plot of the median out-degree/in-degree ratio as a function of out-degree in figure \ref{fig:degree_ratio}.
To further investigate the relationship between nodes in-degree and out-degree we study their joint distribution, considering, for each firm $v$, the pair of values $(k_{in}^{(v)}, k_{out}^{(v)})$. Figure \ref{fig:hist2d_joint_2019} suggests a positive correlation between $k_{in}^{(v)}$ and $k_{out}^{(v)}$, and to estimate its intensity, as in \cite{bacilieri2026firm}, we perform a Total Least Squares (TLS) Regression\footnote{Unlike Ordinary Least Square (OLS), Total Least Squares treats symmetrically the variables under study, minimizing a combination of the residuals for all the variables rather than just for the dependent one. 
For the OLS coefficient of $\log k_{in}$ as a function of $\log k_{out}$ we obtain the value of 0.43.} between the logarithms of the in-degree and out-degree values. 
The result is that the two variables are related by an estimated coefficient of $0.63$. 
Although in the literature the study of the joint distribution of nodes in- and out-degree is less frequent than the analysis of the marginal distributions, the coefficient we find is consistent with the values reported in \cite{bacilieri2026firm}.

Degree centrality fails to account for indirect influences that propagate through the network, as those arising from nodes that are not directly adjacent.
In particular, one might require a centrality measure to assign relevance to nodes that participate in longer value chains or are connected, albeit not directly, to structurally important nodes.
%\footnote{In other words, an in-going connection from a node with a centrality value of 100 should be more important than 10 connections from a nodes with centrality value 1.}
To capture such characteristics the measure needs to {be extended} beyond the immediate neighborhood of a node and take into account its relations with the whole network. 
These measures are referred to as \emph{global} centrality measures, and we describe them in Section \ref{sec:global_centrality}. 

%In particulare The value is similar to other countries, notably...con referenza". Alla fine il paragone con i dati esistenti è uno dei risultati del paper %{Notably, this is the same exact coefficient found for the Hungarian production network <-- footnote?}.
%{Possibile gancio per closeness se lo vogliamo mettere in appendice}.

%\subsection{Assortativity - Connecting with the peers: Assortativity (LB)}
% \subsection{Connecting with Peers: Assortativity}
\subsection{Assortativity}
An interesting property that characterizes the relationship between in-degree and out-degree of nodes in a network is the assortativity; it quantifies the tendency of nodes within a network to be connected to nodes with similar degree. This measure is particularly relevant in the study of shock propagation within economic and financial networks, as it can serve as an indicator of the type of connections the most connected firms have.
Several findings  in the literature (see \cite{newman2003assortativity, fujiwara2010large, bacilieri2026firm}) suggest that production networks exhibit negative assortativity, meaning that firms with a number of connections above the average are mostly linked to firms with a number of connection below the average. 
To formally quantify this property we follow the approach introduced in \cite{newman2003assortativity}, {by} computing the assortativity as the Pearson correlation coefficient between the logarithm of the degrees (both in-going and out-going) of firms that share a link. 
The measure represents a percentage that ranges from -100, indicating a network in which the highest-degree nodes are connected only to the lowest degree nodes, to 100, indicating a network in which the highest degree nodes are connected {only} to other nodes with similar degree values.
A correlation value of 0 suggests the absence of a systematic relationship between the degrees of connected nodes.
Since our network is directed, assortativity has to be evaluated through four distinct measures, each corresponding to a pairwise combination of in-degree and out-degree values.
\begin{table}[H]
    \centering
    \small
    \begin{tabular}{lrr}
      & \multicolumn{1}{c}{Seller in-degree} & \multicolumn{1}{c}{Seller out-degree} \\
    \midrule
    Buyer in-degree  &  $- 5.48$ & $- 7.92$ \\
    Buyer out-degree &  $- 1.35$ & $- 1.60$ \\
    \bottomrule
    \end{tabular}
    \caption{Values for assortativity.}
    \label{tab:assortativity_values}
\end{table}
% \begin{table}[H]
%     \centering
%     \small
%     \begin{tabular}{llrrrr}
%     \toprule
%       Dataset &  Year & $r_{k^{in}, k^{out} }$& $r_{k^{out}, k^{in} }$& $r_{k^{in}, k^{in} }$& $r_{k^{out}, k^{out} }$\\
%     \midrule
%     Italy & 2019 &  -1,35 & -7,92 &  - 5,48 & -1,60 \\
%     \bottomrule
%     \end{tabular}
%     \label{tab:SinkSourcesStats}
% \end{table}
The results of Table \ref{tab:assortativity_values} align with the expectations derived from previous studies, revealing four negative values, whose relative magnitudes are consistent with those reported in the literature, meaning that the highest-degree nodes tend to connect to nodes with degree lower than average. %{, although only mildly so? - DA VEDERE}

% \subsection{A Small World with Hubs: Connectivity}
\subsection{Connectivity}
\label{sec:closeness}

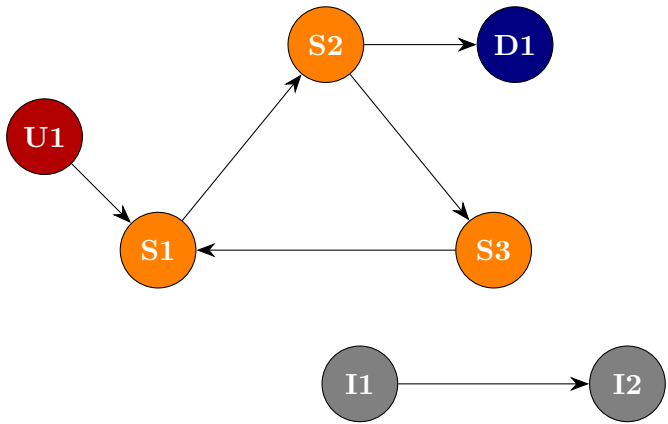
\begin{figure}[H]
    \centering
    \begin{tikzpicture}[
        node distance=2.5cm,
        mynode/.style={circle, draw, fill=#1, text=white, minimum size=1cm, font=\bfseries} 
    ]
    
    \node[mynode=red!50!yellow] (1) {S2};
    \node[mynode=red!50!yellow] (2) [below right=2cm and 1.5cm of 1] {S3};
    \node[mynode=red!50!yellow] (3) [below left=2cm and 1.5cm of 1] {S1};
    \node[mynode=blue!50!black] (4) [right of=1] {D1};
    \node[mynode=red!70!black] (5) [below left=0.5cm and 3cm of 1] {U1};
    \node[mynode=gray] (6) [below left of=2] {I1};
    \node[mynode=gray] (7) [below right of=2] {I2};
    %\node[mynode=gray] (8) [below left of=6] {I3};
    %\node[mynode=gray] (9) [below right of=6] {I4};

    \draw[-{Stealth[scale=1.5]}] (1) -- (2);
    \draw[-{Stealth[scale=1.5]}] (1) -- (4);
    \draw[-{Stealth[scale=1.5]}] (5) -- (3);
    \draw[-{Stealth[scale=1.5]}] (3) -- (1);
    \draw[-{Stealth[scale=1.5]}] (2) -- (3);
    \draw[-{Stealth[scale=1.5]}] (6) -- (7);
    %\draw[-{Stealth[scale=1.5]}] (8) -- (9);

    \end{tikzpicture}
    \caption{The orange nodes represent firms in the strongly connected component: for each pair of nodes there exists a directed path connecting them.  The blue (resp. red) node represents a firm in the downstream (resp. upstream) connected component. Orange, blue and red nodes altogether represent the weakly connected component. The grey nodes represents isolated firms, not connected with the weakly connected component.}
    \label{fig:rete_esempio_componenti}
\end{figure}

%\sout{The links between firms also give information on how connected and close firms are.} %Any firm can be a starting point of a path composed by links Staring from a node, we can follow links and % studied 
%Given a starting firm, the network can be explored by following the links with the other firms both downstream (along the links' direction) and upstream (against the links' direction). The sequence of links connecting two firms is called path. Since the links in our network have a direction, for any pair of firms there may not exist a path connecting them. The largest subset of firms that are pairwise mutually reachable is called strongly connected component.
We can exploit the structure of the network to derive information about the closeness and connectivity of firms.
Given a starting firm, the network can be explored by following the links both downstream (along the links direction, following the flow of invoices {issued by the firm}) and upstream (against the links direction, following the flow of payments).
A sequence of directed edges connecting two firms is called a path, and two firms are connected if such a path exists between them. 
There may exist more than one path connecting two firms, but the length of the shortest one can be used as an indicator of how close the {two} firms are in the production network.
%The subset of all the nodes that are  is called connected component. More precisely, since our network is directed, there are two of connected components: weakly connected components and strongly connected components. A weakly connected component is the subset of nodes such there exists at least one path from one to the other between any pair of nodes. In contrast, a strongly connected component (SCC) is a subgraph in which, for every ordered pair of nodes (u, v), there exists a directed path from u to v and a directed path from v to u. Strong connectivity therefore implies a higher degree of structural cohesion and is often of particular interest in the analysis of flow, influence, or feedback mechanisms within directed networks
The largest subset of nodes for which each element can be reached by any other element with a directed paths is called the \emph{strongly connected component} of the network.\footnote{In principle it is possible to have more than one strongly connected component in a network. In the network of Italian business-to-business commercial relations, as in many networks with a sufficiently large number of nodes, this component is unique.}
In addition to the strongly connected component there are two other types of sub-networks.
We define downstream (resp. upstream) connected component the subset of all nodes reached by a downstream (resp. upstream) path starting from a node in the strongly connected component. 
The union of the strongly connected component, upstream and downstream connected components is called \emph{weakly connected component}, and consists of all nodes of the network which are connected by at least one path.
Firms outside the weakly connected component are called \emph{isolated}.
{Figure \ref{fig:rete_esempio_componenti} provides a schematic illustration of the described structure.}

\begin{figure}[H]
\centering
\begin{minipage}{.5\textwidth}
  \centering
  \includegraphics[width=1\linewidth]{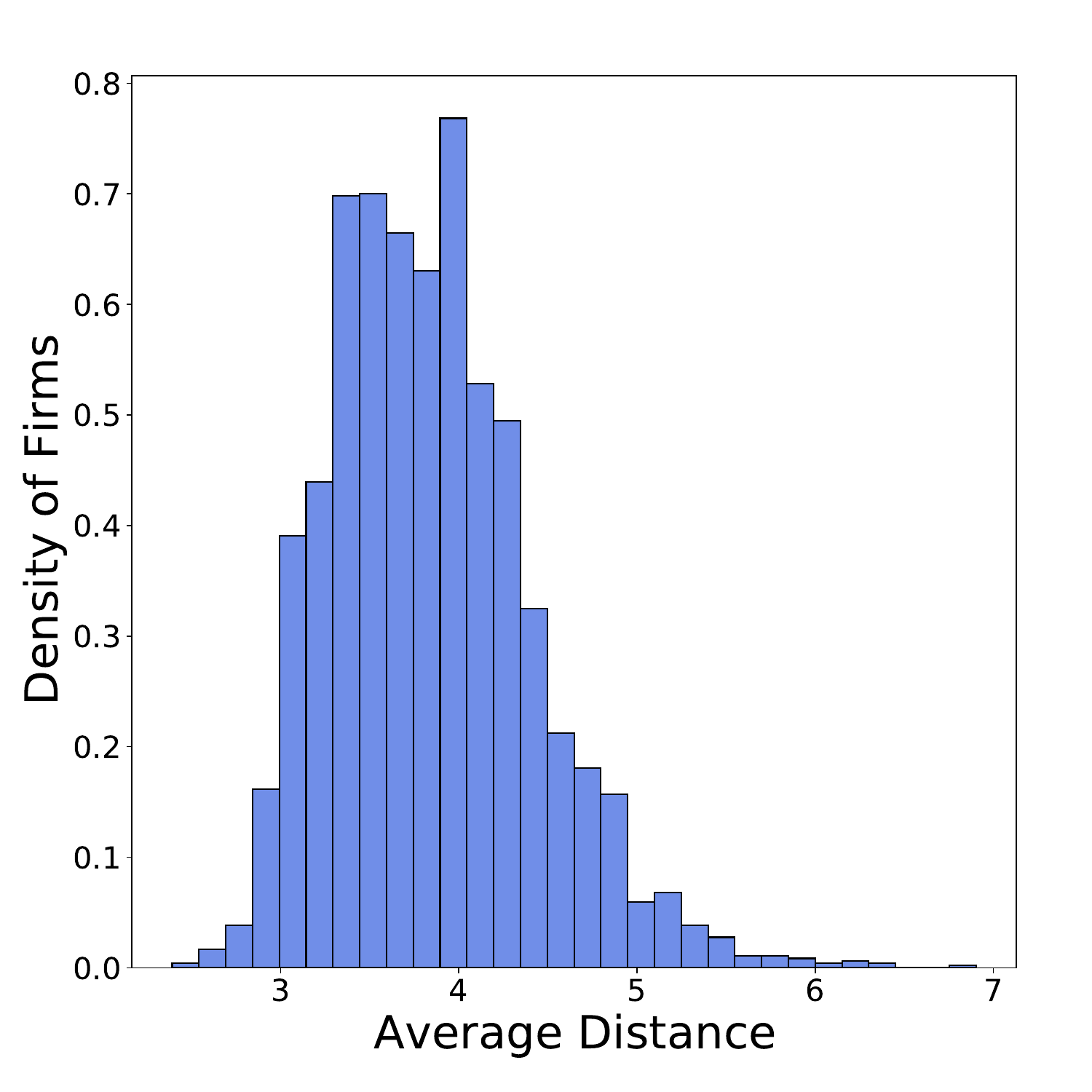}
\end{minipage}%
\begin{minipage}{.5\textwidth}
  \centering
  \includegraphics[width=1\linewidth]{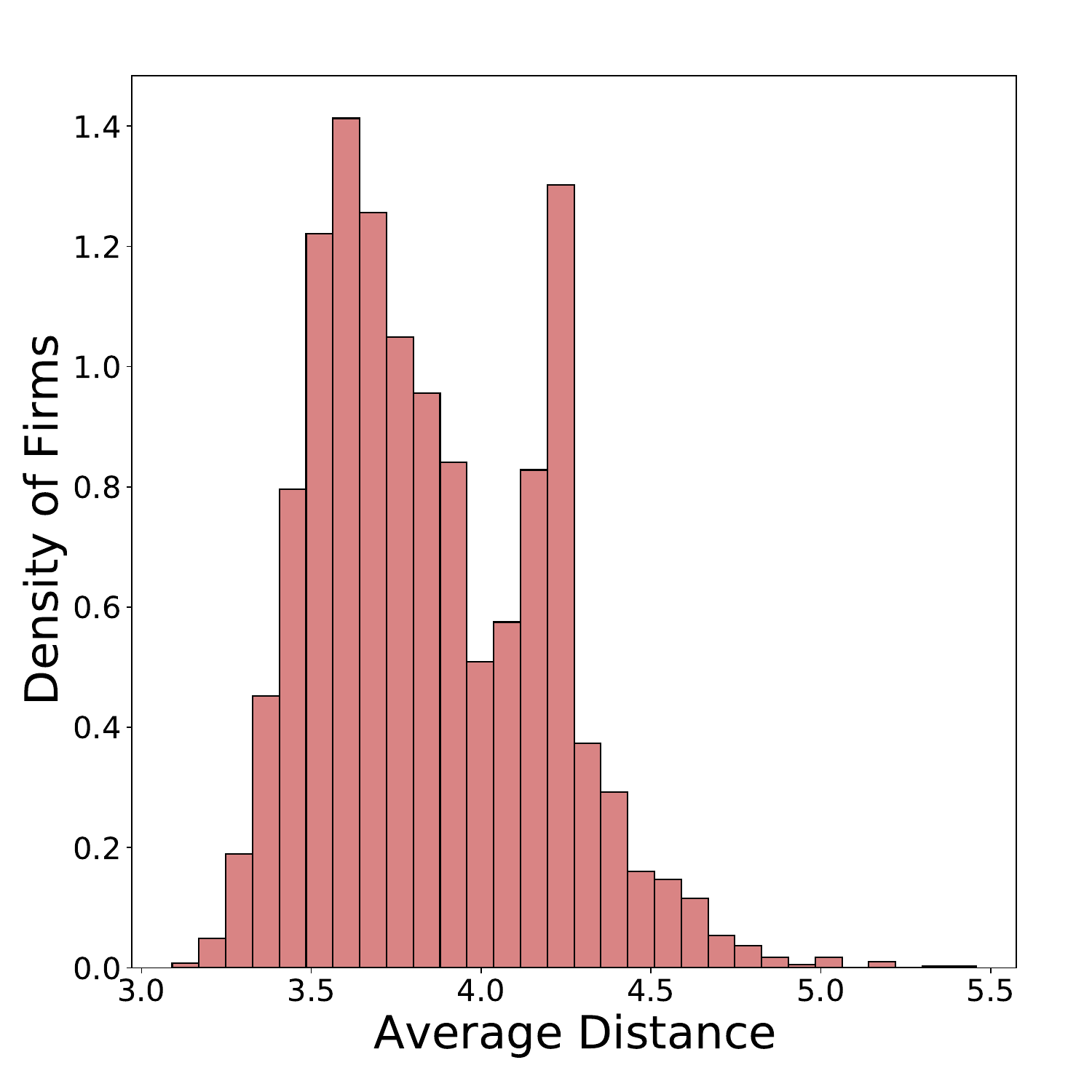}
\end{minipage}
\caption{Histograms for the downstream (left hand side) and upstream (right hand side) distribution of the average distance from the nodes in the connected component. Both distributions have an average value of steps smaller than 4.}
  \label{fig:closeness_average_distance_stream}
\end{figure}

In our network %almost all the nodes are connected in a large component, both upstream and downstream. More precisely
we find a large strongly connected component
%\footnote{ METTI DEFINIZIONE ANCHE QUI, POI RIMANDO APPENDICE See appendix \ref{app:matrix_representation} for a definition of strongly and weakly connected components.}
constituted by 60\% of the firms in the network.
%\footnote{The strongly connected component of a network is a set of nodes, every one of each can be reached from any other node both going upstream and downstream in the network. All the nodes which can be reached going upstream/downstream in the network from the strongly connected component constitute a large upstream/downstream component of the network.}
The upstream connected component is composed of 61\% of the total number of nodes, while the largest component in the downstream direction is composed of 99\% of the total number of nodes. The remaining isolated nodes belong to an extremely scattered disconnected component, composed mainly of isolated pairs of nodes.\footnote{
More in detail, there are 11,108 nodes not reached by the largest component of the downstream network. Most of these are sellers only, however 164 nodes buy from someone not in the largest component of the network. These disconnected sellers are 138, and the total number of nodes in this disconnected component is 295. In this sub-network, 90\% of the sellers have only one buyer.
Similarly 645,952 nodes are not reached by the largest component of the upstream network. The large majority of them are buyers only, but 40,307 nodes form disconnected components, in which 21,852 nodes buy from 21,180 sellers not connected to the largest component of the network. In this sub-network 89\% of the buyers have only one seller.
Hence this disconnected component is itself composed of many small disconnected sub-components.}
For a given node $i$, we can always define its upstream (resp. downstream) connected components, as the set of nodes connected to $i$ by an upstream (resp. downstream) path. It is then possible to compute the average distance of node $i$ to every other node contained in these components:
\begin{equation}
    \delta_i = \frac{1}{N_c} \sum_{j=1}^{N_c} \delta_{ij}
\end{equation}
where $N_c$ is the number of nodes in the connected component, and $\delta_{ij}$ is the length of the shortest directed path connecting node $i$ to node $j$.\footnote{
Another centrality measure found in the literature is the so-called \emph{closeness centrality}, defined as the inverse of the average shortest-path distance from a given node to all other nodes within its connected component. Consequently, a node with a high closeness centrality is, on average, connected via relatively short paths to all the other nodes in its component (\cite{bavelas1950communication}).}
% \va{There are 11108 nodes not reached by the downstream largest component. Of these, 164 buy from someone. The number of sellers in this subnetwork is 138. Of these nodes, 7 are both sellers and buyers in the subnetwork. The total number of nodes in this subnetwork is 295. (We could visualize this subnetwork)}
% \va{There are 645952 nodes not reached by the upstream largest component. Of these, 21180 sell to someone. The number of buyers in this subnetwork is 21852. Of these nodes, 2725 are both sellers and buyers in the subnetwork. The number of nodes in the subnetwork is 40307.}
% {Interessante, potremmo fare dei plot schematizzati}
Figure \ref{fig:closeness_average_distance_stream} shows a distribution of the average distance for the upstream and downstream connected components.
For both components we find a typical average distance of less than 4 steps. 
This values reflect the so called \emph{small-world} property of the network: even though selecting two nodes at random the probability that they are neighbors (connected in one step) is extremely small, there exists a path indirectly connecting them that is composed only of a small number of steps (\cite{watts1998collective}).
This result strongly depends on the presence in the network of nodes with high values of in- and out-degree (namely, number of upstream and downstream links, respectively), called hubs. 
% Reinserire
% The presence of such nodes is typical of \emph{scale-free} networks, i.e. networks with a power-law distribution of nodes degree.\mi{E quindi? Aggiungere referenza? Aggiungere collegamento a statistiche hub? --- Valerio}

%{la figura non si parla con quello scritto nel testo. Direi di fare doppia la figura con media per upstream e downstream e poi farne altre 2 con la scomposizione che ora c'è a destra (che non mi ricordo per cosa era, mi pare grandezza del vicinato) sempre per upstream a doqwn stream. Per questa cosa che la distribuzione della media è bivariata, manca anche una frasetta nel testo per contestualizzare il grafico.}
\begin{figure}[H]
\centering
\begin{minipage}{.5\textwidth}
  \centering
  \includegraphics[width=0.9\linewidth]{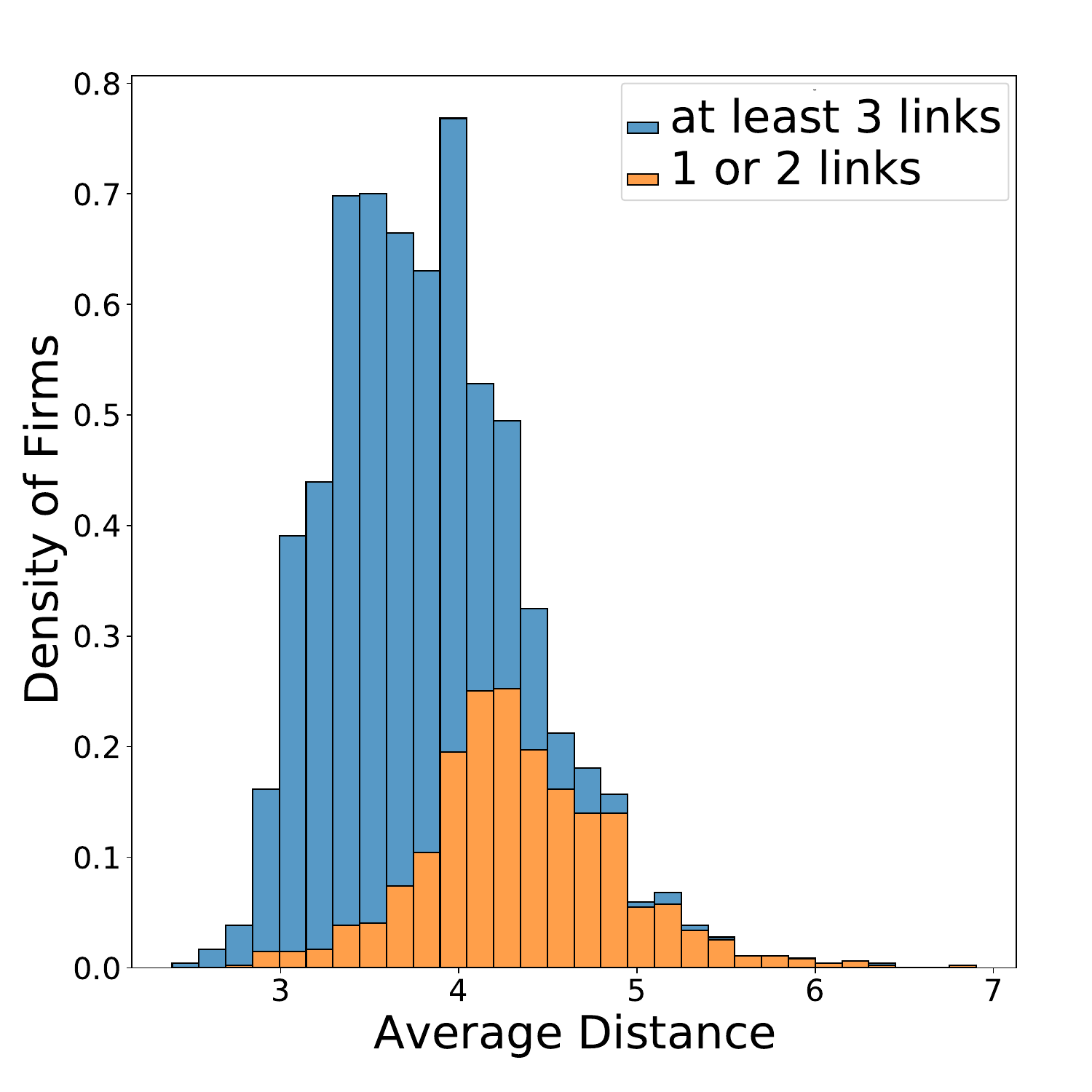}
\end{minipage}%
\begin{minipage}{.5\textwidth}
  \centering
  \includegraphics[width=0.9\linewidth]{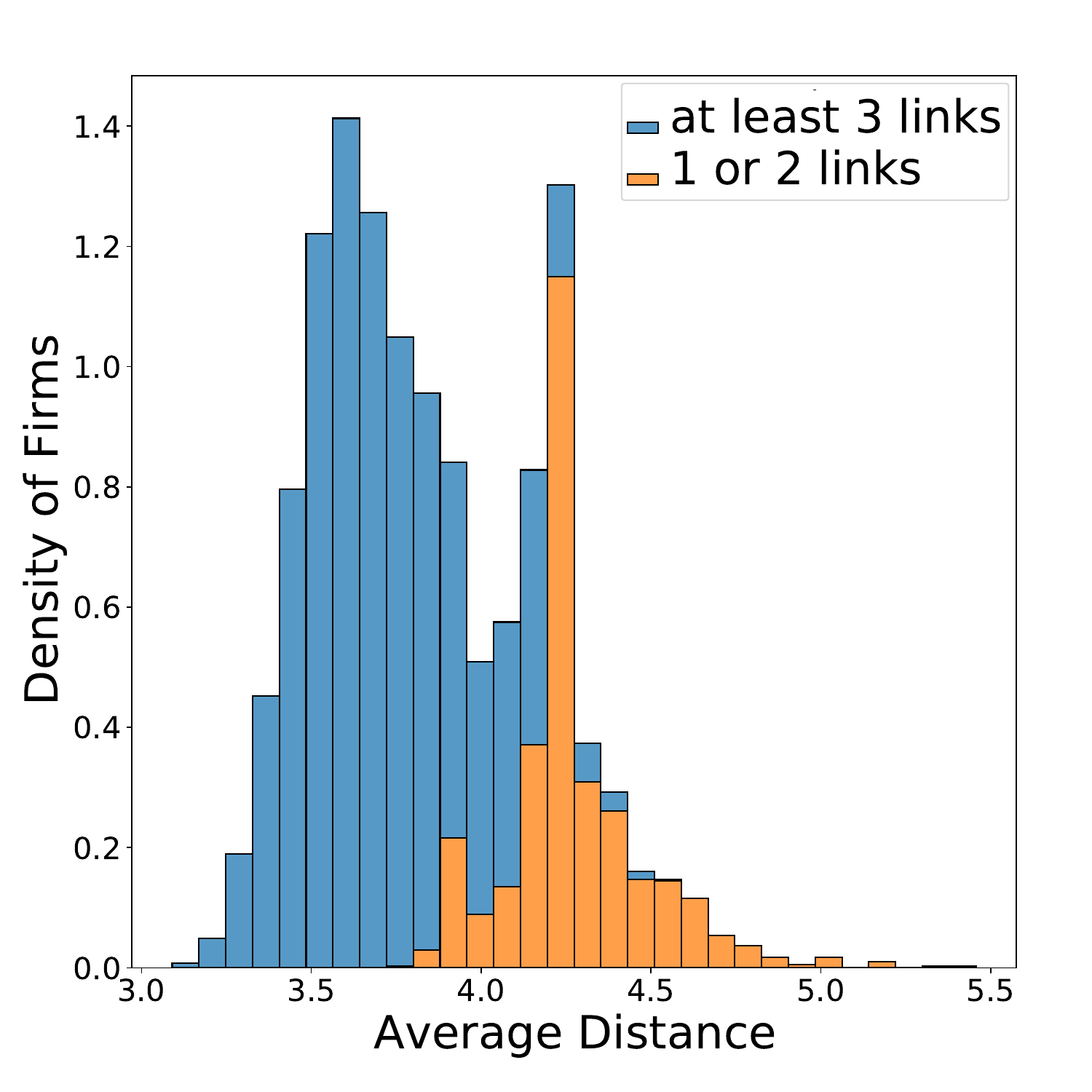}
\end{minipage}
\caption{Histograms for the downstream (left hand side) and upstream (right hand side) distribution of the average distance from the connected component with {a breakdown by the} number of nodes in the neighborhood of the starting point. The second peak in the upstream histogram is {made up almost entirely} of firms with at most 2 neighbors that, on average, are (almost) one step further from the other nodes.}
  \label{fig:closeness_average_distance_fig}
\end{figure}

The shortest path between two random nodes typically passes through one of these hubs, and we see a strong inverse correlation between the number of neighbors of a given node (which is itself proportional to the probability to be close to a hub) and the average distance between the node and its connected component. 
As we can see in Figure \ref{fig:closeness_average_distance_stream}, the bulk of the distributions is concentrated on average path lengths between 3 and 4 (as a comparison a value of 4.3 is found for Japan in \cite{mizuno2014structure}, and values closer to 3 are found for Ecuador and Hungary in \cite{bacilieri2026firm}). 
A second peak is present for the average upstream distances, between 4 and 4.5 steps. 
To understand the bivariate nature of the distribution we consider two disjoint categories of nodes, defined on the basis of the number of neighbors: the first group of nodes has 1 or 2 nodes in the neighborhood, the second group has at least 3. 
We find (Figure \ref{fig:closeness_average_distance_fig}) that the number of nodes in the neighborhood of the starting point is strongly correlated to the inverse average distance from the other nodes: on average nodes with 1 or 2 neighbors (orange in Figure \ref{fig:closeness_average_distance_fig})  are almost 1 step further from the rest of the network, with respect to the other nodes (blue in Figure \ref{fig:closeness_average_distance_fig}).

%\sout{
%consider the number of nodes connected with one step to the starting point of the path (the neighborhood of the starting node). We find}
% {Interessante, potremmo fare dei plot schematizzati}

\subsection{Sectoral and Geographical Distribution of the Most Connected Firms}
\label{sec:degree_distribution_geo_sec}

The heavy-tailed distributions observed in both upstream and downstream degree centrality mark a fundamental structural feature of the network: a limited number of nodes with a large number of links, called \emph{hubs}, in contrast to the large majority of nodes that present a small number of links. The geographical and sectoral identification of these hubs provides critical insights into the locations and economic activities in which the most connected firms operate. To investigate this, we focus on the top 100 firms by out-degree, that we will refer to as \emph{downstream hubs}, and the top 100 firms by in-degree, that we will refer to as \emph{upstream hubs}, and we analyze their geographical and sectoral distribution.

\begin{figure}[H]
\centering
  \begin{subfigure}[b]{0.5\textwidth}
    \centering
    \includegraphics[width=1\linewidth]{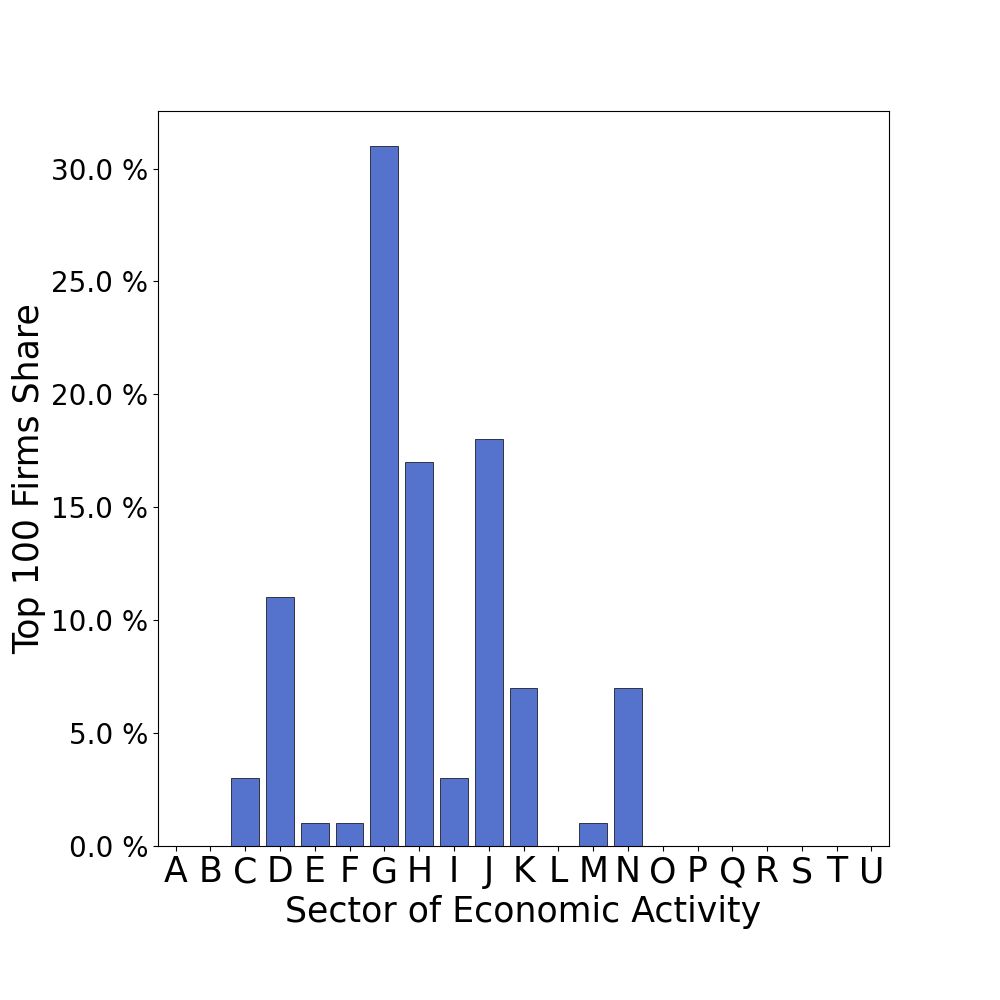}
    \caption{Out-degree Centrality}
  \end{subfigure}%%
  \begin{subfigure}[b]{0.5\textwidth}
    \centering
    \includegraphics[width=1\linewidth]{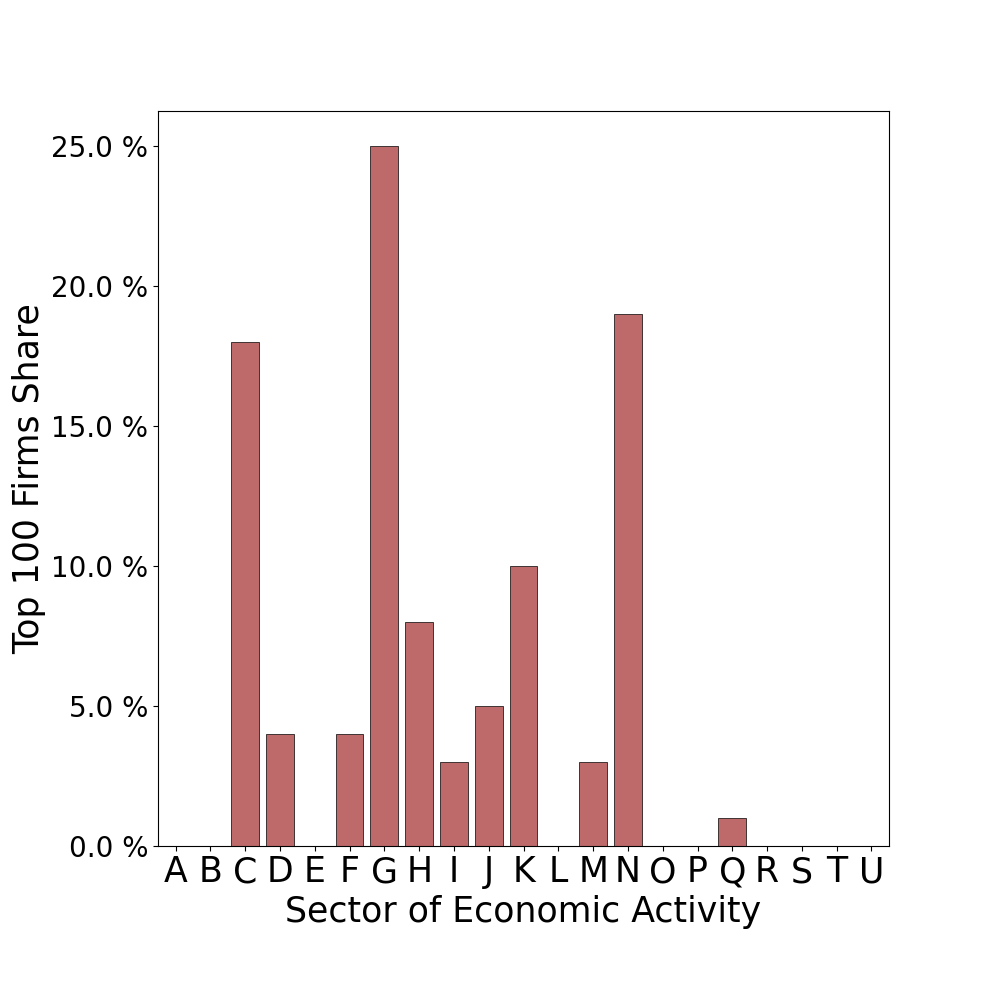}
    \caption{In-degree Centrality}
  \end{subfigure}
  \caption{Distribution of the out-degree (downstream) and in-degree (upstream) centrality shares across the economic sectors of the 100 most connected firms.}
  \label{fig:cat_degree_centrality_100} 
\end{figure}

In terms of sectoral composition, we observe that sector G (wholesale and retail trade; repair of motor vehicles and motorcycles), that is the sector where most Italian firms operate (see Figure \ref{fig:firms_per_category_power}), is well represented among both downstream and upstream hubs. Specifically, approximately 30\% of downstream hubs and 25\% of upstream hubs operate within this sector (see Figure \ref{fig:cat_degree_centrality_100}, left and right hand side, respectively). With regard to downstream hubs, a substantial proportion of highly connected firms operate in industries that provide core services to other businesses, including electricity distribution, logistics, and communication services: around 40\% of downstream hubs are concentrated within sectors D (electricity, gas, steam, and air conditioning supply), H (transport and storage) and J (information and communication services).
In contrast, upstream hubs predominantly operate in manufacturing (sector C), where business models typically depend on a diverse supplier base, and in business support services (sector N: renting, travel agencies, business support activities).
These two sectors account for 40\% of upstream hubs.

From a geographical perspective the distribution of hubs reveals a notably symmetric pattern between upstream and downstream hubs. 
Strikingly, no hub firm is situated south of Rome (RM), which accounts for 22\% of downstream hubs and 15\% of upstream hubs, with the exception of a single upstream hub located in the province of Naples (NA). 
All the remaining hubs are located in provinces within the central-northern regions of the country.

\begin{figure}[H]
\centering
\begin{minipage}{.5\textwidth}
  \centering
  \includegraphics[width=1\linewidth]{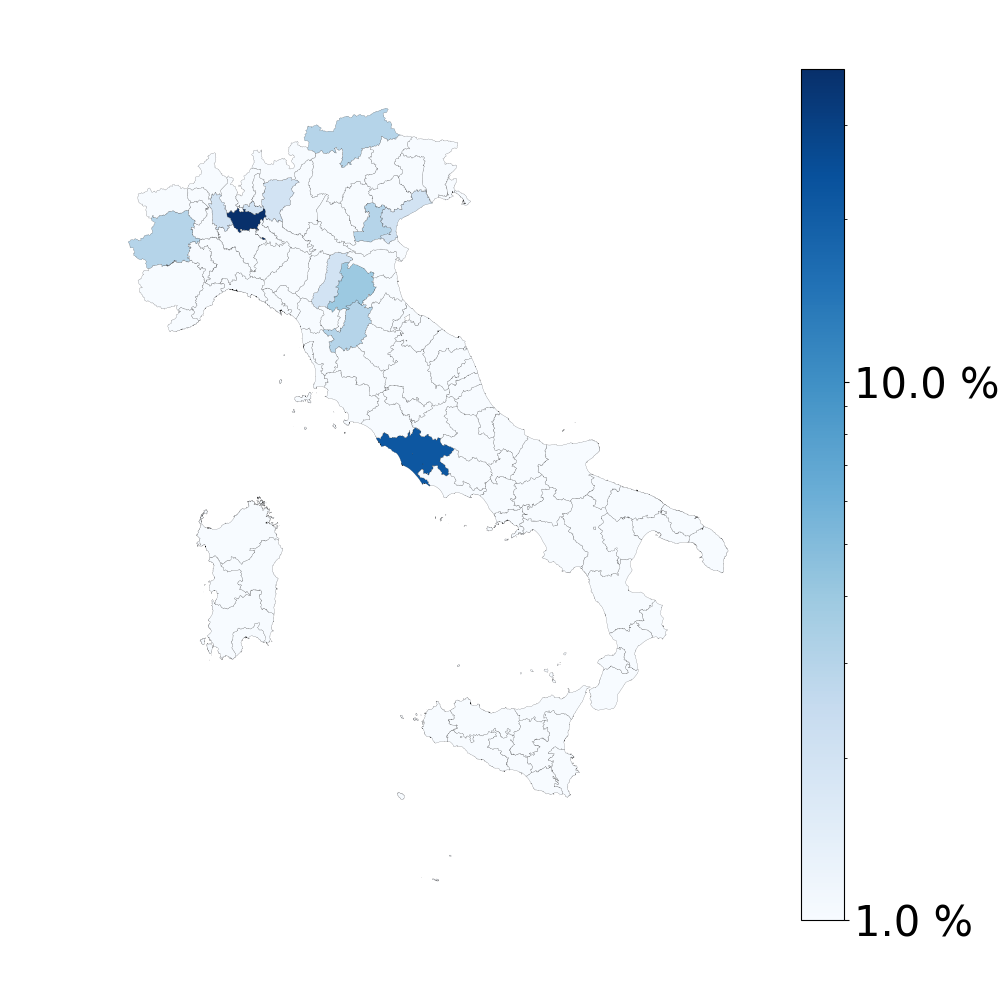}
\end{minipage}%
\begin{minipage}{.5\textwidth}
  \centering
  \includegraphics[width=1\linewidth]{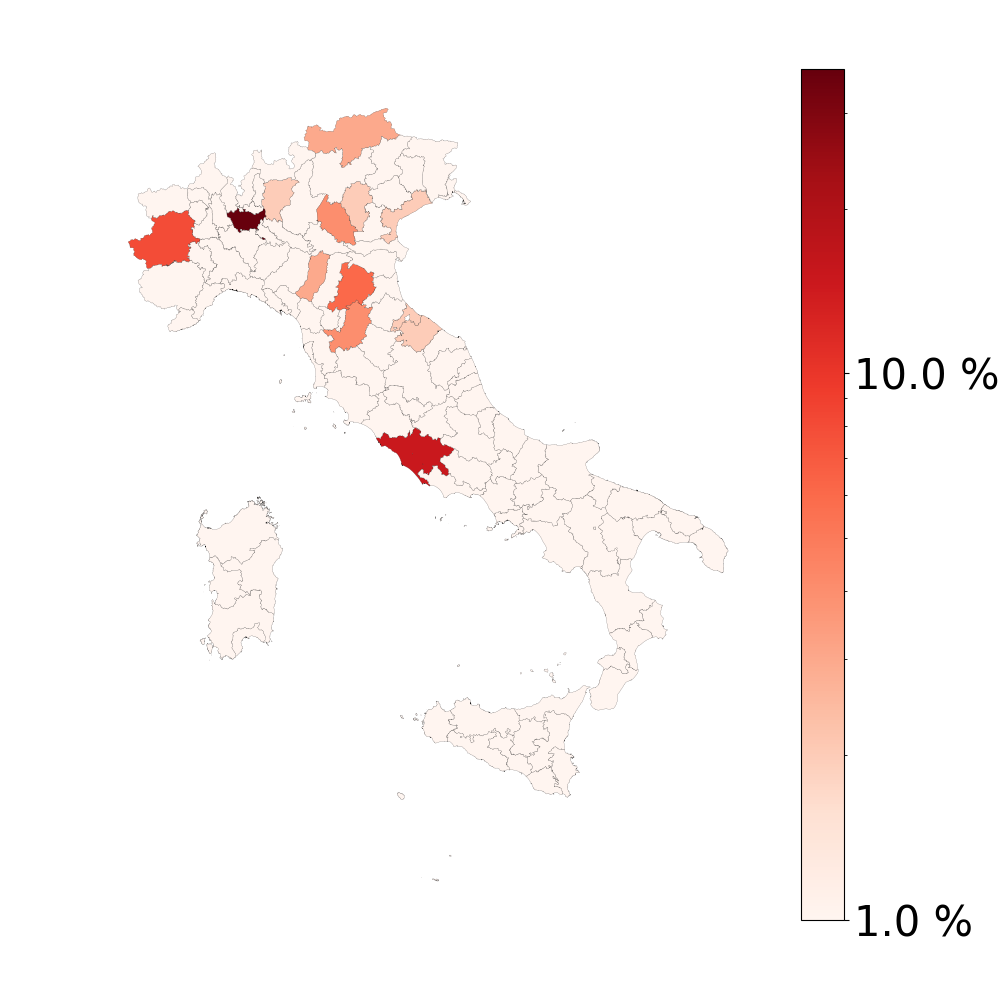}
\end{minipage}
\caption{Distribution of the out-degree (downstream) and in-degree (upstream) centrality shares across the provinces of registered office of the 100 most connected firms.}
  \label{fig:grado_provincie_top_100}
\end{figure}

\section{Importance of Firms in the Network: Global Centrality Measures} 
\label{sec:global_centrality}
%\lb{cambio titolo}
The concept of network centrality quantifies the importance of a node in relation to the rest of the network and has many applications, in economics and beyond. 
As we saw in Section \ref{sec:degree_distribution}, the degree of a node can be seen as a basic, local measure of centrality; multiple extensions were considered in the network analysis literature. 
One kind of measure focuses on the average distance of a given node from the rest of the network; we briefly described this type of centrality in Section \ref{sec:closeness}.
The most studied form of centrality, though, quantifies the effects on the whole network of shocks impacting a given node.
These measures are functions of the matrix describing the network,\footnote{See Appendix \ref{app:matrix_representation} for the description of an example network in {matrix form}.} and they are global, meaning that the centrality of each node takes into account the relations of the node with the whole network, not only with its neighbors. 
In this section we focus on this second type of centrality.
% In economics, in \cite{acemoglu2012network} the relation of node centrality to economic importance is explored, while in \cite{banerjee2013diffusion, banerjee2014gossip} the authors analyzed the role of network centrality in information diffusion and financial participation.
% The most direct application of the concept of centrality is in the analysis of shock propagation: in \cite{acemoglu2010cascades, acemoglu2013network} the importance of the network structure and in particular of the most central nodes in the amplification of economic shocks is analyzed.
% In \cite{haldane2011systemic, cont2013network} the authors investigate the central role of the network structure in the propagation of shocks in financial systems. 
% Finally, see \cite{watts1998collective, albert2000error} and \cite{motter2002cascade} for the analysis of shock propagation and the role of central nodes in more general networks. 
% Perhaps the most famous application of node centrality has been the introduction by Google of the PageRank algorithm to rank web pages (\cite{brin1998anatomy, page1999pagerank}).
% This measure is equivalent to the probability for a user randomly clicking on hyperlinks to land on a given page, and it proved extremely efficient in the classification of web pages.
% There are many centrality measures available in the literature (see for example \cite{das2018study} for a review), and each of them captures a different notion of importance of the node considered.  
These quantities often scale with the dimension of the network; for this reason centrality measures are often useful in {relative terms}, to compare different nodes of the same network.  
All the metrics considered {below} show considerable concentration: a negligible fraction of firms contributes to a significant portion of the sum of the centrality scores in the network. 
% \sout{This is a consequence of the power-law distribution of the centrality metrics, and in particular of the low exponents we find in the tail of the distributions: the power-law nature of these tails implies the presence in the network of nodes having centrality values many standard deviations above the mean, which would be impossible with normal distributions.} 
Large outliers have centrality values orders of magnitude larger than the typical node, and influence significantly the aggregates. 
Similarly to the number of connections we find that the distributions of downstream centrality values are wider - i.e. have lower tail exponents - than the distributions of upstream centrality scores. 
In other words, we find much more variability in the importance of the nodes as sellers, than as buyers. 

% \footnote{As an example, probably a node must be considered central if it has 200 connections in a network with 1000 nodes, but not as much if the network is composed of 1 million nodes.}.
% \sout{In section \ref{sec:degree_distribution} we already described the degree distribution of the nodes in our network. This distribution is one of the simplest and yet most informative centrality measure one can consider (in this context it is often called \emph{degree centrality}). }{già detto 10 volte}
Given the fact that we are working with a directed network, we will have two versions for each centrality measure; we will consider both an upstream centrality and a downstream centrality for the in-going and out-going links respectively. 
We consider here the \emph{eigenvector}, \emph{Katz} and \emph{PageRank Centrality}.
The features and limitations of each of these measures will be described in the following sections.

\subsection{Eigenvector Centrality}
% \subsection{From Local to Global Importance: Eigenvector Centrality}
\label{sec: eigenvector_centrality}
In Section \ref{sec:degree_distribution} we presented the degree centrality as a local importance measure for the nodes in our network.
Its local nature can be deduced by the definition of its value. 
As defined in equation \eqref{eq: def_degree_centrality}, the degree centrality of node $i$ is given by:
% \va{non è indispensabile ripeterla, ma evita la scocciatura di dover risalire per guardarla...}
% \footnote{In this section all definitions and results are described for the { incoming }connections of a node. Identical expressions apply to the {out-going} connections.} 
\begin{equation}
    d_i \propto \sum_{j = 1}^{N_i} 1 \nonumber
\end{equation}

%\footnote{The neighborhood of a node $j$ is the set of nodes reachable from $j$ with a single step. The downstream neighborhood is the set of $j$'s buyers, while its upstream neighborhood is composed by its sellers.}
%here $n_i$ is the neighborhood of node $i$, and the proportionality is given by the fact that the final centrality vector is considered up to a normalization.

By this definition it is clear that every node in the neighborhood of node $i$ is weighted equally in the evaluation of the centrality, independently of its importance. 
In other words, we can see that the evaluation of the centrality score of a node is independent of the centrality of any other node: hence the \emph{local} nature of this measure. 

To take into account the importance of the neighbors of node $i$ in the evaluation of its centrality, we can modify Equation \eqref{eq: def_degree_centrality} weighting each neighboring node by its importance:
\begin{equation}
    \label{eq: eigenvector_centrality}
    e_i \propto \sum_{j = 1}^{N_i} e_j,
\end{equation}
{where $e_i$ is the new centrality value assigned to node $i$.}
Using the adjacency matrix $A_{ij}$, defined in equation \eqref{eq:adjacency_matrix}, this centrality can be equivalently defined as:
\begin{equation}
    \label{eq: eigenvector_centrality_matrix}
    e_i \propto \sum_{j=1}^{N} A_{ij}\,e_j
\end{equation}
where now the summation is performed over all the nodes in the network.
The definition \eqref{eq: eigenvector_centrality_matrix} coincides with an eigenvalue equation for the adjacency matrix $A$, hence the name \emph{eigenvector centrality} (\cite{bonacich1987power}) for this measure.
For a matrix with non-negative entries the only eigenvector having all positive values is the one associated to the largest eigenvalue (\cite{perron1907theorie, frobenius1912matrizen}), so the components of this eigenvector can be associated to the centrality values of the nodes.

The implications of the global nature of the eigenvector centrality (as opposed to the local nature of degree centrality) can be seen from a description of how shocks propagate in the network. 
Applying a shock to a given node $i$, the number of nodes directly impacted is given by the sum:
\begin{equation}
        e_{i,1} = \sum_{j=1}^N A_{ij}
\end{equation}
i.e. the degree centrality of node $i$.
The impact of the shock at second order is given by the number of two-steps neighbors of node $i$:
\begin{equation}
    e_{i,2} = \sum_{j=1}^N \sum_{k=1}^N A_{ik} A_{kj} = \sum_{j=1}^N  \left(A^2 \right)_{ij} 
\end{equation}
To consider the long-term effects of the shock we can iterate the propagation to subsequent steps. 
The resulting expression is given by Equation \eqref{eq: eigenvector_centrality_matrix}, the eigenvector centrality associated to node $i$.

\begin{figure}[h]
\centering
  \includegraphics[width=1\linewidth]{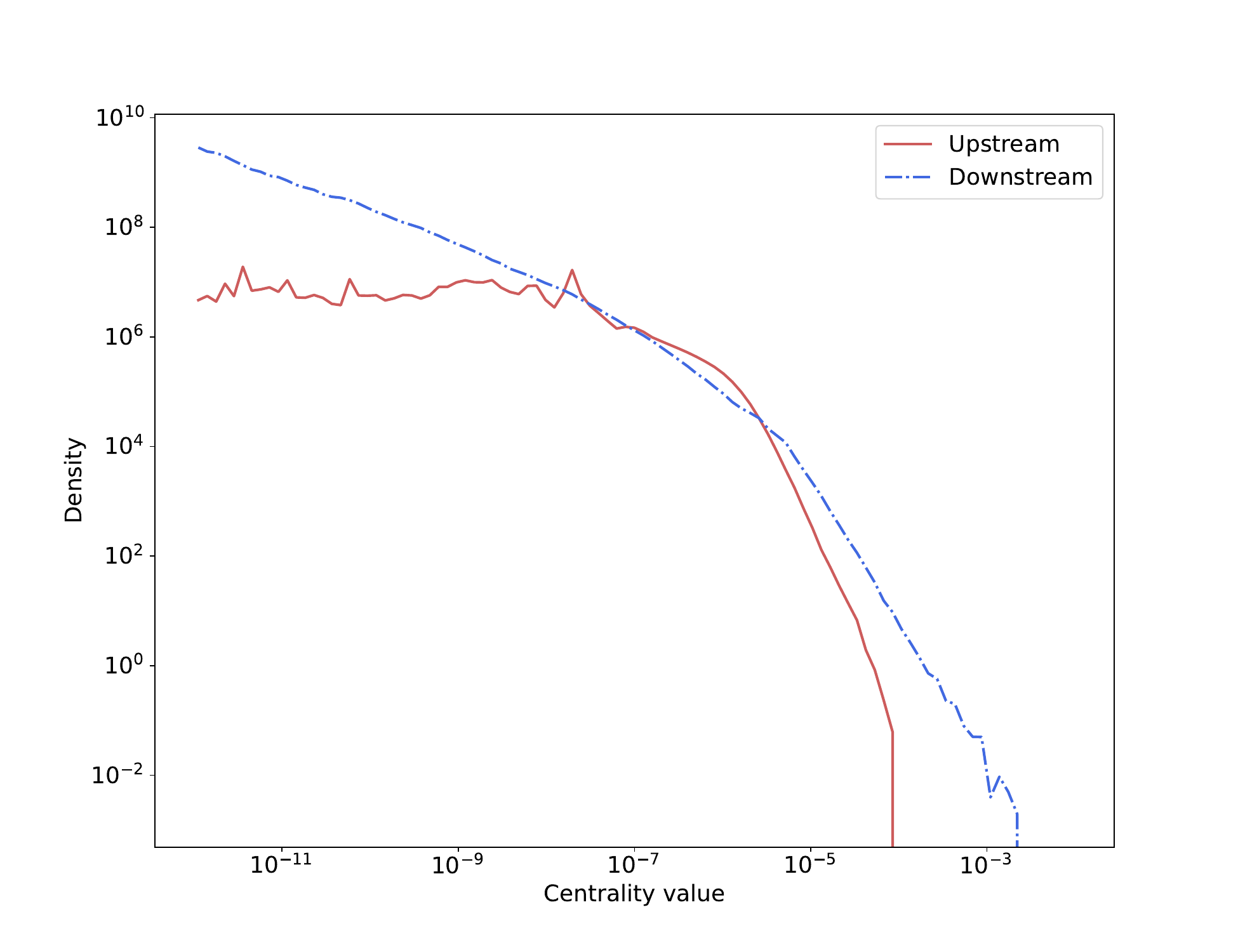}
  \caption{Density histograms of upstream and downstream eigenvector centrality. The x-axis is partitioned into 100 logarithmically spaced bins, while the y-axis reports the density of firms falling within each bin.}
  % \va{Alberto propone di commentare il plateau, ma io non saprei come, e comunque dipende dal fatto che la scala è logaritmica...}
  \label{fig:eigenvector_centrality}
\end{figure}

We plot the distribution of eigenvector centrality values for the nodes of the network, both in the upstream and downstream direction,\footnote{As mentioned in Section \ref{sec:closeness} there are approximately 11,000 firms which are sellers only, and approximately 600,000 which are buyers only. These firms have null {upstream} and {downstream} eigenvector centrality respectively. In Figure \ref{fig:eigenvector_centrality} we plot the distribution of values only for firms with centrality greater than zero.} in Figure \ref{fig:eigenvector_centrality}.
% We can see, similarly to the degree and sales distributions, that the centrality values span several orders of magnitude\va{commento alberto: \rev che significa"? Proviamo a spiegarglielo, o lo togliamo?}, both for the upstream and the downstream distribution. 
Confirming expectations in the literature (see \cite{bacilieri2026firm}), we notice that the downstream centrality distribution has a much greater dispersion than the upstream one.
The former is different from zero over almost ten orders of magnitude (from $10^{-12}$ to $10^{-2}$).
Upstream centrality values, on the other hand, are spread over eight orders of magnitude (from $10^{-12}$ to $10^{-4}$). 
This difference is confirmed by the tail exponents of the two distributions: the (right) tail exponent of the upstream eigenvector centrality distribution is equal to $\gamma_u = 3.58$ $(0.03)$, while the downstream one is equal to $\gamma_d = 2. 62\, (0. 01)$.
% \va{abbiamo aggiunto spiegazioni}
% \footnote{The error bars on the exponents represent the value of one standard deviation of the distribution of the estimated values.} \va{sezione troppo ostica, valorizzarla con resoconto non tecnico (???)}

% \begin{figure}[H]
% \centering
%   \includegraphics[width=0.7\linewidth]{plots/grafici_centralita/2024_12_23_adiacenze_soldi.pdf}
%   \caption{}
%   \label{fig:upstream_adjacency_centrality}
% \end{figure} 

% \begin{figure}[H]
% \centering
%   \includegraphics[width=0.7\linewidth]{plots/grafici_centralita/2024_12_23_adiacenze_merce.pdf}
%   \caption{}
%   \label{fig:downstream_adjacency_centrality}
% \end{figure}  

\subsection{Katz Centrality}
% \subsection{Dealing with Zero-Centrality Nodes: Katz Centrality}
The eigenvector centrality defined in the previous section takes into account the relative importance of all the nodes in the chain of connections terminating to the evaluated node, hence correctly representing the global properties of the network connectivity.
It has however a subtle shortcoming when a large number of sinks and sources are present in the network, as in our case (see Section \ref{sec:degree_distribution} and Appendix \ref{subsection:SinkSources}).
{These nodes have null eigenvector centrality.}
To see how this can introduce distortions in the centrality values of other nodes, it is sufficient to consider for example a node which has a large number of out-going connections, all of which are going to sink nodes. 
By definition \eqref{eq: eigenvector_centrality_matrix} sink nodes have zero {downstream} eigenvector centrality (they have no out-going links), hence they will not contribute to the eigenvector centrality of the nodes they are connected to. 
This implies that a node could have a large number of out-going connections, but a very small centrality if a large share of these connections go to zero-centrality nodes.
% \va{Non sono solo i sink. Il problema si propaga anche ad altri nodi.} 

\begin{figure}[h]
\centering
  \includegraphics[width=1\linewidth]{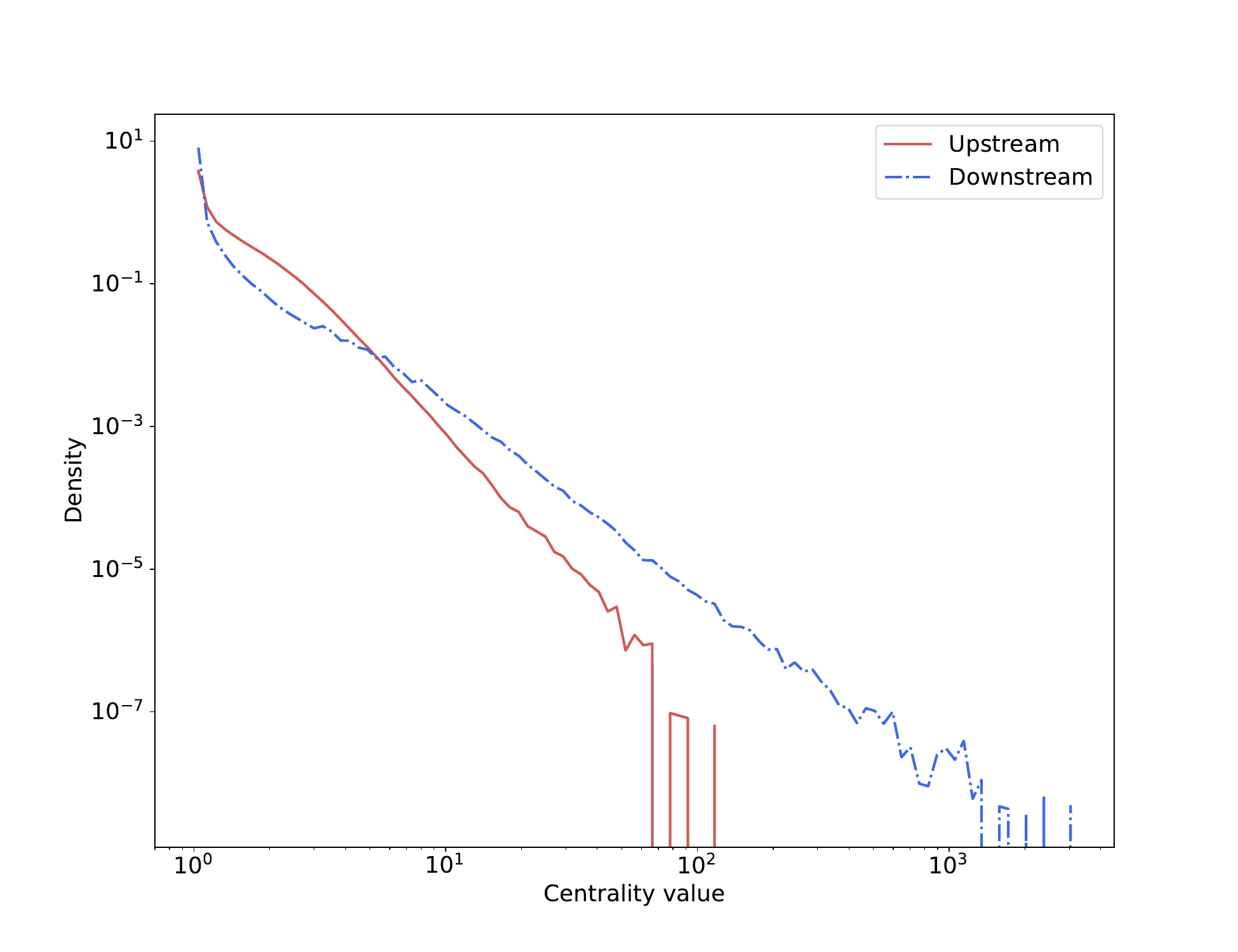}
  \caption{ Density histograms of upstream and downstream Katz centrality. The x-axis is partitioned into 100 logarithmically spaced bins, while the y-axis reports the density of firms falling within each bin.}
  \label{fig:katz_centrality}
\end{figure}

To obviate this limitation a small value of centrality is assigned to every node, independently of the number of {its} neighbors. 
% \va{Commento di alberto: \rev ???e questa parte descrittiva è molto importante, perchè i dettagli tecnici sotto sono piuttosto ostici. Di nuovo, è importante accompagnare la parte tecnica con una narrazione comprensibile a un lettore mediano di BI senza laurea in matematica... Sfugge inoltre i link tra quest para e il precedente sui sink"}
The resulting quantity - called \emph{Katz centrality} (\cite{katz1953new}) - can be defined as:
% \begin{equation}
%     \label{eq: eigenvector_centrality_matrix}
%     e_i \propto \sum_{j=1}^{N} A_{ij}\,e_j
% \end{equation}

\begin{equation}
    \label{eq: katz_centrality}
    k_i = \alpha \sum_{j=1}^{N} A_{ij}\,k_j + \beta
\end{equation}
The two parameters $\alpha$ and $\beta$ determine the relative importance of the standard eigenvector centrality ($\alpha$) and the minimum centrality value assigned to all nodes ($\beta$).
Given that we are interested only in the relative values of $k_i$, only the ratio between $\alpha$ and $\beta$ is meaningful, and we can redefine Katz centrality as:
\begin{equation}
    \label{eq: katz_centrality_1}
    k_i = \alpha \sum_{j=1}^{N} A_{ij}\,k_j + 1 %k_i = \alpha \sum_{j}^{N} A_{ij}\,k_j + 1
\end{equation}
This equation can be inverted to obtain: % (in vector form):
% \begin{equation}
% \label{eq: vec_katz}
%     \vec{k} = \left( \hat{\mathbb{1}} - \alpha \hat{A} \right)^{-1} \vec{1}
% \end{equation}
\begin{equation}
\label{eq: vec_katz}
    k_i = \sum_{j=1}^{N} \left( {\mathbb{1}} - \alpha A \right)^{-1}_{ij} 
\end{equation}
where $\mathbb{1}$ is the identity matrix and $A$ is the adjacency matrix. %, and $\vec{1}$ is the vector of all ones. 
The maximum value of the parameter $\alpha$ for the right hand side of equation \eqref{eq: vec_katz} to be defined is the inverse of the maximum eigenvalue $\lambda_{\text{max}}^{A}$ of the adjacency matrix ${A}$. 
It is possible to deduce from equation \eqref{eq: katz_centrality_1} that a value of $\alpha$ close to its maximum gives more weight to the standard eigenvector centrality, and assigns a small base-value to zero-centrality nodes.
On the other hand, for $\alpha$ close to zero, we obtain a centrality close to $1$ for all nodes.  
In the following we chose a value of $\alpha$ close to its maximum, $\alpha = 0.85/\lambda_{\text{max}}^{A}$.\footnote{As we will see in the following, this value coincides with the one used by Google Search in its original implementation of PageRank centrality.}
% \va{The maximum value compatible with the existence of equation 15}
The distribution of Katz centrality values is shown in Figure \ref{fig:katz_centrality}, for the upstream and downstream networks. 
As can be seen from the figure, the behavior for low centrality values is radically different from the eigenvector centrality: while there are nodes in the network presenting arbitrarily small eigenvector centrality, the introduction of a minimal value in equation \eqref{eq: katz_centrality} implies that the minimum Katz centrality value is 1 (or, if a different normalization is considered, a constant). 
In addition the distribution of Katz centralities is much more concentrated, spanning respectively two and three orders of magnitude, as opposed to the eight and ten of eigenvector centrality.
The tail exponents of the upstream and downstream Katz centrality distributions are $\gamma_u = 3.87\, (0.02)$ and $\gamma_d = 2.70\, (0.01)$.

% \begin{figure}[H]
% \centering
%   \includegraphics[width=0.7\linewidth]{plots/grafici_centralita/2025_01_08_katz_soldi.pdf}
%   \caption{Updated upstream Katz centrality}
%   \label{fig:upstream_katz_centrality}
% \end{figure} 

% \begin{figure}[H]
% \centering
%   \includegraphics[width=0.7\linewidth]{plots/grafici_centralita/2025_01_08_katz_merce.pdf}
%   \caption{Updated downstream Katz centrality}
%   \label{fig:downstream_katz_centrality}
% \end{figure} 

% \begin{figure}[H]
% \centering
%   \includegraphics[width=0.7\linewidth]{plots/grafici_centralita/2024_12_23_katz_soldi.pdf}
%   \caption{}
%   \label{fig:upstream_katz_centrality}
% \end{figure} 

% \begin{figure}[H]
% \centering
%   \includegraphics[width=0.7\linewidth]{plots/grafici_centralita/2024_12_23_katz_merce.pdf}
%   \caption{}
%   \label{fig:downstream_katz_centrality}
% \end{figure}  

\subsection{PageRank Centrality}
\begin{figure}[h]
\centering
  \includegraphics[width=1\linewidth]{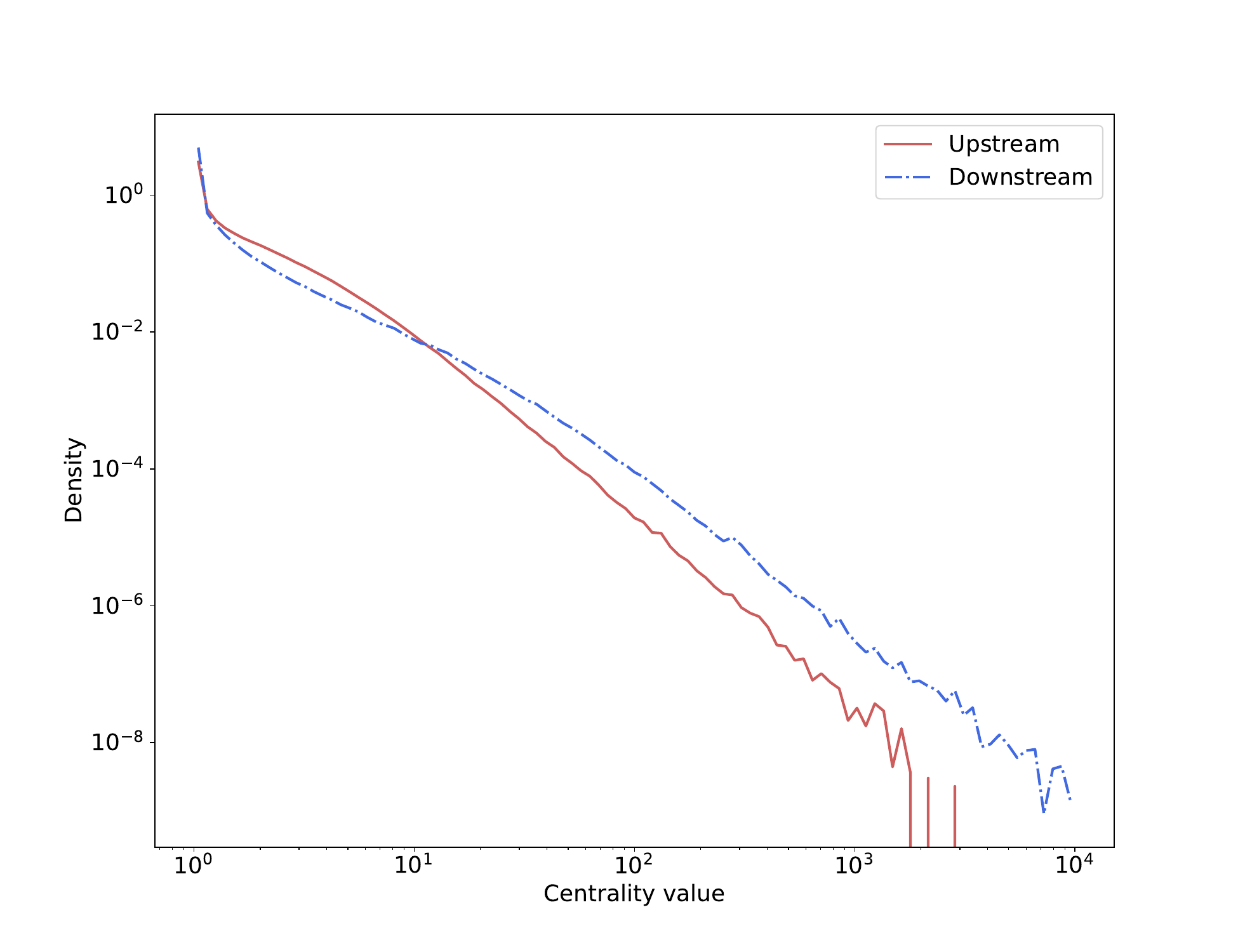}
  \caption{Density histograms of upstream and downstream PageRank centrality. The x-axis is partitioned into 100 logarithmically spaced bins, while the y-axis reports the density of firms falling within each bin.}
  \label{fig:pagerank_centrality}
\end{figure}

% \subsection{Less is More: PageRank Centrality}
The last centrality measure we take into consideration is the so-called \emph{PageRank} centrality, introduced by Google Search to classify search results by relevance (\cite{brin1998anatomy, page1999pagerank}).
The main difference with respect to Katz centrality is a normalization of the columns of the adjacency matrix, which in this case takes into account the number of out-going links from each node.  
As we saw in Section \ref{sec: eigenvector_centrality}, a node has a high centrality value if it has many connections with nodes with high centrality. 
Still, not all nodes with high centrality are equal. 
In particular the idea leading to PageRank centrality is that an out-going link to a central node with few incoming links should weight more than an out-going link to a central node with many incoming links.
% \footnote{The number of links in our network is roughly comparable with the number of Internet links over which the original PageRank algorithm was tested, and the number of iterations to convergence is similar (see \cite{page1999pagerank}).
% Paraphrasing the description of the authors of the algorithm: \rev Nodes that have perhaps only one link from a very important node are also generally worth visiting." 
% }  
The definition of PageRank centrality can be written as:
\begin{equation}
    \label{eq: pagerank_centrality}
    p_i = \alpha \sum_{j=1}^{N} \frac{A_{ij}}{A_j}\,p_j + \beta \quad , \quad A_j = \sum _{i=1}^{N} A_{ij}
\end{equation}
where $A_{ij}$ is the adjacency matrix element from node $i$ to node $j$, and $A_j$ is the sum of these matrix elements over $i$, or in other words the number of incoming link to node $j$.
In practice this translates into using a matrix $M_{ij} = \frac{A_{ij}}{A_j}$ which is column-stochastic.\footnote{A matrix is defined column-stochastic (or row-stochastic) if its elements are non-negative and all its columns (rows) sum to one.}
% The rationale for using a columns-stochastic matrix in the original formulation of PageRank was to decrease the importance associated to links coming from web-pages with many out-going links.
% As an example, given that Google links to hundreds of billions of pages, having an incoming link from Google should not increase significantly the centrality of a page, even though Google has a high centrality score.} 
% \va{commento alberto: \rev mettere nel testo principale ultima frase per aiutare il lettore a capire il razionale di questa metrica?" - MESSA}} 
The rationale for using a columns-stochastic matrix in the original formulation of PageRank was to decrease the importance associated to links coming from web-pages with many out-going links. 
As an example, given that Google links to hundreds of billions of pages, having an incoming link from Google should not increase significantly the centrality of a page, even though Google has a high centrality score.
As in the case of Katz centrality, the value of $\beta$ is set to 1, hence \eqref{eq: pagerank_centrality} can be written as:
\begin{equation}
    p_i = \alpha \sum_{j=1}^{N} {M_{ij}}\,p_j + 1 \quad 
\end{equation}
A different interpretation of this centrality measure can be given in terms of a probability distribution over the nodes of the network. 
In particular when PageRank centrality is normalized to one, it corresponds to the probability of visiting a given node after a large number of random steps through the network (\cite{brin1998anatomy}).
% \va{Commento alberto: \rev ecco, questo è un bell'esempio di come "far capire" le cose a un lettore mediano BI"}
The distributions of PageRank centrality for the upstream and downstream networks are shown in Figure \ref{fig:pagerank_centrality}.
As can be seen in the figure there are many similarities with Katz centrality, but PageRank centrality is less concentrated, and correspondingly there are many more nodes of disproportionate centrality by this measure. 
Similarly to the other centrality measures, the upstream centrality distribution is more concentrated than the downstream one, with respective tail exponents of $\gamma_u = 2.75\, (0.02)$ and $\gamma_d = 2.37\, (0.01)$. 
% \va{commento alberto: \rev mi è rimasto molto poco di queste sezioni 4.1-4.2-4.3. Sarebbe utile un recap finale per sintetizzare i main takeaways concettuali (i.e. senza formalismi)"}
% Alla fine mi sembra di aver capito che alberto si fosse convinto che queste sono sezioni tecniche, quindi non serviva spiegare troppo di più 
% \begin{figure}[H]
% \centering
%   \includegraphics[width=0.7\linewidth]{plots/grafici_centralita/2025_02_03_pagerank_adiacenze_soldi.pdf}
%   \caption{}
%   \label{fig:upstream_pagerank_centrality}
% \end{figure} 

% \begin{figure}[H]
% \centering
%   \includegraphics[width=0.7\linewidth]{plots/grafici_centralita/2025_02_03_pagerank_adiacenze_merce.pdf}
%   \caption{}
%   \label{fig:downstream_pagerank_centrality}
% \end{figure} 

%\subsection{Weighted links}
\subsection{Weighted Links}
\label{sec:weighted_links}

\begin{figure}[h]
\centering
  \includegraphics[width=1\linewidth]{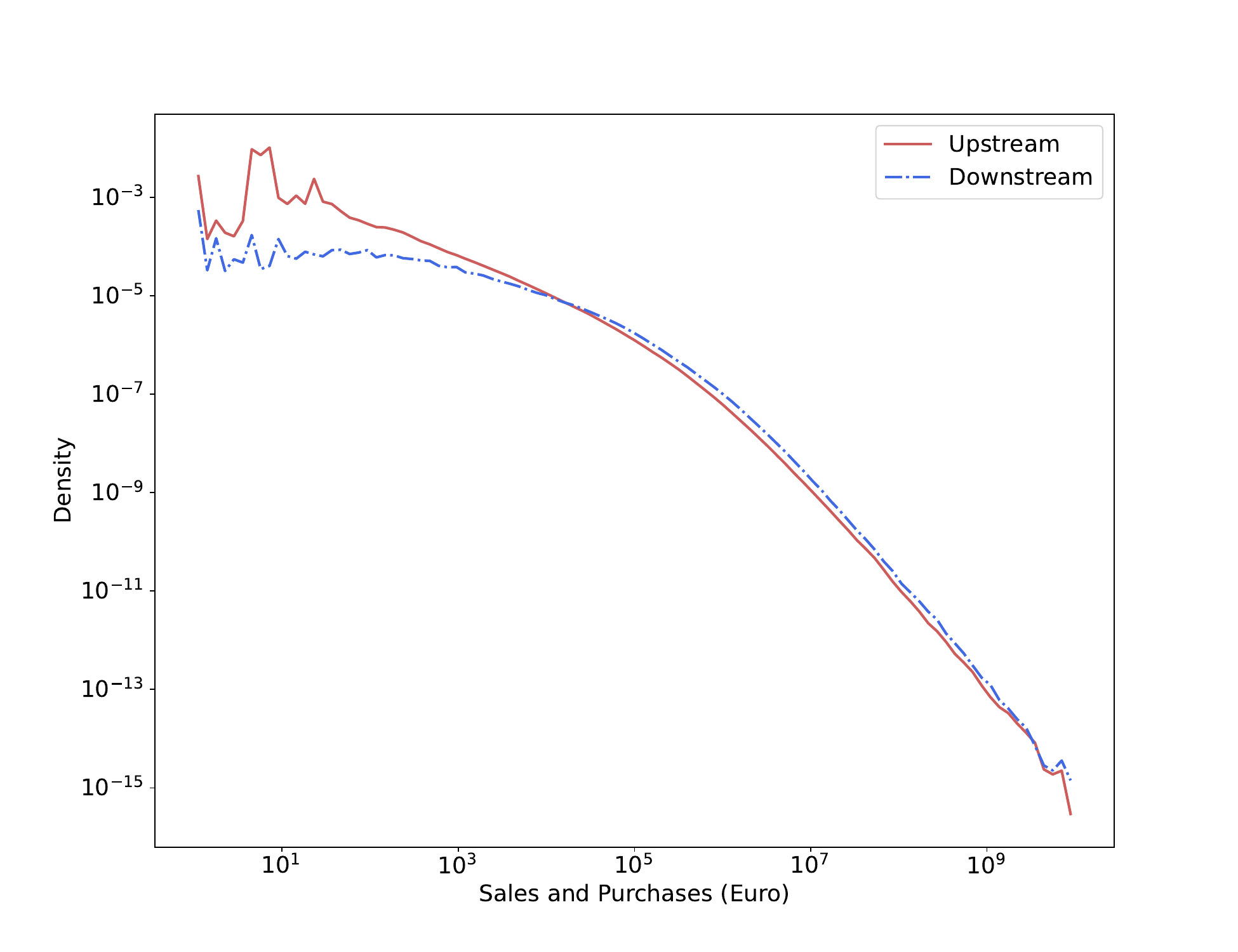}
  \caption{Density of sales (downstream) and purchases (upstream). The x-axis is partitioned into 100 logarithmically spaced bins and the y-axis reports the density of amounts falling within each bin.}
  \label{fig:sales_distribution}
\end{figure}

\begin{figure}[h]
\centering
  \includegraphics[width=1\linewidth]{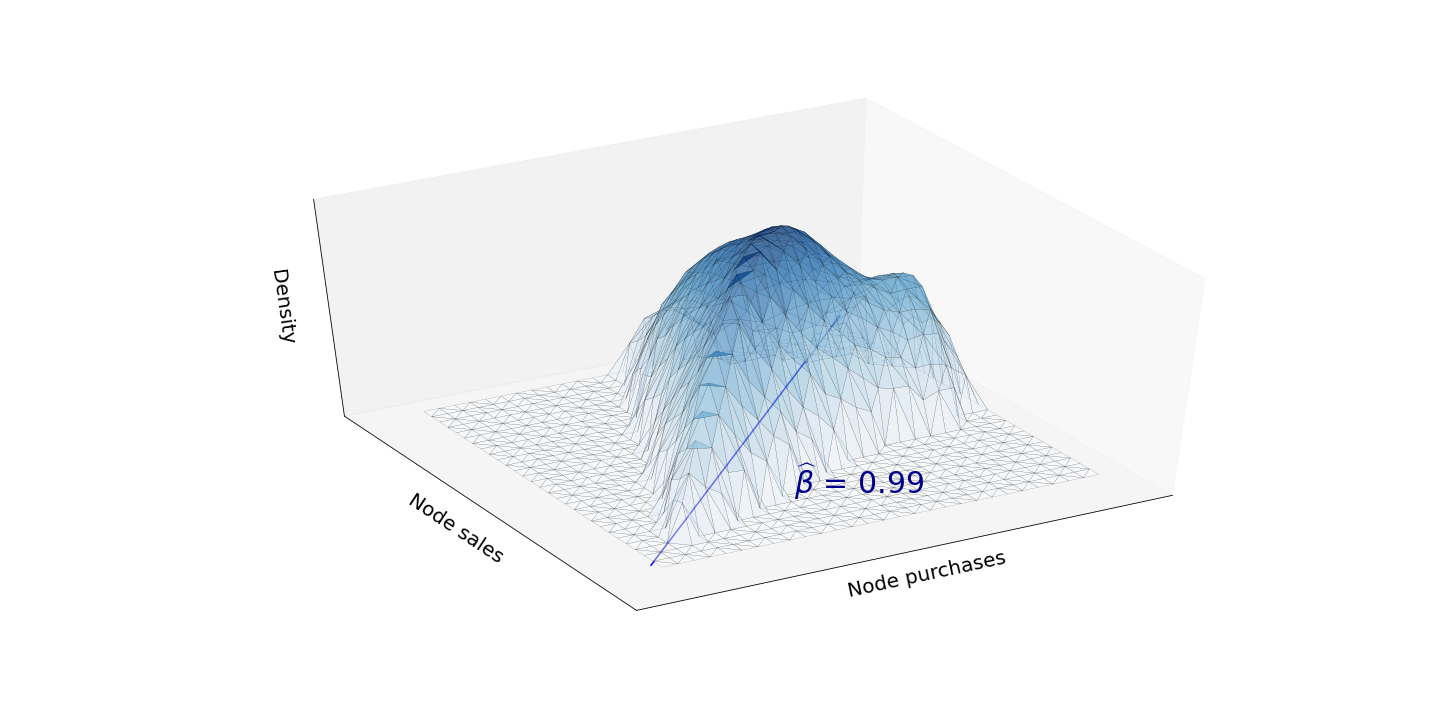}
  \caption{3D surface plot of the joint distribution of out-strength (sales) and in-strength (purchases) of the firms. We divide the joint distribution domain in 625 logarithmically spaced squares and the surface reports the number of nodes falling in each square. The blue line report the Total Least Squares estimated coefficient between the two quantities.}
  \label{fig:hist2d_weights_joint_2019}
\end{figure}

\begin{figure}[h]
\centering
  \begin{subfigure}[b]{0.49\textwidth}
    \centering
    \includegraphics[width=1\linewidth]{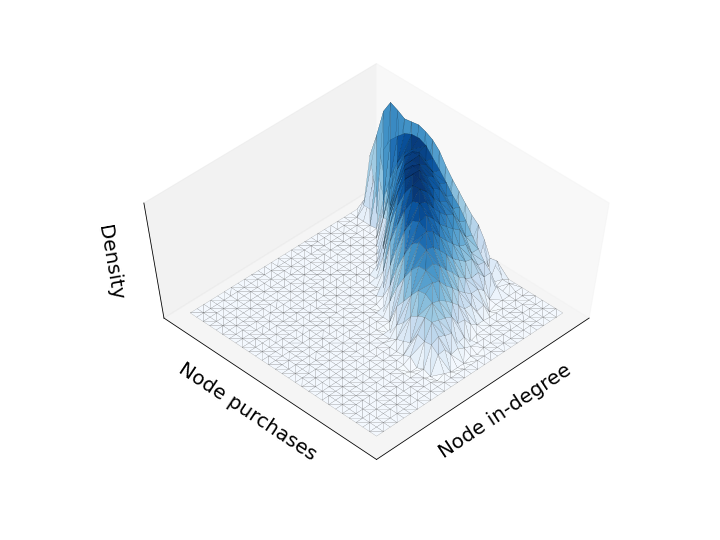}
    \caption{Purchases versus number of incoming links.} 
    \label{fig:deg_str_in_degree_in_strength}
    %\vspace{4ex}
  \end{subfigure}%% 
  \begin{subfigure}[b]{0.49\textwidth}
    \centering
    \includegraphics[width=1\linewidth]{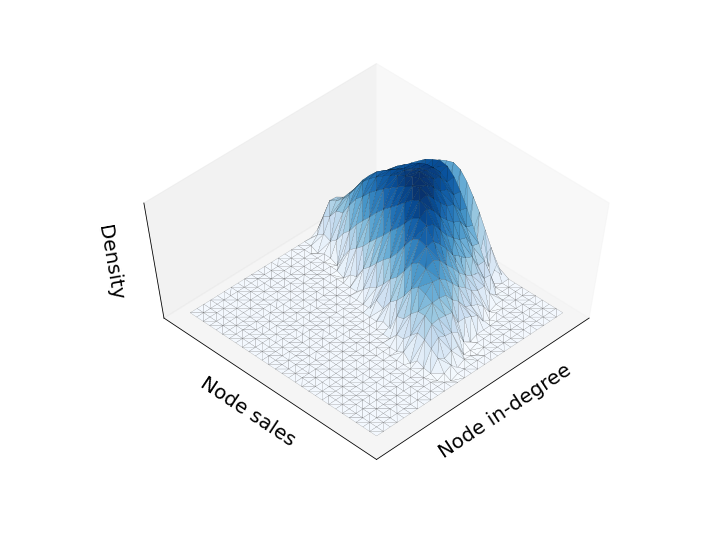}
    \caption{Sales versus number of incoming links.} 
    \label{fig:deg_str_in_degree_out_strength} 
    %\vspace{4ex}
  \end{subfigure} 
  \begin{subfigure}[b]{0.49\textwidth}
    \centering
    \includegraphics[width=1\linewidth]{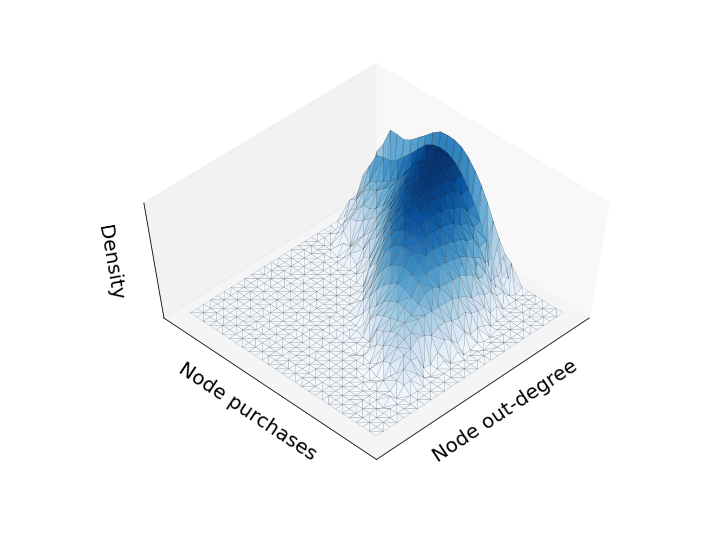}
    \caption{Purchases versus number of out-going links.} 
    \label{fig:deg_str_out_degree_in_strength} 
  \end{subfigure}%%
  \begin{subfigure}[b]{0.49\textwidth}
    \centering
    \includegraphics[width=1\linewidth]{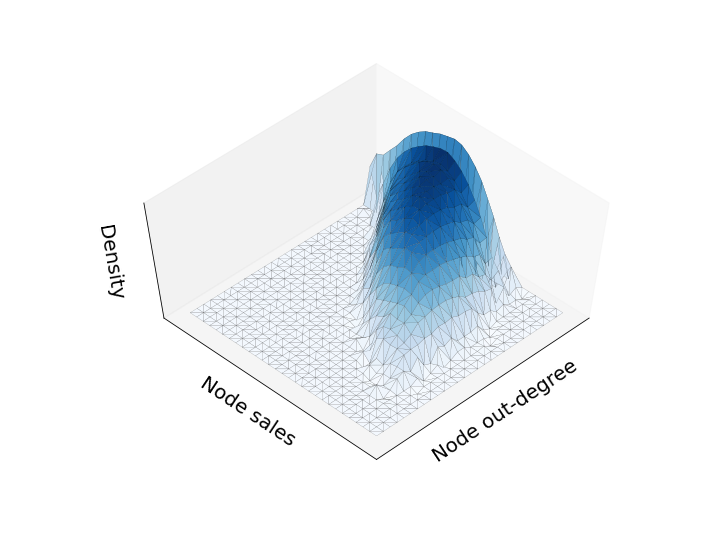}
    \caption{Sales versus number of out-going links.} 
    \label{fig:deg_str_out_degree_out_strength} 
  \end{subfigure} 
  \caption{3D surface plot of the joint distributions of degrees and links intensity. We divide the domain in 625 logarithmically spaced squares and the surface reports the number of node falling in each square.}
  \label{fig:sales_degree_correlation} 
\end{figure}

In the previous sections we considered centrality measures defined by the adjacency matrix.
%, with entries equal to 1 if the link corresponding to the matrix element is present, and 0 otherwise.  
These measures take into account the connectivity structure of the network, but not the amounts exchanged over the links. 
In this section we consider the weights associated to the links. 
A matrix describing the weighted network can be defined in a similar way to equation \eqref{eq:adjacency_matrix}:
\begin{equation} 
\label{eq:weighted_adjacency_matrix}
 S = (S_{ij})_{u_i,u_j \in \mathcal{N}} = \left\{
\begin{array}{cc}
      w_{ij} & (u_i,u_j) \in \mathcal E\\
       & \\
      0 & \text{otherwise}
      \end{array}
\right.
\end{equation}
where $\mathcal N$ and $\mathcal E$ are the sets of nodes and links in the network, defined in Section \ref{sec:FromDataToNetwork}, and $w_{ij}$ is the total amount of invoices issued by firm $u_i$ to firm $u_j$.
The exchanged amounts in the network, as many other quantities, are spread over many orders of magnitude.
The minimum values are smaller than one euro, while the largest ones are of the order of billions of euro. 
% This can be seen in Figure \ref{fig:exchanged_amounts}: 
This implies that a large fraction of the exchanged amount in the network is concentrated in few, important links.
We report summary statistics of the distribution of exchanged amounts in Table \ref{tab:exchanged_amounts_stats}, and plot it in Appendix \ref{app:additional_figures}, Figure \ref{fig:exchanged_amounts}.
\begin{table}[H]
    \centering
    \small
   %\begin{tabular}{lrrrrrr}
   %%\toprule 
   %{} & \multicolumn{1}{c}{Min} &\multicolumn{1}{c}{p25} &\multicolumn{1}{c}{p50} &\multicolumn{1}{c}{p75} &\multicolumn{1}{c}{Max (billion)} &\multicolumn{1}{c}{Average}\\
   %\midrule
   %Exchanged amounts & 0.01 & 161 & 774 & 4,000 & 6.85 & 28,599 \\
   %\bottomrule
   %\end{tabular}

   \begin{tabular}{lrrrrrr}
    %\toprule 
    {} &\multicolumn{1}{c}{p01} &\multicolumn{1}{c}{p25} &\multicolumn{1}{c}{p50} &\multicolumn{1}{c}{p75} &\multicolumn{1}{c}{p99} &\multicolumn{1}{c}{Average}\\
    \midrule
    Exchanged amounts & 8 & 161 & 774 & 4,000 & 311,176 & 28,599 \\
    \bottomrule
    \end{tabular}

    \caption{Exchanged amounts summary statistics. All values are expressed in euro.
    The standard deviation is not reported, in that it is not meaningful for the distribution.}
    \label{tab:exchanged_amounts_stats}
\end{table}
A similar behavior is found in the distribution of sales and purchases of each firm (i.e. the aggregation by firm of the exchanged amounts, upstream and downstream), as can be seen in Figure \ref{fig:sales_distribution}.
The exchanged amount distribution has a power-law right tail with exponent $\gamma_{w} = 2.23\, (0.02)$, while the total sales and total purchases per firm have distributions with right tails exponents $\gamma_d = 2.10\, (0.01)$ for the sales, $\gamma_u = 2.11\, (0.01)$ for the purchases.
% \va{commento alberto: \rev questa nota merita di andare nel testo principale, come "utilizzo economico" delle cose che avete fatto. Va rimpolpata la parte descrittiva/narrativa e chiarito un po' (tipo: cos'è il suppression factor?) Magari un esempio aiuta?"} 
These exponents play a crucial role in connecting shocks to single firms and aggregate fluctuations of the production system output. 
A common expectation in the economics literature, derived from the assumption that relevant quantities are normally distributed, was that aggregate fluctuations in the production system output should be attenuated by a factor $\sqrt{N}$, with $N$ the number of firms in the system (see \cite{lucas1977understanding, carvalho2019production}). 
The presence of a heavy-tailed distribution of firms' sales, on the other hand, implies that the combination of shocks to single firms - even assuming independence - does not result in an attenuation of the aggregate fluctuations proportional to the square root of the number of firms. 
The expected attenuation factor in this case is proportional to $N^{(\gamma_d - 2)/(\gamma_d - 1)} \approx N^{\text{0.09}}$ (see \cite{gabaix2011granular} for a detailed derivation).
The sales and purchases of each firm show an almost perfect proportionality: as can be seen from Figure \ref{fig:hist2d_weights_joint_2019}, the best linear fit of the relation between sales and purchases has an estimated coefficient very close to 1. 
Similarly, sales and purchases show a strong correlation with the in- and out-degree of the firms, as can be seen from Figure \ref{fig:sales_degree_correlation}.
% \begin{figure}[H]
% \centering
%   \includegraphics[width=1\linewidth]{plots/importi_log_scale_2.pdf}
%   \caption{
%   \va{Le x non sono più espresse in euro} Density histograms of the invoices amounts. The x-axis is partitioned into 100 logarithmically spaced bins and the y-axis reports the density of invoices falling within each bin.}
%   \label{fig:exchanged_amounts}
% \end{figure} 
In turn, as described in Section \ref{sec:degree_distribution}, in- and out-degree are themselves strongly correlated, hence a strong dependence can be found also between sales and in-degree and purchases and out-degree.
The fact that exchanged amounts are distributed over many orders of magnitude is reflected on the properties of the weighted network.  
In particular the weighted network is effectively much sparser than the adjacency one. 
As shown in Table \ref{tab:amounts_concentration}, the exchanged amounts are concentrated on a small number of links of the network: only 3\% of the links are needed to account for more than 80\% of the aggregate exchanged value.
%weighting each link of the network by the amount exchanged over that link, we obtain a network which is effectively much sparser than the one considered in the previous sections. 
% For example transactions larger than 100.000 euro represent less than 3\% of the number of links of the weighted network, but their aggregate amount exceeds 80\% of the total amount exchanged.
% \vspace{-0.1cm}
\begin{table}[h]
    \centering
    \small
    \begin{tabular}{lcccccc}
    %\toprule 
    {Threshold (euro)} & \multicolumn{1}{c}{$10^3$} &\multicolumn{1}{c}{$10^4$} &\multicolumn{1}{c}{$10^5$} &\multicolumn{1}{c}{$10^6$} &\multicolumn{1}{c}{$10^7$} &\multicolumn{1}{c}{$10^8$}\\
    \midrule
    Fraction of links   & 0.46 & 0.15 & 0.03 & $3.1\times10^{-3}$ & $2.3\times10^{-4}$ & $1.4\times10^{-5}$ \\
    Fraction of total amount    & 0.99 & 0.97 & 0.83 & 0.59 & 0.34 & 0.17 \\
    \bottomrule
    \end{tabular}
    \caption{Concentration of the exchanged amounts. 
    For each amount listed, the first row shows the fraction of links that involve at least that amount.
    The second row shows the fraction of the aggregate transaction value that comes from transactions of at least that amount.
    As an example, amounts larger than 100,000 euro represent 3\% of the number of links, but their aggregate amount exceeds 80\% of the total amount exchanged in the network.}
    % amounts larger than 100 million euro account for $1,4\times10^{-5}$ of the total number of links, and 0,17 of the total value exchanged in the network.}
    \label{tab:amounts_concentration}
\end{table}
% \va{commento alberto: \rev ma più che un esempio, non sarebbe utile una tabellina per \rev 
% visualizzare" questa joint distribution?"}
% \va{Io ho messo una tabellina sopra, ma ovviamente non è della joint distribution - che non ricordo come volevamo rappresentare.}
As described in the previous sections, centrality measures are defined in terms of sums over rows of the matrix describing the network.
If a given row presents a value much larger than all the others, every summation of the type defined in equations \eqref{eq: eigenvector_centrality} and \eqref{eq: katz_centrality} will be dominated by this value. 
%\sout{, and the presence of smaller matrix elements will be inconsequential.}
%\va{alberto ha cancellato questa frase, ma senza perde di senso tutto il paragrafo.}
Hence the centrality values will be determined by the few links corresponding to the largest exchanged amounts in the network. 
For this reason centrality measures derived from the weighted matrix, without row- or column-wise normalization, are less interesting than the one derived from the {unweighted} adjacency matrix.
The previous reasoning does not apply to PageRank centrality, defined in Equation \eqref{eq: pagerank_centrality}, and for which the summations are performed over column-normalized matrix elements. 
This implies that all matrix elements are comprised between zero and one, and it is quite possible to have rows with many elements of comparable magnitude. 
% \sout{Here we consider PageRank centrality for the weighted network, which as we have seen is derived considering a transition matrix with normalized columns.}
This matrix has also an interesting interpretation in terms of the flow of money (or goods) through the network: 
suppose we want to model the evolution of an initial distribution of purchases, described as a vector with one entry for each firm in the network, representing the total expenses of the firm. 
The evolution of this distribution is given by multiplication for the column-normalized weighted matrix describing the downstream network.\footnote{The downstream matrix has column index labeling the buyers, hence we can sum all the purchases from a given seller multiplying by the downstream matrix to the right.}
As an example, let's consider as initial distribution the vector with 1 euro for a given firm, and 0 for all the others. 
The distribution of this amount after one year is identical to the normalized vector of yearly expenses of this firm, i.e. the normalized column associated to the downstream weighted matrix. 
If $W_{ij}$ are the elements of the normalized matrix, they represent the fraction of the purchases from buyer $j$ to seller $i$ in one year. 
The element of the matrix power $\left(W^n\right)_{ij}$ represents the fraction of the purchases from buyer $j$ to seller $i$ after $n$ (identical) years.

\begin{figure}[h]
\centering
  \includegraphics[width=1\linewidth]{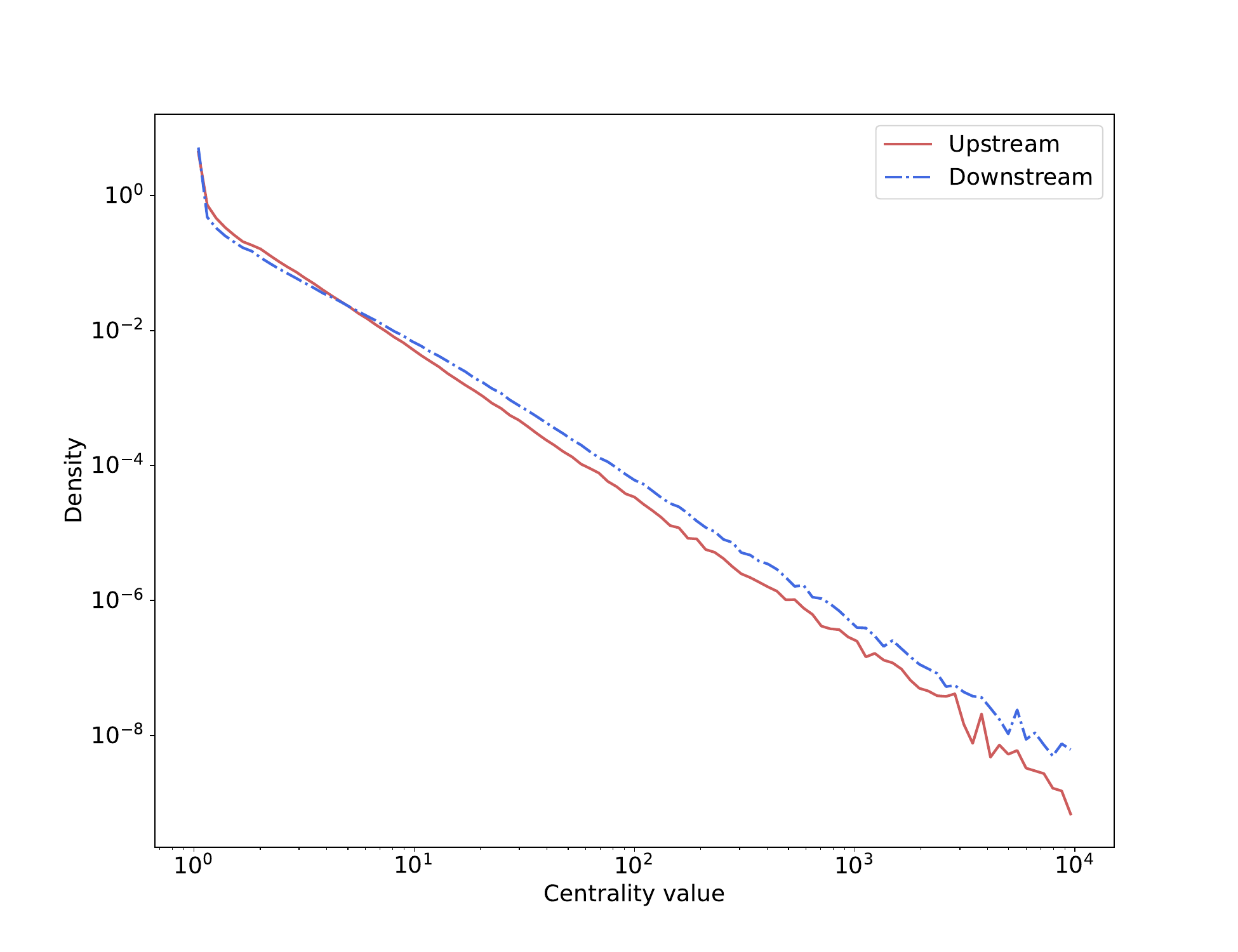}
  \caption{Density histograms of upstream and downstream weighted PageRank centrality. The x-axis is partitioned into 100 logarithmically spaced bins, while the y-axis reports the density of firms falling within each bin.}
  \label{fig:weighted_pagerank_centrality}
\end{figure}

The definition of PageRank centrality for the downstream flow can be written as:
\begin{equation}
    \label{eq:eigenvector_centrality}
    p_i = \alpha \sum_{j} \frac{S_{ij}}{S_j}\,p_j + \beta
\end{equation}
where $S_{ij}$ is the yearly sales value from node $i$ to node $j$, and $S_j$ are the total yearly sales to node $j$ (or, equivalently, the total purchases of firm $j$).
In practice this translates into using a matrix $W_{ij} = \frac{S_{ij}}{S_j}$ which is column-stochastic.
The distribution of PageRank centrality values over the nodes of the network is shown in Figure \ref{fig:weighted_pagerank_centrality}. 
% This fact is easy to understand given the heterogeneous distribution of exchanged amounts over the links of a given node: considering only the adjacency values is equivalent to evaluating a PageRank centrality in a network with equal amounts exchanged over each link. 
% Given that in the real network the distribution over the links for each node can be highly concentrated, also centrality values are more concentrated. 
The main difference with respect to the previously considered measures is a clear power-law behavior of a larger portion of the distribution, and the lower tail exponents $\gamma_u = 2.18\, (0.01)$, $\gamma_d = 2.10\, (0.01)$.
The PageRank centrality measure for the weighted network is the most suited to assess shock propagation in the network. 
In particular the {downstream} PageRank centrality value of a firm approximates the long-term effect on the firm of a random demand-side shock on the network. Analogously, the {upstream} PageRank centrality represents the effect on a given firm of a random supply-side shock on the network. 
% \va{commento alberto: \rev ottimo linguaggio per parlare al lettore mediano in BI. Valorizzare, usando fig 20 per dire, in questi termini, cosa trovate. Chiarire meglio "effect on a given firm": effect on its sales? on its purchases? Visto che la cosa è appealing e che dalle distribuzioni con bin logartmici e densità si capisce poco, aggiungerei tavolina con i quartili, min, max e media (separato per upstream e down)"}
On the other hand the centrality values of a firm quantify how much a shock to that firm will impact the network as a whole {in the reversed direction}.\footnote{See Appendix \ref{app:matrix_representation} for a short review of the relationship between centrality and shock propagation.}
The lower tail exponents of the PageRank centrality distributions indicate a greater concentration of centrality in the right-tail.
% \va{Non mi è chiaro come dovrei citare i debiti commerciali qui...}
This fact, in turn, implies the presence of a small number of firms which are on one hand sensitive to diffused shocks to the network, and on the other very effective at transferring direct, concentrated shocks to the rest of the network. 
We show a set of descriptive statistics for the PageRank centrality distributions in Table \ref{tab:stats_pagerank}.
\begin{table}[h]
    \centering
    \small
    %\begin{tabular}{lrrrrrr}
    %%\toprule 
    %{} & \multicolumn{1}{c}{Min} &\multicolumn{1}{c}{p25} &\multicolumn{1}{c}{p50} &\multicolumn{1}{c}{p75} &\multicolumn{1}{c}{Max} &\multicolumn{1}{c}{Average}\\
    %\midrule
    %Upstream   & 1.00 & 1.01 & 1.19 & 2.22 & 25,408  & 3.85 \\
    %Downstream & 1.00 & 1.00 & 1.09 & 2.32 & 153,348 & 6.60 \\
    %\bottomrule
    %\end{tabular}

     \begin{tabular}{lrrrrrrrr}
    %\toprule 
    {} & \multicolumn{1}{c}{p01} & \multicolumn{1}{c}{p25} &\multicolumn{1}{c}{p50} &\multicolumn{1}{c}{p75} & \multicolumn{1}{c}{p99} &\multicolumn{1}{c}{Average}\\
    \midrule
    Upstream   & 1.00 & 1.01 & 1.19 & 2.22  & 33.29 & 3.85 \\
    Downstream & 1.00 & 1.00 & 1.09 & 2.32  & 58.72 & 6.60 \\
    \bottomrule
    \end{tabular}
    
    \caption{PageRank centrality summary statistics. Standard deviations are not reported because they are not meaningful for these distributions.}
    \label{tab:stats_pagerank}
\end{table}

\begin{figure}[H]
\centering
  \includegraphics[width=1\linewidth]{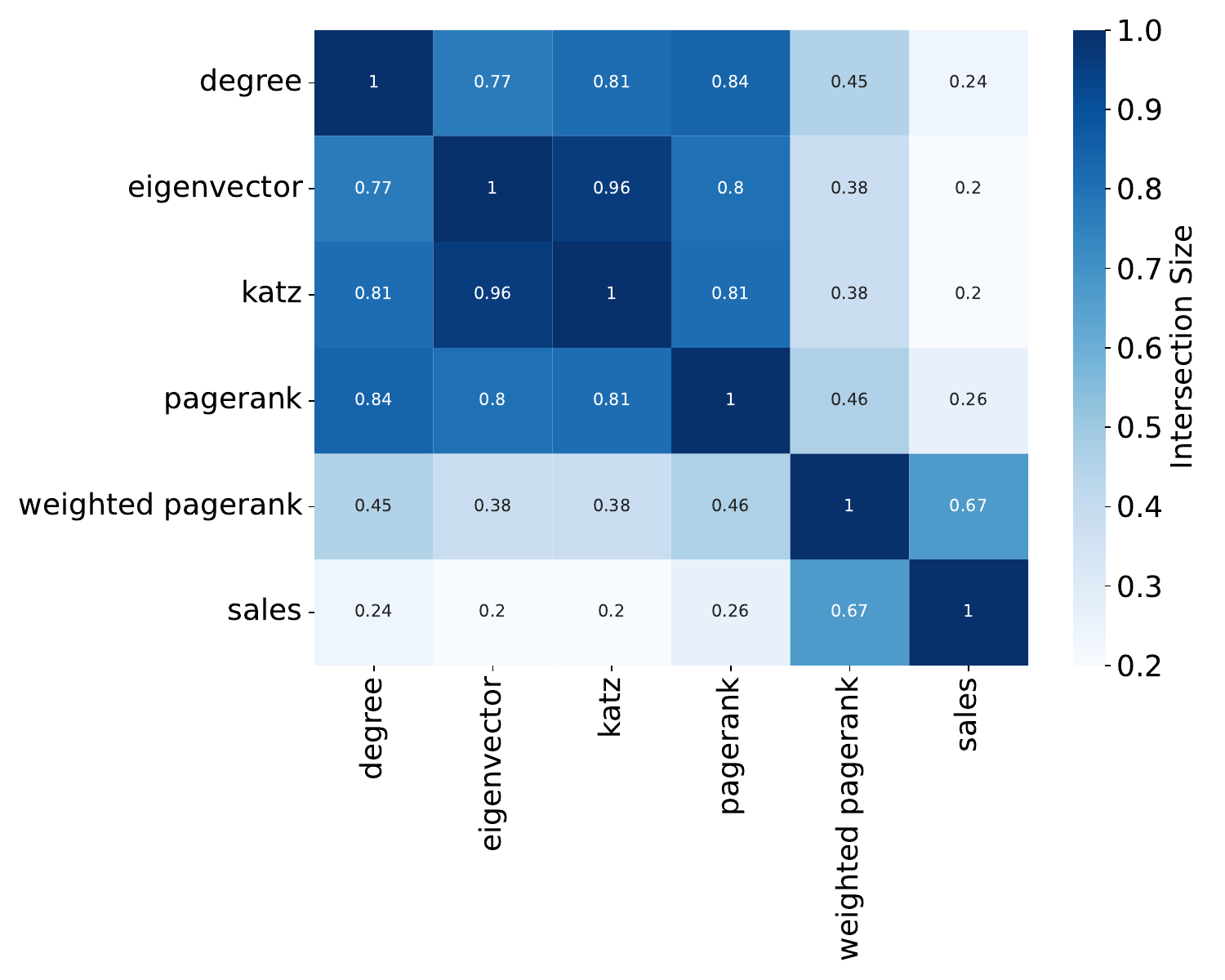}
  \caption{Correspondence between different downstream centrality measures, and firms' sales. The intersection size in each square is given by the number of firms in the intersection between the 100 most central firms by the centrality measure in the x-axis, and the 100 most central firms by the centrality measure in the y-axis, normalized by the maximum value (100).}
  \label{fig:sovrapposizioni_cedenti}
\end{figure}

\begin{figure}[H]
\centering
  \includegraphics[width=1\linewidth]{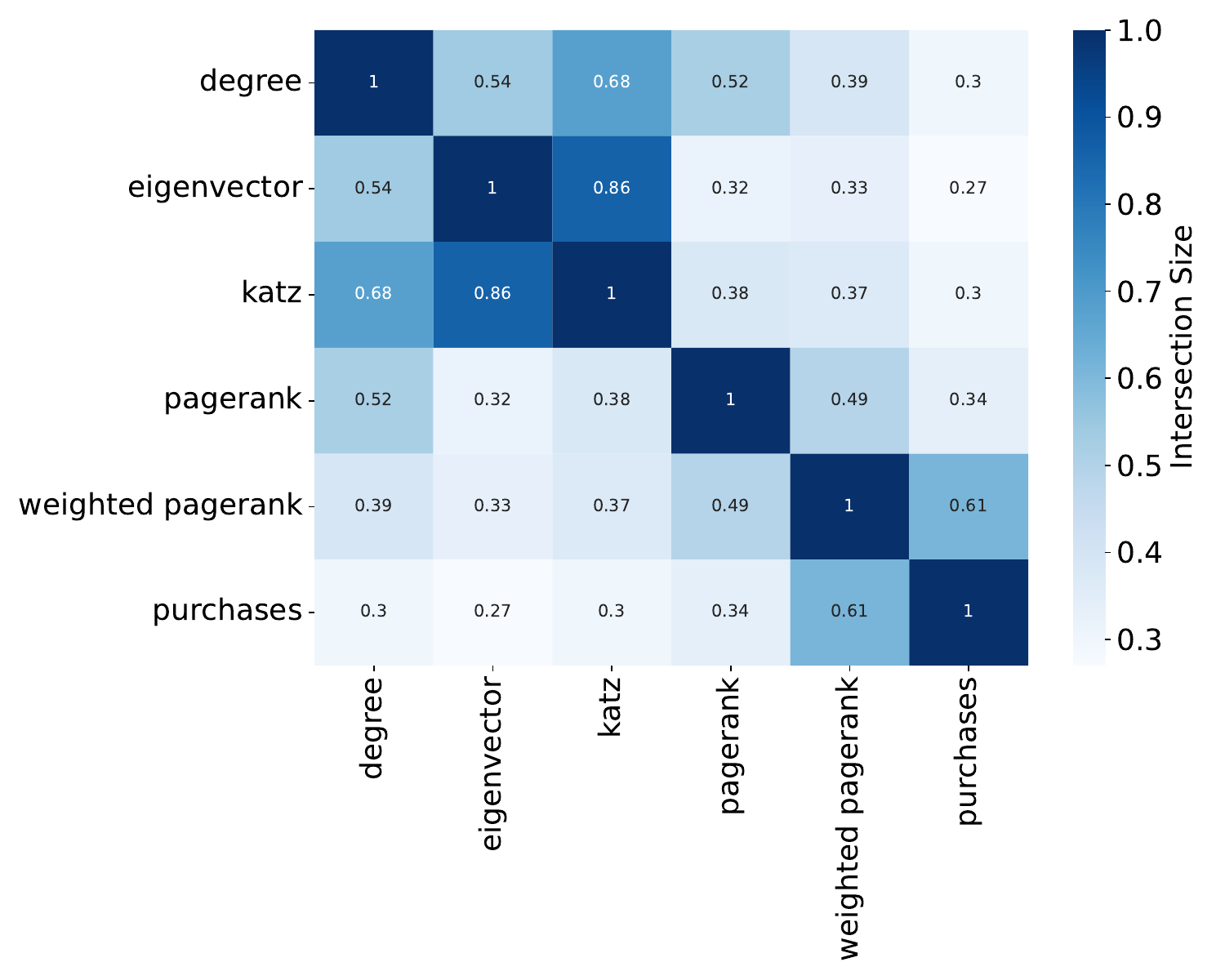}
  \caption{Correspondence between different upstream centrality measures, and firms' purchases. The intersection size in each square is given by the number of firms in the intersection between the 100 most central firms by the centrality measure in the x-axis, and the 100 most central firms by the centrality measure in the y-axis, normalized by the maximum value (100).}
  \label{fig:sovrapposizioni_cessionari}
\end{figure}

Most of the centrality measures presented in the previous sections are correlated to a certain degree. 
% We present a comparison between the different measures in Appendix \ref{sec:appendix_correlation_centrality}.
Taking into account the fact that the meaning of the measures is mostly relative - only the relation between the centrality of two firms is relevant - to compare different centrality measures we describe how they rank the most important firms.
% \va{In questo confronto si, ma ovviamente anche se una firm è mille volte più centrale di un'altra è rilevante.} 
In particular we select the 100 most central firms according to the measures introduced in previous sections: degree, eigenvector, Katz, PageRank and weighted PageRank centrality, and compare them with sales and purchases. 
Among the 100 most central firms for each measure, we count the number of firms in the pairwise intersections. 
Given the large number of firms in our dataset, if the centrality measures were assigned at random the typical size of these intersections would be zero. 
Instead we find sizable superpositions between all the measures. 
Furthermore, for the pairs of measures more strongly related (e.g. eigenvector and Katz centrality) these superpositions are close to $100\%$.
We show the results of these comparisons in figures \ref{fig:sovrapposizioni_cedenti} and \ref{fig:sovrapposizioni_cessionari}.
% \va{L'interpretazione di questi risultati non è così clear-cut, e ce li hanno fatti spostare in appendice dal main text (giusto?)}

\subsection{Global Centrality: Sectoral and Geographical Breakdown}
\begin{figure}[h]
\centering
  \begin{subfigure}[b]{0.49\textwidth}
    \centering
    \includegraphics[width=1\linewidth]{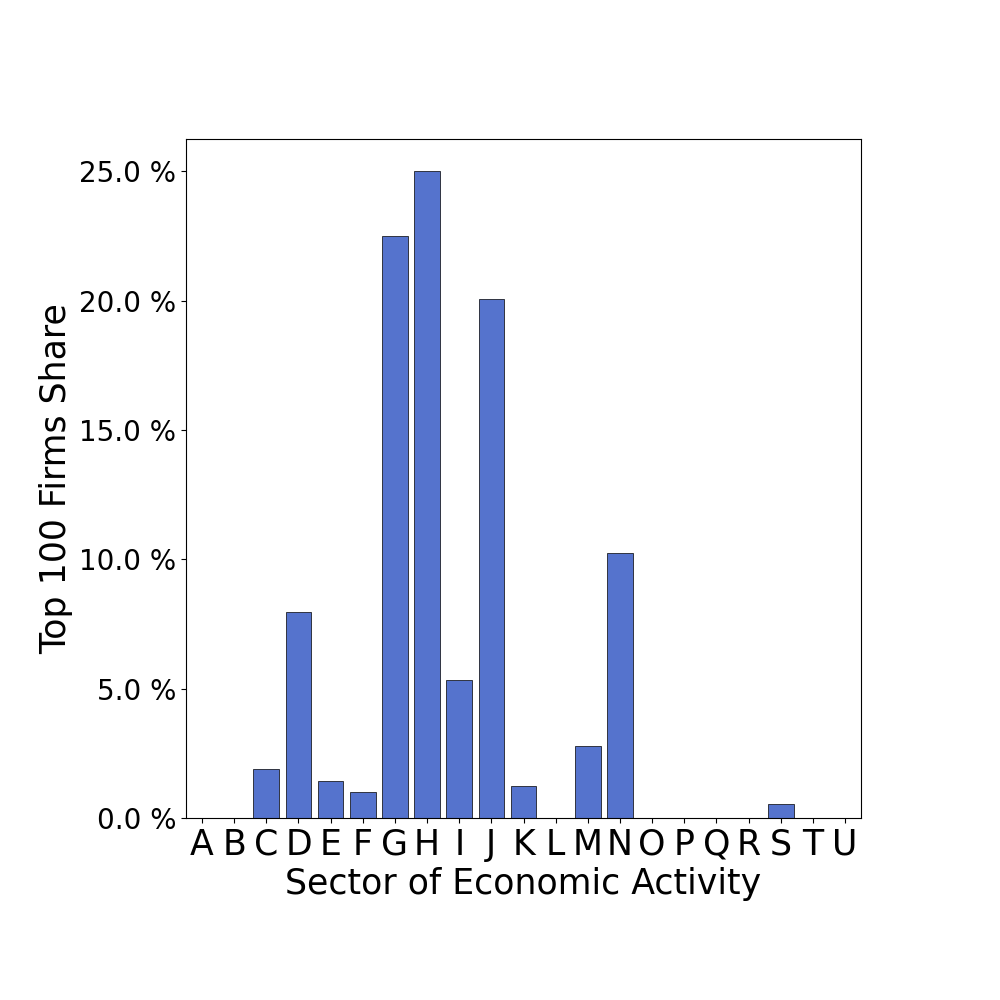}
    \caption{Eigenvector Centrality}
    %\vspace{4ex}
  \end{subfigure}%% 
  \begin{subfigure}[b]{0.49\textwidth}
    \centering
    \includegraphics[width=1\linewidth]{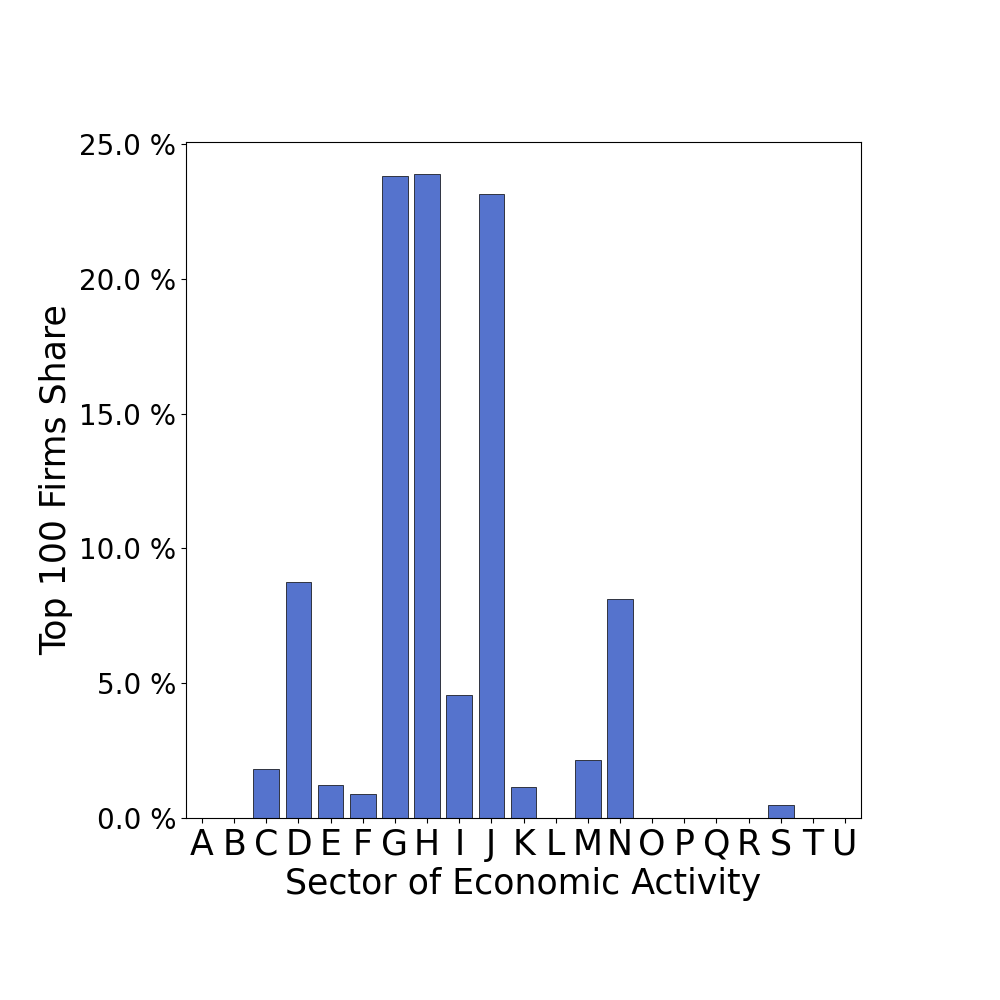}
    \caption{Katz Centrality}
    %\vspace{4ex}
  \end{subfigure} 
  \begin{subfigure}[b]{0.49\textwidth}
    \centering
    \includegraphics[width=1\linewidth]{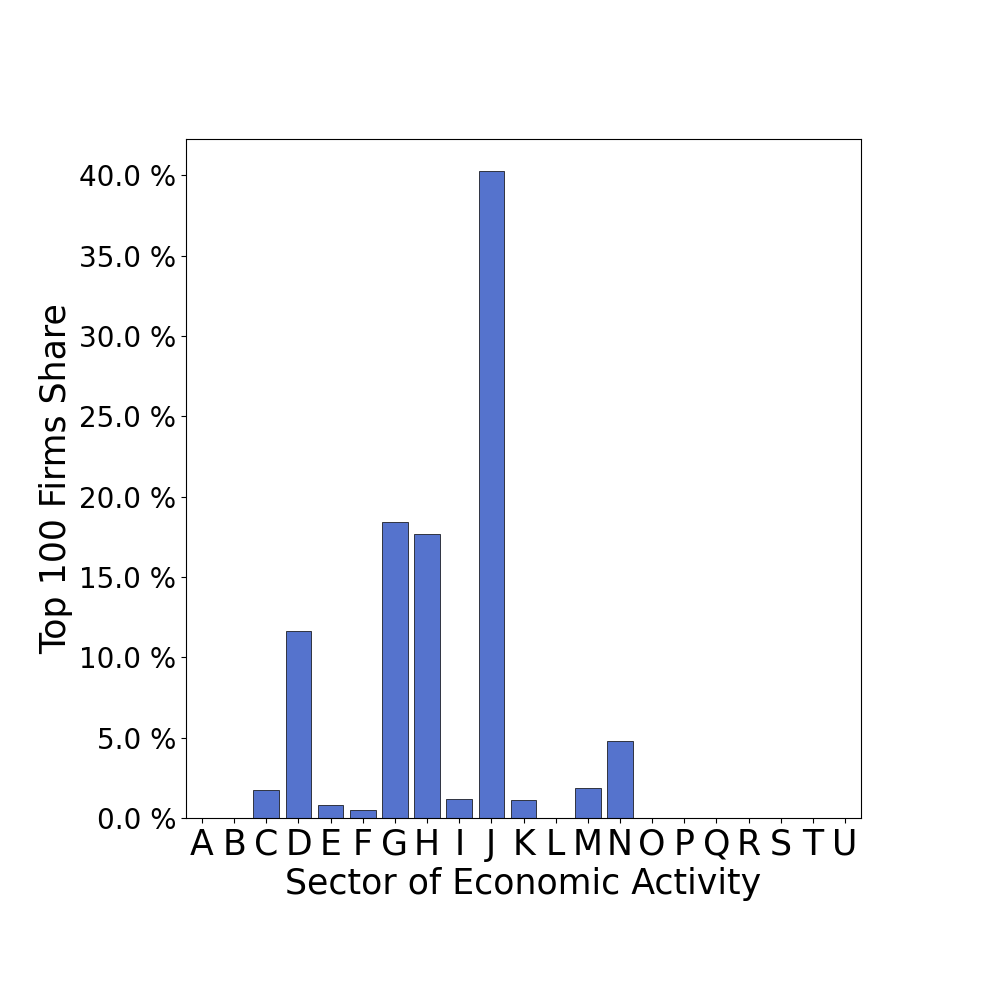}
    \caption{PageRank Centrality}
  \end{subfigure}%%
  \begin{subfigure}[b]{0.49\textwidth}
    \centering
    \includegraphics[width=1\linewidth]{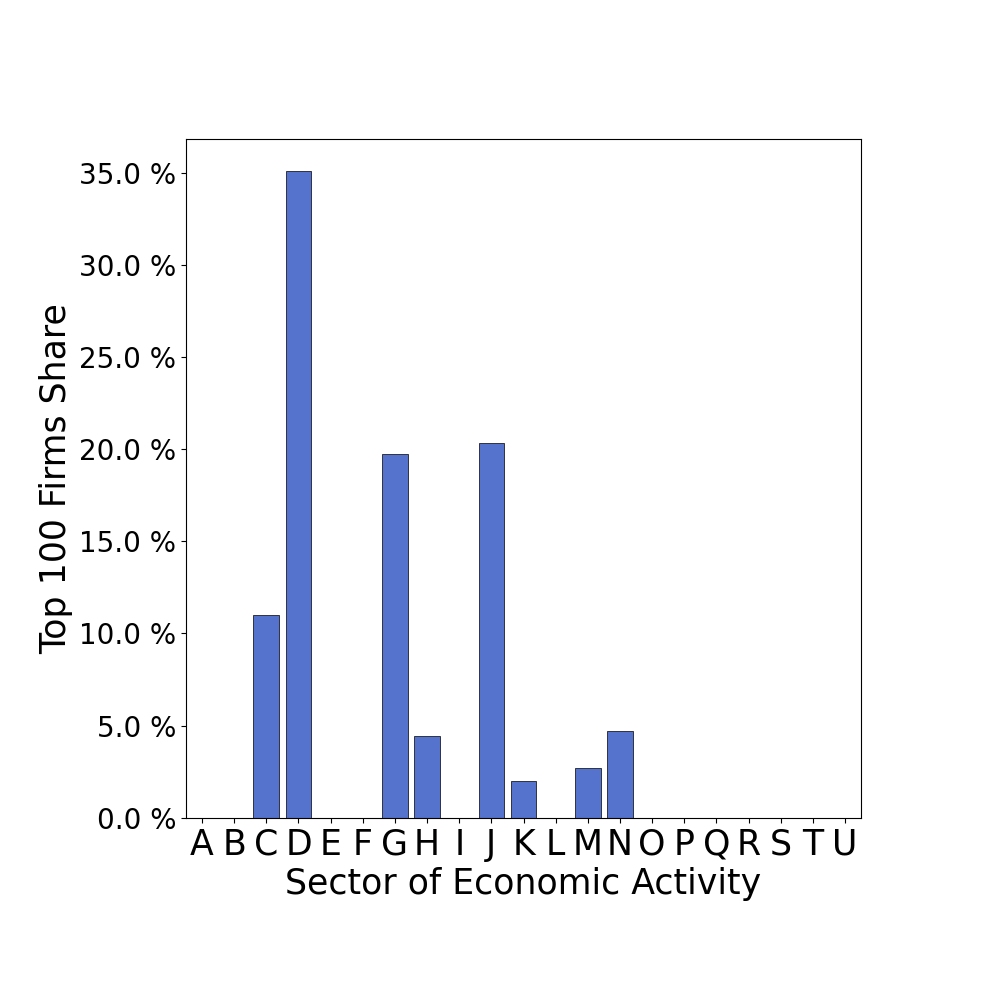}
    \caption{Weighted PageRank Centrality}
  \end{subfigure} 
  \caption{Distribution of the downstream centrality share across the sectors of the 100 most important firms.}
  \label{fig:downstream_global_cat_100} 
\end{figure}

\begin{figure}[h]
\centering
  \begin{subfigure}[b]{0.49\textwidth}
    \centering
    \includegraphics[width=1\linewidth]{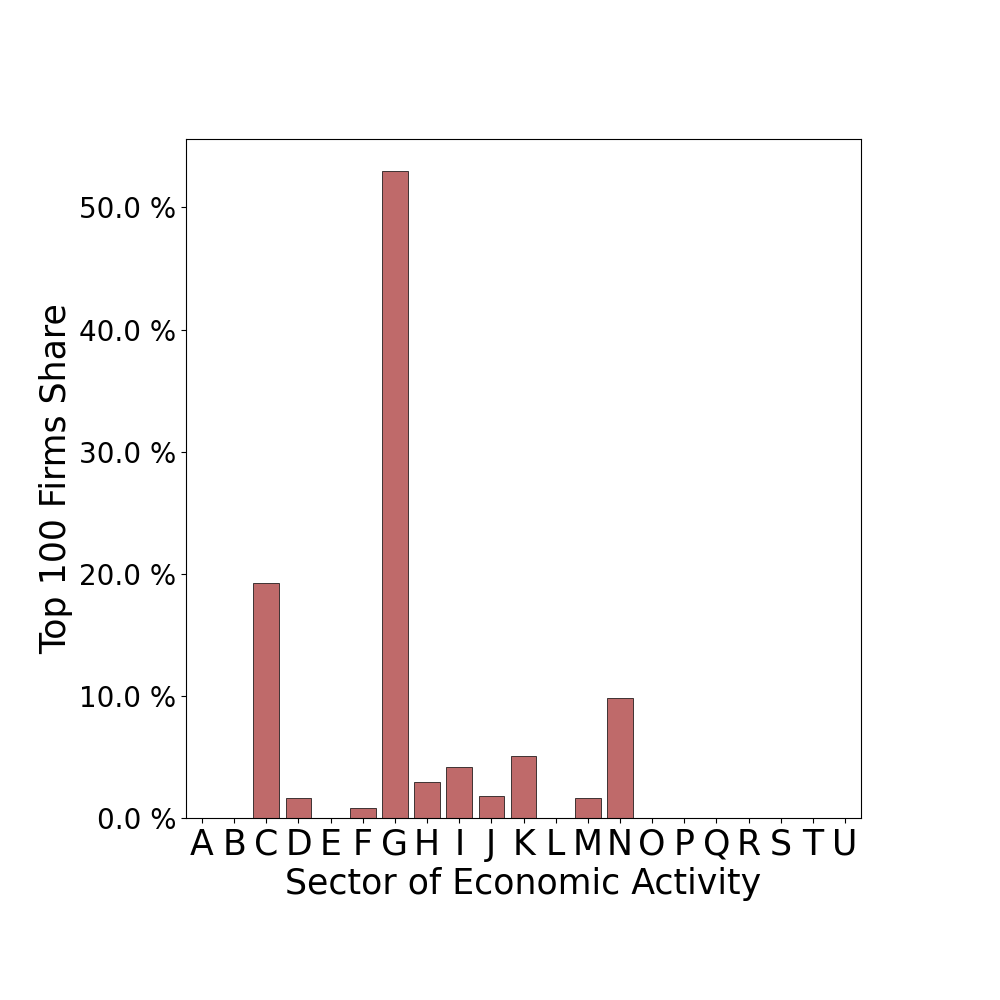}
    \caption{Eigenvector Centrality}
    %\vspace{4ex}
  \end{subfigure}%% 
  \begin{subfigure}[b]{0.49\textwidth}
    \centering
    \includegraphics[width=1\linewidth]{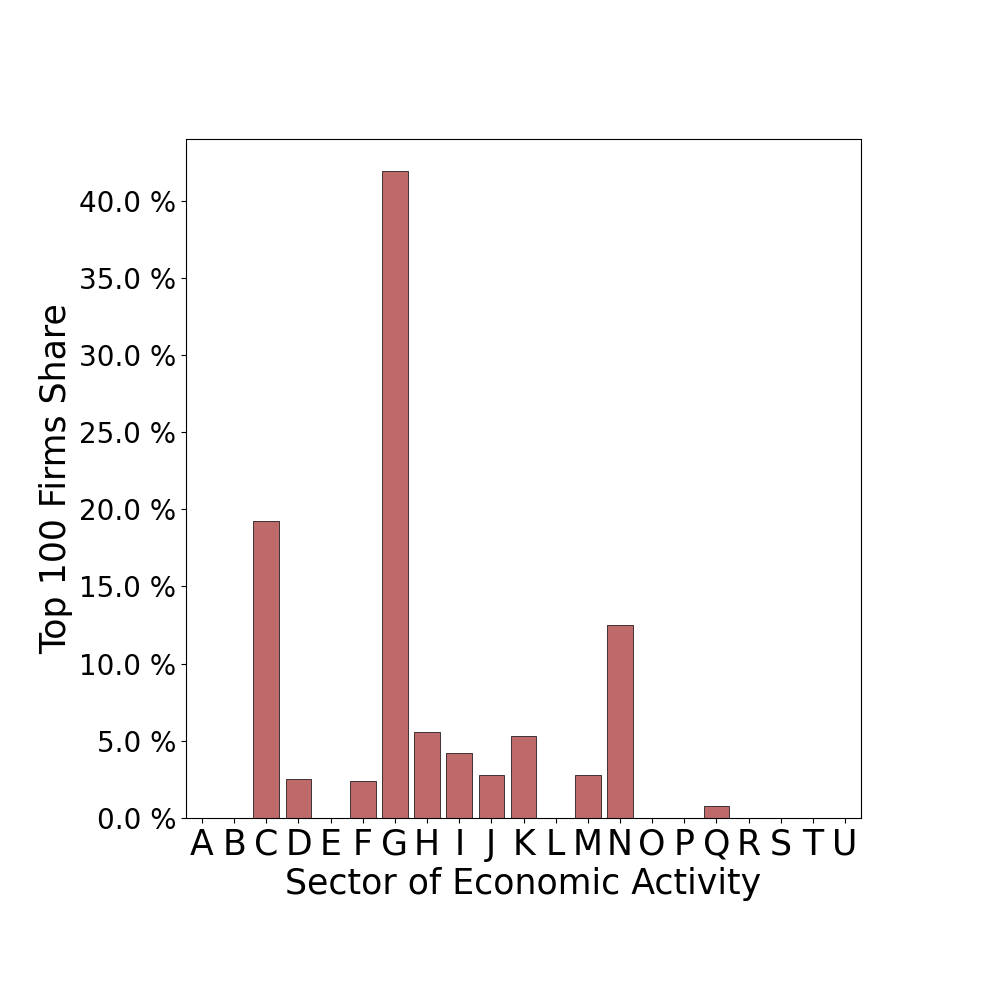}
    \caption{Katz Centrality}
    %\vspace{4ex}
  \end{subfigure} 
  \begin{subfigure}[b]{0.49\textwidth}
    \centering
    \includegraphics[width=1\linewidth]{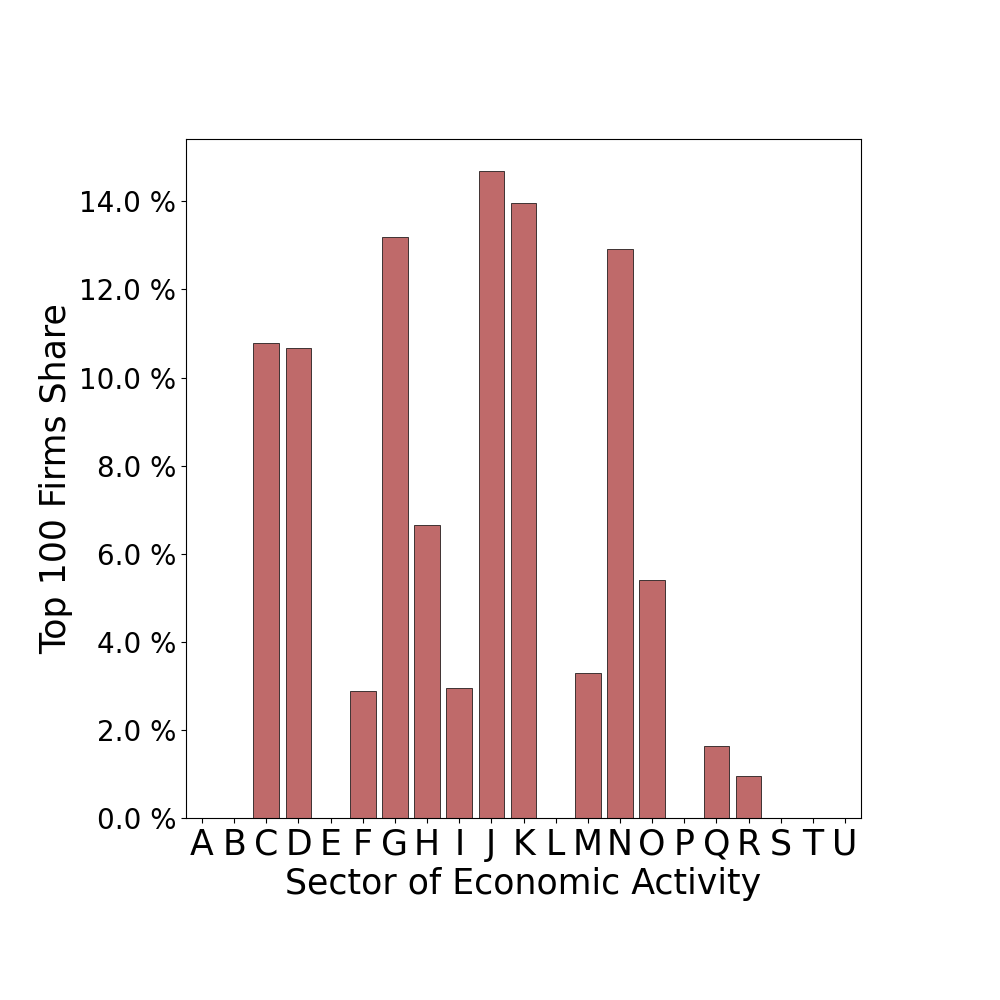}
    \caption{PageRank Centrality}
      \end{subfigure}%%
  \begin{subfigure}[b]{0.49\textwidth}
    \centering
    \includegraphics[width=1\linewidth]{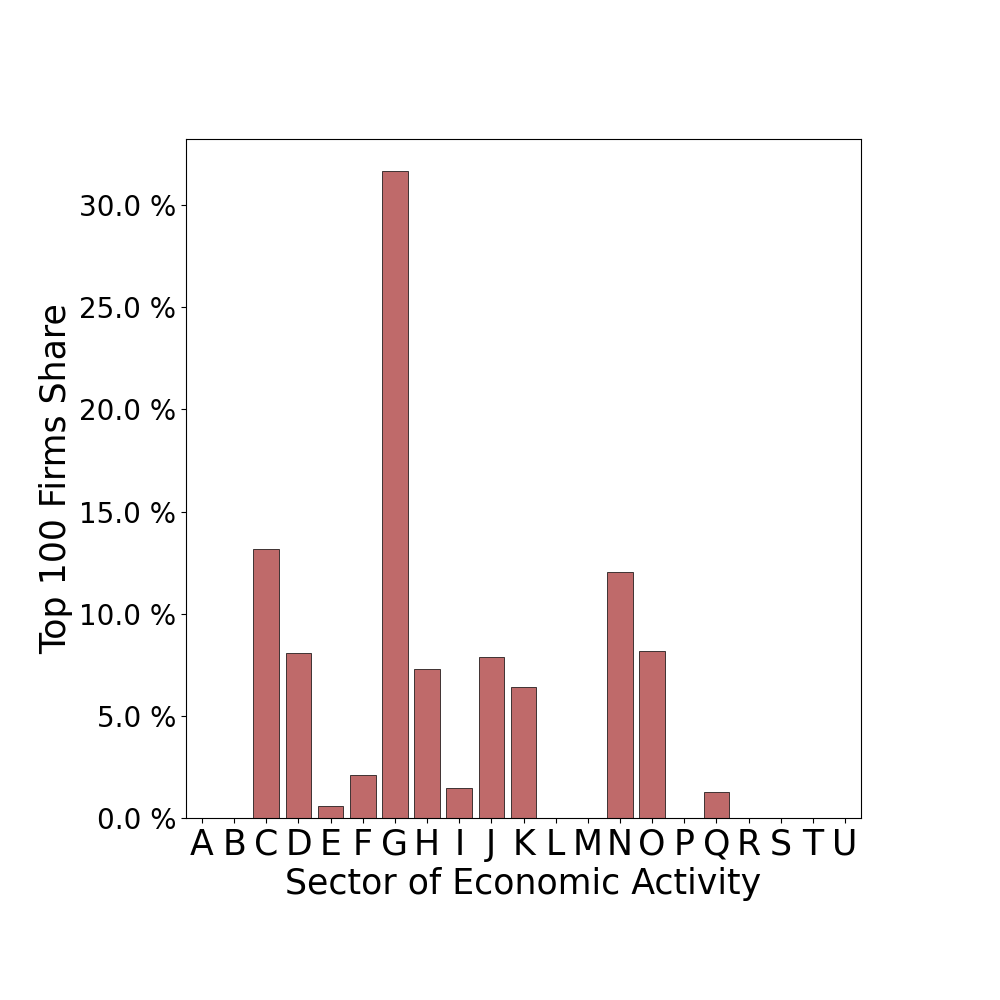}
    \caption{Weighted PageRank Centrality}
  \end{subfigure} 
  \caption{Distribution of the upstream centrality share across the sectors of the 100 most important firms.}
  \label{fig:upstream_global_cat_100} 
\end{figure}

The distribution of centrality measures by province and sector of economic activity is heavily influenced by the distribution of the number of firms. 
We evaluate the centrality of a given province (or sector of activity) as the sum of the single firms' centrality values in that province (or sector).
It is therefore straightforward that provinces and sectors of activity containing many firms are also comparatively more central in the network.
We leave the distributions of sectoral and regional centrality measures in Appendix \ref{subsec:appendix_plot_global_centrality}, and focus here on the distribution of the most central firms in the network.
%\sout{The global centrality measures breakdowns by province and category of economic activity are consistent with the distribution of firms. They, indeed, show concentration in those provinces and categories with the highest numbers of firms. We therefore report these histograms in the appendix (see Section \ref{subsec:appendix_plot_global_centrality}) and carry out further analysis.}
For each centrality measure we consider the 100 firms with the highest value, and study their geographical and sectoral distributions.
% \footnote{In Appendix \ref{sec:appendix_correlation_centrality} we also perform an analysis on how coherent such measures are, computing the intersection between this sets of 100 firms.}
Firms with high values of centrality play a crucial role in the network, often more important than the multitude of low-centrality firms in the bulk of the distribution. 
For this reason it is interesting to study how these highly central nodes are distributed. 
\begin{table}[h]
    \centering
    \small
    \begin{tabular}{lrrrr}
      & \multicolumn{1}{c}{Eigenvector} &  \multicolumn{1}{c}{Katz} & \multicolumn{1}{c}{PageRank} & \multicolumn{1}{c}{\makecell{Weighted PageRank}} \\
    \midrule
    Downstream & 5.21 & 2.51  & 12.32  & 18.62  \\%& 623.914 & 37,98\% \\
    Upstream &  0.43  & 0.18  &  1.48  & 8.19 \\%& 623.914 & 37,98\% \\
    \bottomrule
    \end{tabular}
    \caption{Share of centrality held by the top 100 firms (values in percentage).}
    \label{tab:centrality_values}
\end{table}
% \footnote{Simply considering the average would introduce the opposite problem: groups with a large number of firms contains also a disproportionate fraction of small firms, which have the effect of lowering the group averages.}
% We then express the result as a share of the total centrality of the group considered.
% Provinces and categories of activity with a more concentrated distribution of centrality have correspondingly a higher share of centrality in the most central firms.   
%\va{la soluzione più ovvia sarebbe considerare la media invece che la somma. Il problema mi sa che poi va nell'altra direzione però.}
%\sout{To better understand the characteristics of the most central firms, we plot the aggregated centrality values by province and category, of the 100 firms with the highest centrality values, the 100 most “important” firms based on the various centrality definitions.}
% \va{Tentativo: Selecting a fixed number of firms for each group we exclude the effect of the number of firms on the group's centrality value.}
% These results also are quite aligned with the overall firm distribution \va{non mi è chiaro in che senso}; however, some additional noteworthy patterns emerge, offering further insight into the structural relevance of specific provinces and categories.
% Number of connections
% "J": "INFORMATION AND COMMUNICATION SERVICES",
% importi D:
% "D": "ELECTRICITY, GAS, STEAM, AND AIR CONDITIONING SUPPLY",
% Energetic
The breakdowns by sector of economic activity of downstream centrality measures are shown in Figure \ref{fig:downstream_global_cat_100}.
We observe that firms operating in sector J (Information and Communication Services) are among the most central by all measures. For eigenvector and Katz centrality firms in sectors G (Wholesale and Retail Trade) and H (Transport and Storage) are also relevant.
When considering weighted connections, taking into account the monetary value of transactions, the most central firms are in sector D (Electricity, Gas, Steam, and Air Conditioning Supply).
Thus, service-sector firms appear more central when considering the number of connections, while utility companies emerge as the most influential nodes in terms of the monetary volume exchanged.
With regard to the number of sellers (upstream centrality, Figure \ref{fig:upstream_global_cat_100}), centrality measures are concentrated mainly in sector G, which is also the sector with the highest weighted PageRank centrality, thus the sector with the most important firms in terms of the invoice amount. 
%In the case of PageRank centrality, the leading sectors include J, K (Financial and Insurance Activities), and D.
In the case of PageRank centrality, several sectors have firms with elevated centrality scores. %\ma{in questo caso risultati molto diversi...}
In addition to sectors G, C (Manufacturing) and N (Renting, Travel Agencies, Business Support Activities), the most significant for the other measures, here also sectors D, J, and K (Financial and Insurance Activities) emerge as relevant.
% While sectors G, and C consistently account for the largest shares of network centrality, noteworthy deviations arise.
While centrality values tend to scale with the number of firms in each sector, some significant deviations from this pattern arise.
In particular, sector I (Hotels and Restaurants) exhibits both downstream and upstream above-average eigenvector centrality, while sector J shows above-average downstream PageRank centrality in both its weighted and unweighted specifications, despite both sectors being characterized by a below-average number of firms. %Furthermore, sectors D and N exhibit above-average PageRank centrality—downstream and upstream, respectively—despite having a below-average number of firms.
In contrast, sector F (Construction), which frequently ranks among the sectors with the highest centrality and number of firms, exhibits an average ranking for downstream eigenvector centrality and downstream PageRank centrality. Also sector L (Real Estate Activities), which ranks fourth in terms of number of firms, displays a below-average centrality across nearly all measures.
% upstream no pagerank G,C,N
% upstream pagerank J,K,C,D,N
% upstream importi
% "N": "RENTING, TRAVEL AGENCIES, BUSINESS SUPPORT ACTIVITIES",
All plots in Figures \ref{fig:downstream_global_cat_100} and \ref{fig:upstream_global_cat_100} report on the vertical axis the share of centrality accounted for by each sector of the 100 most central firms.
Table \ref{tab:centrality_values} reports the share of centrality held by the 100 most central firms with respect to the total amount of centrality.
%{SCRIVI CHE NEI GRAFICI c'è LA SHARE TRA LE PRIME 100 E SCRIVI SHARE DELLE PRIME 100 RISPETTO A TUTTE LE ALTRE, che poi è la X di figura 25. mettiamo lineare? FORSE SI}:
% eigenvector_upstream 0.43063982873572687
% katz_upstream 0.17940207274401923
% pagerank_upstream 1.476800163062956
% pagerank_importi_upstream 8.189857871732183
% eigenvector_downstream 5.21379447768265
% katz_downstream 2.5124575869013923
% pagerank_downstream 12.315934167477211
% pagerank_importi_downstream 18.615525401437
% \begin{table}[H]
%     \centering
%     \small
%     \begin{tabular}{lrrrr}
%       & \multicolumn{1}{c}{Eigenvector} &  \multicolumn{1}{c}{Katz} & \multicolumn{1}{c}{PageRank} & \multicolumn{1}{c}{\makecell{Weighted PageRank}} \\
%     \midrule
%     Downstream & 5,21 & 2,51  & 12,32  & 18,62  \\%& 623.914 & 37,98\% \\
%     Upstream &  0,43  & 0,18  &  1,48  & 8,19 \\%& 623.914 & 37,98\% \\
%     \bottomrule
%     \end{tabular}
%     \caption{Share of centrality held by the top 100 firms. Values are in percentage.}
%     \label{tab:centrality_values}
% \end{table}
% The total centrality share held by the 100 most central firms increases across different measures. Katz centrality has the lowest value, followed by eigenvector, PageRank, and weighted PageRank that has the highest value. This hold for both upstream and downstream.
Both downstream and upstream centrality are concentrated towards the top: the share of centrality held by the 100 most central firms represent a small portion of the total for Katz and eigenvector centrality, while for PageRank and weighted PageRank the top 100 firms account for up to almost 19\% of the total centrality.
% \sout{This behavior correlates with the tail exponents observed in the distributions of centrality values. As illustrated in Figure \ref{fig:plot_esponenti_cumulate_share_100}, a lower tail exponent (in absolute value) is associated to a greater concentration of centrality among the top firms, suggesting that fewer firms collect a larger share of the overall centrality. Lower exponents (in absolute value) correspond to heavier-tailed distributions, thus implying a broader dispersion and more extreme centrality values.}
% \begin{figure}[h]
% \centering
%   \includegraphics[width=1\linewidth]{plots/plot_esponenti_cumulate_share_100.pdf}
%   \caption{Relation between concentration as described by the share of the total centrality held by the 100 most central firms (x axis) and the tail exponents of the centrality distributions (y axis). From left to right, in both plots, we find Katz centrality, eigenvector centrality, PageRank and weighted PageRank centrality. \va{Valutare se tenere questi grafici.}}
%   \label{fig:plot_esponenti_cumulate_share_100}
% \end{figure} 

\begin{figure}[h]
\centering
  \begin{subfigure}[b]{0.49\textwidth}
    \centering
    \includegraphics[width=1\linewidth]{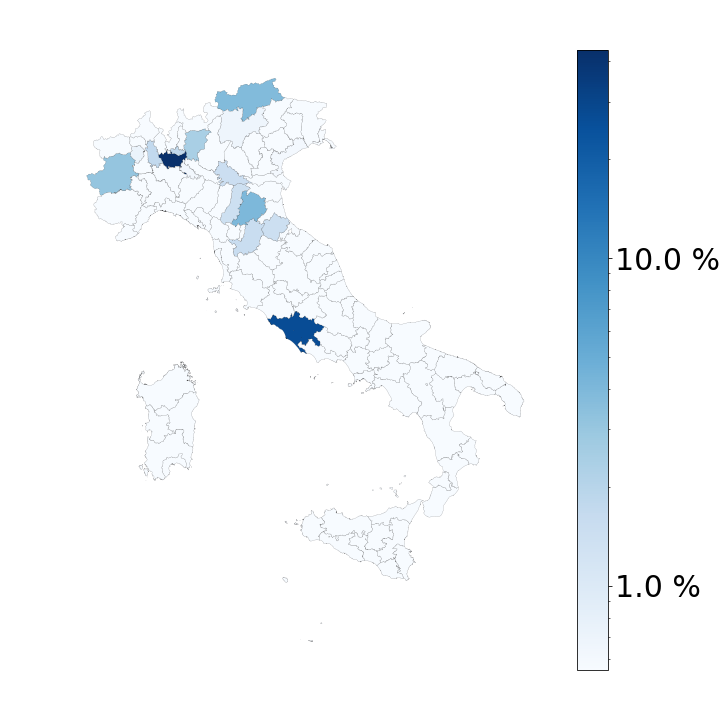}
    \caption{Eigenvector Centrality}
    \label{fig:eigenvector_downstream_provincie_top_100}
    %\vspace{4ex}
  \end{subfigure}%% 
  \begin{subfigure}[b]{0.49\textwidth}
    \centering
    \includegraphics[width=1\linewidth]{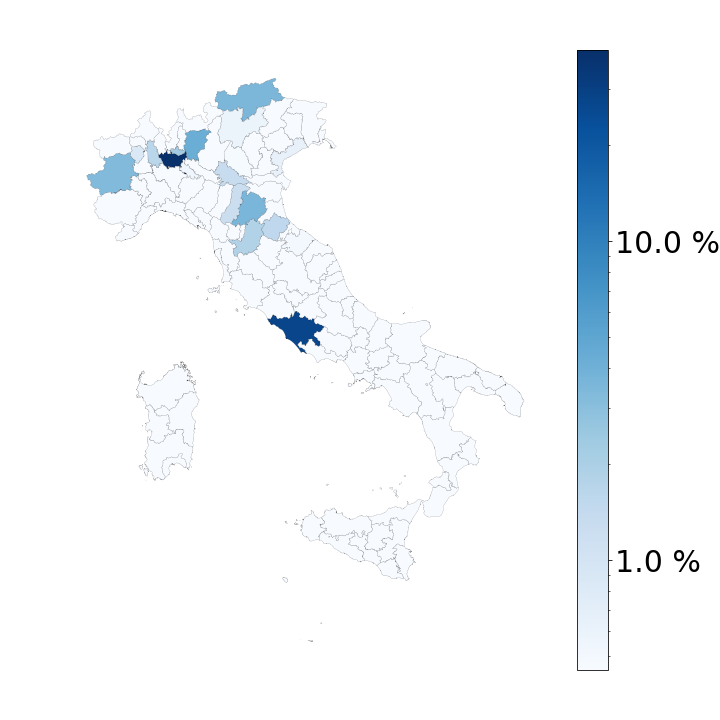}
    \caption{Katz Centrality}
    \label{fig:katz_downstream_provincie_top_100} 
    %\vspace{4ex}
  \end{subfigure} 
  \begin{subfigure}[b]{0.49\textwidth}
    \centering
    \includegraphics[width=1\linewidth]{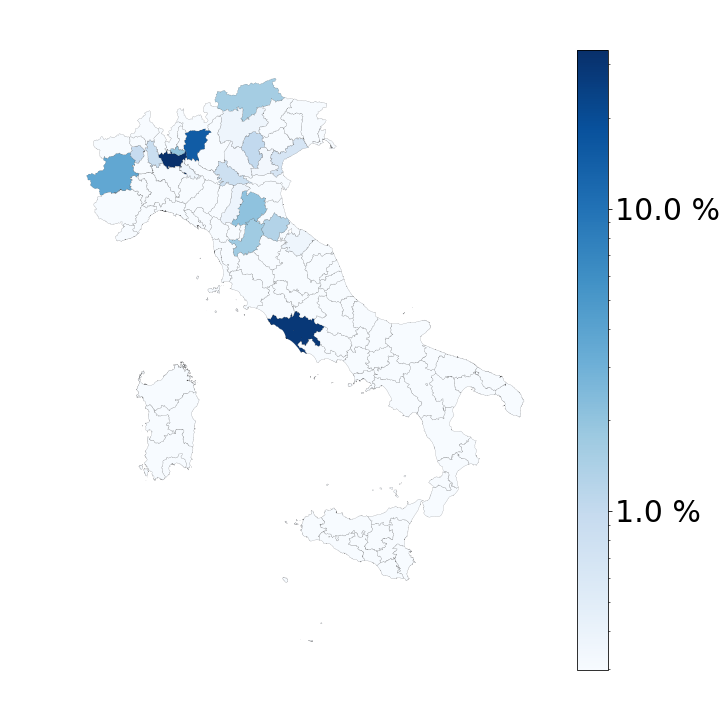}
    \caption{PageRank Centrality}
    \label{fig:pagerank_downstream_provincie_top_100} 
  \end{subfigure}%%
  \begin{subfigure}[b]{0.49\textwidth}
    \centering
    \includegraphics[width=1\linewidth]{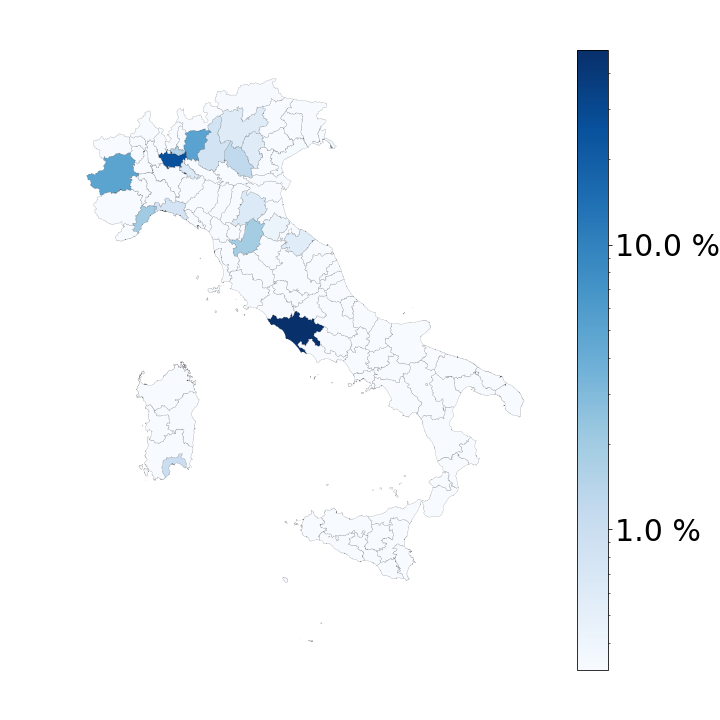}
    \caption{Weighted PageRank Centrality}
    \label{fig:pagerank_importi_downstream_provincie_top_100} 
  \end{subfigure} 
  \caption{Downstream global centrality of the 100 most central firms grouped by province of registered office.} 
  \label{fig:centrality_downstream_provincie_top_100} 
\end{figure}

\begin{figure}[h]
\centering
  \begin{subfigure}[b]{0.49\textwidth}
    \centering
    \includegraphics[width=1\linewidth]{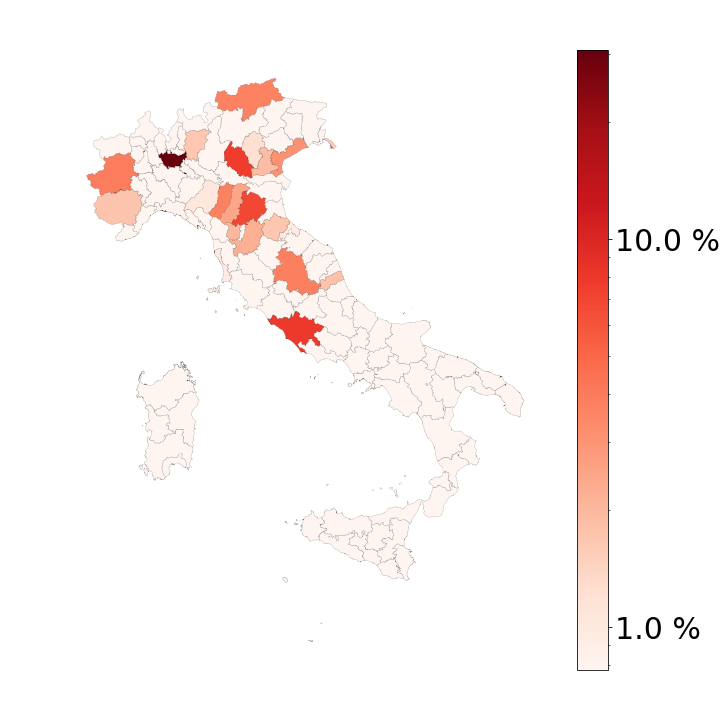}
    \caption{Eigenvector Centrality}
    \label{fig:eigenvector_upstream_provincie_top_100}
    %\vspace{4ex}
  \end{subfigure}%% 
  \begin{subfigure}[b]{0.49\textwidth}
    \centering
    \includegraphics[width=1\linewidth]{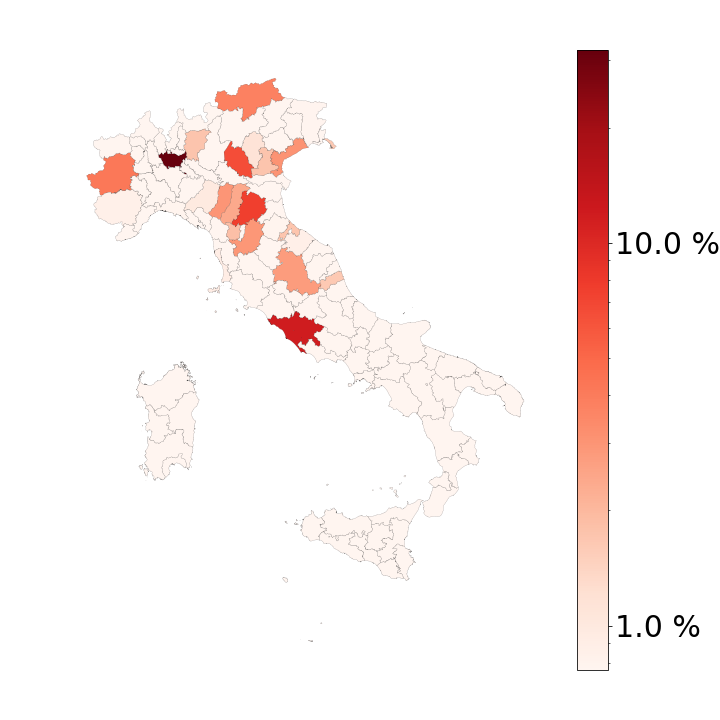}
    \caption{Katz Centrality}
    \label{fig:katz_upstream_provincie_top_100} 
    %\vspace{4ex}
  \end{subfigure} 
  \begin{subfigure}[b]{0.49\textwidth}
    \centering
    \includegraphics[width=1\linewidth]{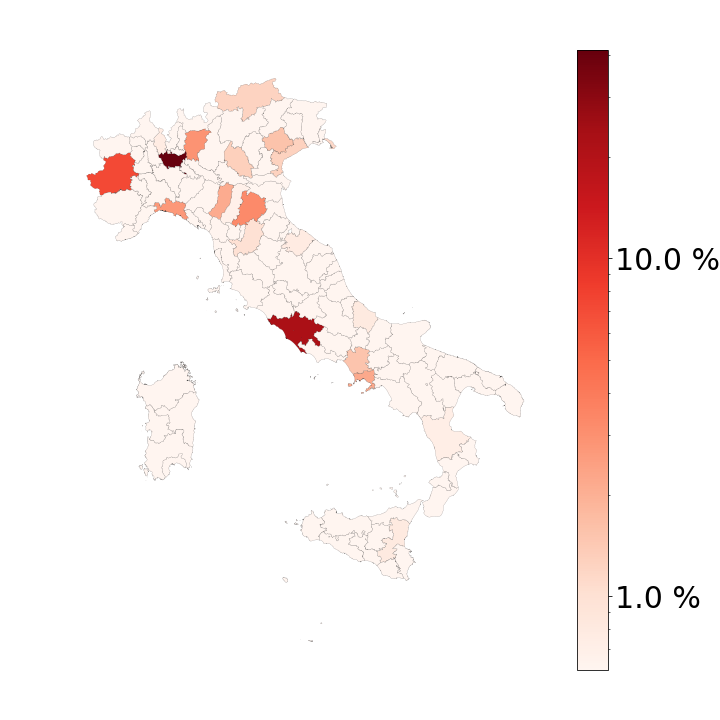}
    \caption{PageRank Centrality}
    \label{fig:pagerank_upstream_provincie_top_100} 
  \end{subfigure}%%
  \begin{subfigure}[b]{0.49\textwidth}
    \centering
    \includegraphics[width=1\linewidth]{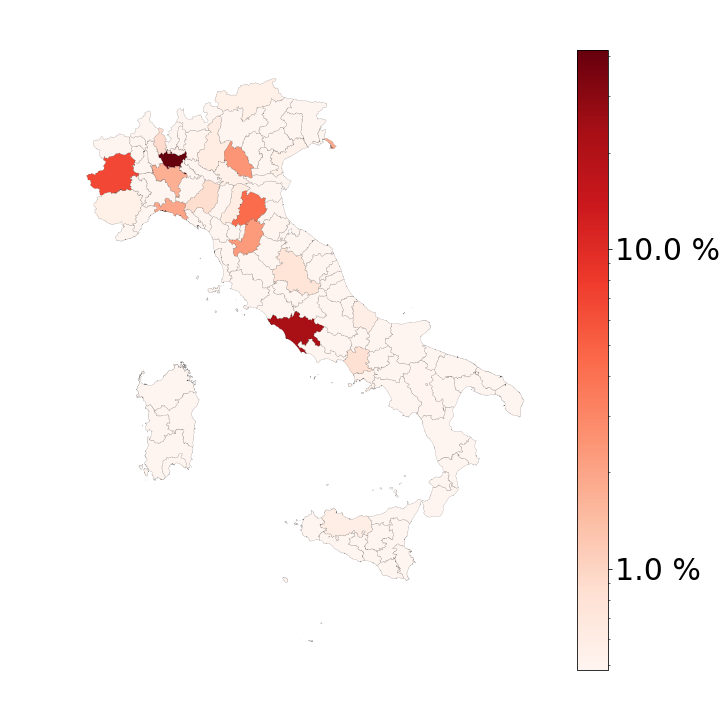}
    \caption{Weighted PageRank Centrality}
    \label{fig:pagerank_importi_upstream_provincie_top_100} 
  \end{subfigure} 
  \caption{Upstream global centrality of the 100 most central firms grouped by province of registered office.}
  \label{fig:centrality_upstream_provincie_top_100} 
\end{figure}

The geographical breakdown, similarly to the case of degree centrality (see Section \ref{sec:degree_distribution_geo_sec}), highlights a disparity between the southern regions of the country and the central-northern areas.
This is clearly observable in terms of the number of customers and in terms of weighted downstream links (see Figure \ref{fig:centrality_downstream_provincie_top_100}).
% \sout{However, in this case, the disparity is less pronounced, although it remains clearly observable.
% In particular, together with Milan and Rome, the provinces of Turin (TO) and Bologna (BO) consistently occupy central positions within the network, both in terms of the number of buyers (see Figure \ref{fig:centrality_downstream_provincie_top_100}) and sellers (see Figure \ref{fig:centrality_upstream_provincie_top_100}).}
%, and of the weighted connections.
%{Aggiungi frase che riassume per macrozone i risultati che presenti dopo, inoltre scrivi del divario nord-sud}
% \sout{Alongside these, other provinces emerge as relevant depending on the specific centrality measure considered. For for eigenvector and Katz upstream centrality, the province of Perugia (PG) appears relevant, primarily due to the presence of construction firms, while Bergamo (BG) mostly for PageRank downstream centrality, and Florence (FI) for the downstream centralities and the weighted PageRank upstream centrality, reflecting the presence of manufacturing and real estate firms.}
However, in the case of the maps by number of suppliers and weighted upstream links (see Figure \ref{fig:centrality_upstream_provincie_top_100}), the heterogeneity is slightly less pronounced. %\ma{anche qui andrebbe spiegato cosa significa, cosa potrebbe implicare, if any}%, mainly due to {aggiungere settori --- Daniele (up to you)}.
%{Alongside these, other provinces seems to emerge as relevant depending on the specific centrality measure considered. This is primarily due to the presence in this provinces of a  concentration of firms operating in a specific economic sector}.
Although broadly correlated with the extensive nature of the number of firms in a region, the total centrality measures also reveal features of the production network depending on firms' quality.
Lombardia and Lazio consistently lead the rankings both in terms of firm counts and across centrality indicators.
Firms located in Campania, Emilia-Romagna, and Piedmont exhibit above-average numbers of customers and higher downstream centrality, while firms in Veneto, Emilia-Romagna, and Friuli-Venezia Giulia display above-average numbers of suppliers and elevated upstream centrality.

\clearpage
\section{Conclusions}
\label{sec:conclusion}

In this paper we provide the first analysis of the micro structure of the Italian production network, as seen through business-to-business electronic invoicing data.
Our analysis of the network topology reveals a complex structure, characterized by heterogeneous distributions of the number of links, with most firms being connected to a limited number of suppliers and buyers, and a small fraction of firms exhibiting very high connectivity, serving as central hubs in the production network.
In particular we study the distributions of the numbers of buyers and sellers per firm, obtaining a power-law behavior in the right tail, consistent with previous findings in the production networks literature.  
%, firms providing core services to other businesses (such as electricity distribution,logistics, and communication services), and business support services (manufacturing and X).
% Most of the distributions analyzed show a power-law behavior in the right tail, consistent with previous findings in the production networks literature. 
This behavior implies the presence of a small number of highly connected nodes - playing an important role in ensuring the network’s high level of connectivity despite a relatively low typical number of links between firms.

In addition to the large heterogeneity in the connectivity structure, the heavy-tailed distributions suggest that shocks affecting highly connected and more central firms could potentially propagate diffusely through the economy, pointing to the importance of identifying and monitoring these systemic firms. 
We perform also a study of the most widely known centrality measures on the network.
These measures describe the importance of each node in the network, in terms of propagation of shocks and vulnerability to external perturbations.  
For each firm we compute eigenvector, Katz and PageRank centrality. 
These centrality values display heavy tailed distributions as well, implying that the few most central nodes have a significant portion of the total centrality scores in the network.
An evaluation of the structure of direct and indirect connections between firms reveals that most firms are close to the rest of the network, meaning that most pairs of nodes can be connected with a small number of steps, less than 4 on average. 

We also carry out a geographical and sectoral analysis of all firms present in the dataset.
Invoice exchanges are primarily concentrated in the provinces of Rome, Milan, and Naples - where more than 25\% of firms are located, and among firms operating in wholesale and retail trade or in manufacturing - which together account for more than 45\% of the exchanged amount.
%However, the analysis of average flows reveals key exchanges also in provinces with less firms. \sout{such as manufacturing firms in Savona for invoices issued, and energy firms in Trieste for invoices received.}
A study of the very extreme right tails of the distributions of centrality reveals a pronounced geographical concentration of the most central firms in the central and northern regions - as observed for hubs, albeit less markedly. 
Finally, we carry out an additional analysis focusing on the 100 most central firms, specifically quantifying the proportion of overall network centrality they account for. 
% The findings reveal a clear positive correlation with the tail exponent of the power-law distribution characterizing the centrality distributions.

The static nature of our analysis does not capture the evolution of the production network over time, which could reveal important trends in economic integration and specialization.
This suggests a potential extension of this study, in which data for different years are used to study the dynamics of the network structure. 
Moreover, while we have thoroughly mapped the structure of the network, we have not modeled how shocks might propagate through this structure, which would require additional economic data on production functions. 
Future research could build upon our structural findings to develop more sophisticated modelling of economic resilience and vulnerability.
Additionally, our focus on domestic connections leaves open questions about how this national production network is embedded within global value chains. Extending this analysis to include international connections would provide a more complete picture of economic inter-dependencies and vulnerabilities.
% This detailed mapping of the Italian production network contributes to the understanding of the complex network of relationships underpinning economic activity.
% to enhance economic resilience or stimulate growth in strategic sectors. As production networks continue to evolve under the pressures of technological change, globalization, and external shocks, such network-based analyses will {provide} increasingly valuable tools for economic research and policy design.

% The methodological approach and findings presented here can inform policymakers about the structure of economic relationships, potentially guiding targeted {policy measures}.

% \section{References}
\clearpage
\bibliographystyle{chicago}
\bibliography{bibliography}

@article{brin1998anatomy,
  title={The anatomy of a large-scale hypertextual web search engine},
  author={Brin, Sergey and Page, Lawrence},
  journal={Computer networks and ISDN systems},
  volume={30},
  number={1-7},
  pages={107--117},
  year={1998},
  publisher={Elsevier}
}

@techreport{page1999pagerank,
  title={The PageRank citation ranking: Bringing order to the web.},
  author={Page, Lawrence and Brin, Sergey and Motwani, Rajeev and Winograd, Terry},
  year={1999},
  institution={Stanford infolab}
}

@article{newman2003assortativity,
  title = {Mixing patterns in networks},
  author = {Newman, M. E. J.},
  journal = {Phys. Rev. E},
  volume = {67},
  issue = {2},
  pages = {026126},
  numpages = {13},
  year = {2003},
  month = {Feb},
  publisher = {American Physical Society},
  doi = {10.1103/PhysRevE.67.026126},
  url = {https://link.aps.org/doi/10.1103/PhysRevE.67.026126}
}

@article{acemoglu2012network,
  title={The network origins of aggregate fluctuations},
  author={Acemoglu, Daron and Carvalho, Vasco M and Ozdaglar, Asuman and Tahbaz-Salehi, Alireza},
  journal={Econometrica},
  volume={80},
  number={5},
  pages={1977--2016},
  year={2012},
  publisher={Wiley Online Library}
}

@book{banerjee2014gossip,
  title={Gossip: Identifying central individuals in a social network},
  author={Banerjee, Abhijit V and Chandrasekhar, Arun G and Duflo, Esther and Jackson, Matthew O},
  number={w20422},
  year={2014},
  publisher={National Bureau of Economic Research}
}

@article{banerjee2013diffusion,
  title={The diffusion of microfinance},
  author={Banerjee, Abhijit and Chandrasekhar, Arun G and Duflo, Esther and Jackson, Matthew O},
  journal={Science},
  volume={341},
  number={6144},
  pages={1236498},
  year={2013},
  publisher={American Association for the Advancement of Science}
}

@techreport{acemoglu2010cascades,
  title={Cascades in networks and aggregate volatility},
  author={Acemoglu, Daron and Ozdaglar, Asuman and Tahbaz-Salehi, Alireza},
  year={2010},
  institution={National Bureau of Economic Research}
}

@techreport{acemoglu2013network,
  title={The network origins of large economic downturns},
  author={Acemoglu, Daron and Ozdaglar, Asuman and Tahbaz-Salehi, Alireza},
  year={2013},
  institution={National Bureau of Economic Research}
}

@article{haldane2011systemic,
  title={Systemic risk in banking ecosystems},
  author={Haldane, Andrew G and May, Robert M},
  journal={Nature},
  volume={469},
  number={7330},
  pages={351--355},
  year={2011},
  publisher={Nature Publishing Group UK London}
}

@article{cont2013network,
  title={Network structure and systemic risk in banking systems},
  author={Cont, Rama and Moussa, Amal and Santos, Edson B.},
  journal={Handbook on systemic risk},
  volume={5},
  year={2013},
  publisher={Cambridge University Press Cambridge, UK}
}

@article{watts1998collective,
  title={Collective dynamics of ‘small-world’networks},
  author={Watts, Duncan J and Strogatz, Steven H},
  journal={Nature},
  volume={393},
  number={6684},
  pages={440--442},
  year={1998},
  publisher={Nature Publishing Group}
}

@article{albert2000error,
  title={Error and attack tolerance of complex networks},
  author={Albert, R{\'e}ka and Jeong, Hawoong and Barab{\'a}si, Albert-L{\'a}szl{\'o}},
  journal={Nature},
  volume={406},
  number={6794},
  pages={378--382},
  year={2000},
  publisher={Nature Publishing Group UK London}
}

@article{motter2002cascade,
  title={Cascade-based attacks on complex networks},
  author={Motter, Adilson E and Lai, Ying-Cheng},
  journal={Physical Review E},
  volume={66},
  number={6},
  pages={065102},
  year={2002},
  publisher={APS}
}

@article{das2018study,
  title={Study on centrality measures in social networks: a survey},
  author={Das, Kousik and Samanta, Sovan and Pal, Madhumangal},
  journal={Social network analysis and mining},
  volume={8},
  pages={1--11},
  year={2018},
  publisher={Springer}
}

@article{bonacich1987power,
  title={Power and centrality: A family of measures},
  author={Bonacich, Phillip},
  journal={American journal of sociology},
  volume={92},
  number={5},
  pages={1170--1182},
  year={1987},
  publisher={University of Chicago Press}
}

@article{katz1953new,
  title={A new status index derived from sociometric analysis},
  author={Katz, Leo},
  journal={Psychometrika},
  volume={18},
  number={1},
  pages={39--43},
  year={1953},
  publisher={Springer-Verlag}
}

@article{perron1907theorie,
  title={Zur theorie der matrices},
  author={Perron, Oskar},
  journal={Mathematische Annalen},
  volume={64},
  number={2},
  pages={248--263},
  year={1907},
  publisher={Springer}
}

@article{frobenius1912matrizen,
  title={Über Matrizen aus nicht negativen Elementen},
  author={Frobenius, Georg},
  journal={K{\"o}nigliche Akademie der Wissenschaften Berlin},
  year={1912},
  pages={456-477}
}

@article{bavelas1950communication,
  title={Communication patterns in task-oriented groups},
  author={Bavelas, Alex},
  journal={Journal of the acoustical society of America},
  year={1950},
  publisher={Acoustical Society of American}
}

@article{gabaix2011granular,
  title={The granular origins of aggregate fluctuations},
  author={Gabaix, Xavier},
  journal={Econometrica},
  volume={79},
  number={3},
  pages={733--772},
  year={2011},
  publisher={Wiley Online Library}
}

@article{carvalho2014micro,
  title={From micro to macro via production networks},
  author={Carvalho, Vasco M},
  journal={Journal of Economic Perspectives},
  volume={28},
  number={4},
  pages={23--48},
  year={2014},
  publisher={American Economic Association 2014 Broadway, Suite 305, Nashville, TN 37203-2418}
}

@article{baqaee2019macroeconomic,
  title={The macroeconomic impact of microeconomic shocks: Beyond Hulten's theorem},
  author={Baqaee, David Rezza and Farhi, Emmanuel},
  journal={Econometrica},
  volume={87},
  number={4},
  pages={1155--1203},
  year={2019},
  publisher={Wiley Online Library}
}

@book{leontief1941structure,
    author = {Leontief, Wassily W},
    title = {The structure of American economy, 1919-1929: An empirical application of equilibrium analysis},
    publisher = {Harvard University Press},
    year = {1941}
}

@article{leontief1991economy,
  title={The economy as a circular flow},
  author={Leontief, Wassily},
  journal={Structural change and economic dynamics},
  volume={2},
  number={1},
  pages={181--212},
  year={1928},
  publisher={Elsevier}
}

@article{padgett1993robust,
  title={Robust Action and the Rise of the Medici, 1400-1434},
  author={Padgett, John F and Ansell, Christopher K},
  journal={American journal of sociology},
  volume={98},
  number={6},
  pages={1259--1319},
  year={1993},
  publisher={University of Chicago Press}
}

@article{albert1999diameter,
  title={Diameter of the world-wide web},
  author={Albert, R{\'e}ka and Jeong, Hawoong and Barab{\'a}si, Albert-L{\'a}szl{\'o}},
  journal={Nature},
  volume={401},
  number={6749},
  pages={130--131},
  year={1999},
  publisher={Nature Publishing Group UK London}
}

@article{pagani2013power,
  title={The power grid as a complex network: a survey},
  author={Pagani, Giuliano Andrea and Aiello, Marco},
  journal={Physica A: Statistical Mechanics and its Applications},
  volume={392},
  number={11},
  pages={2688--2700},
  year={2013},
  publisher={Elsevier}
}

@article{gonzalez2008understanding,
  title={Understanding individual human mobility patterns},
  author={Gonzalez, Marta C and Hidalgo, Cesar A and Barabasi, Albert-Laszlo},
  journal={Nature},
  volume={453},
  number={7196},
  pages={779--782},
  year={2008},
  publisher={Nature Publishing Group UK London}
}

@article{bernard2019production,
  title={Production networks, geography, and firm performance},
  author={Bernard, Andrew B and Moxnes, Andreas and Saito, Yukiko U},
  journal={Journal of Political Economy},
  volume={127},
  number={2},
  pages={639--688},
  year={2019},
  publisher={The University of Chicago Press Chicago, IL}
}

@article{dhyne2021trade,
  title={Trade and domestic production networks},
  author={Dhyne, Emmanuel and Kikkawa, Ayumu Ken and Mogstad, Magne and Tintelnot, Felix},
  journal={The Review of Economic Studies},
  volume={88},
  number={2},
  pages={643--668},
  year={2021},
  publisher={Oxford University Press}
}

@techreport{criscuolo2024estonia,
  title={Estonia’s firm-level production network: Lessons for industrial policy},
  author={Criscuolo, Chiara and Dechezlepr{\^e}tre, Antoine and Guillouet, Louise and Lalanne, Guy and Manaresi, Francesco},
  year={2024},
  institution={OECD Publishing}
}

@article{fujiwara2010large,
  title={Large-scale structure of a nation-wide production network},
  author={Fujiwara, Yoshi and Aoyama, Hideaki},
  journal={The European Physical Journal B},
  volume={77},
  pages={565--580},
  year={2010},
  publisher={Springer}
}

@article{clauset2009power,
  title={Power-law distributions in empirical data},
  author={Clauset, Aaron and Shalizi, Cosma Rohilla and Newman, Mark EJ},
  journal={SIAM review},
  volume={51},
  number={4},
  pages={661--703},
  year={2009},
  publisher={SIAM}
}

@article{jeong2001lethality,
  title={Lethality and centrality in protein networks},
  author={Jeong, Hawoong and Mason, Sean P and Barab{\'a}si, A-L and Oltvai, Zoltan N},
  journal={Nature},
  volume={411},
  number={6833},
  pages={41--42},
  year={2001},
  publisher={Nature Publishing Group UK London}
}

@article{redner1998popular,
  title={How popular is your paper? An empirical study of the citation distribution},
  author={Redner, Sidney},
  journal={The European Physical Journal B-Condensed Matter and Complex Systems},
  volume={4},
  number={2},
  pages={131--134},
  year={1998},
  publisher={Springer}
}

@article{mizuno2014structure,
  title={The structure and evolution of buyer-supplier networks},
  author={Mizuno, Takayuki and Souma, Wataru and Watanabe, Tsutomu},
  journal={Plos one},
  volume={9},
  number={7},
  pages={e100712},
  year={2014},
  publisher={Public Library of Science San Francisco, USA}
}

@article{afrouzi2020growing,
  title={Growing by the masses-revisiting the link between firm size and market power},
  author={Afrouzi, Hassan and Drenik, Andres and Kim, Ryan},
  year={2020},
  journal={CESifo Working Paper}
}

@article{bernard2022origins,
  title={The origins of firm heterogeneity: A production network approach},
  author={Bernard, Andrew B and Dhyne, Emmanuel and Magerman, Glenn and Manova, Kalina and Moxnes, Andreas},
  journal={Journal of Political Economy},
  volume={130},
  number={7},
  pages={1765--1804},
  year={2022},
  publisher={The University of Chicago Press Chicago, IL}
}

@article{carvalho2019production,
  title={Production networks: A primer},
  author={Carvalho, Vasco M and Tahbaz-Salehi, Alireza},
  journal={Annual Review of Economics},
  volume={11},
  number={1},
  pages={635--663},
  year={2019},
  publisher={Annual Reviews}
}

@article{oberfield2018theory,
  title={A theory of input--output architecture},
  author={Oberfield, Ezra},
  journal={Econometrica},
  volume={86},
  number={2},
  pages={559--589},
  year={2018},
  publisher={Wiley Online Library}
}

@inproceedings{lucas1977understanding,
  title={Understanding business cycles},
  author={Lucas, Robert E},
  booktitle={Carnegie-Rochester Conference Series on Public Policy},
  volume={5},
  number={1},
  pages={7--29},
  year={1977},
  organization={Elsevier}
}

@article{mcnerney2022production,
  title={How production networks amplify economic growth},
  author={McNerney, James and Savoie, Charles and Caravelli, Francesco and Carvalho, Vasco M and Farmer, J Doyne},
  journal={Proceedings of the National Academy of Sciences},
  volume={119},
  number={1},
  pages={e2106031118},
  year={2022},
  publisher={National Academy of Sciences}
}

@inproceedings{eikmeier2017revisiting,
  title={Revisiting power-law distributions in spectra of real world networks},
  author={Eikmeier, Nicole and Gleich, David F},
  booktitle={Proceedings of the 23rd ACM SIGKDD international conference on knowledge discovery and data mining},
  pages={817--826},
  year={2017}
}

@article{bacilieri2026firm,
  title={Firm-level production networks: what do we (really) know?},
  author={Bacilieri, Andrea and Borsos, Andr{\'a}s and Astudillo-Estevez, Pablo and Hoefer, Mads and Lafond, Fran{\c{c}}ois},
  journal={Journal of Economic Dynamics and Control},
  volume={187},
  pages={105313},
  year={2026},
  publisher={Elsevier}
}

@article{dhyne2016three,
  title={Three regions, three economies?},
  author={Dhyne, Emmanuel and Duprez, C{\'e}dric},
  journal={Economic Review},
  number={iii},
  pages={59--73},
  year={2016},
  publisher={National Bank of Belgium}
}
\clearpage

\appendix
\renewcommand\thefigure{\thesection.\arabic{figure}}  
\renewcommand{\thetable}{\thesection.\arabic{table}}

\section{Matrix Representation of a Network}
\setcounter{figure}{0}  
\setcounter{table}{0}
\label{app:matrix_representation}
In the paper we make large use of a matrix representation of the network of commercial exchanges between the firms in our dataset. 
Here we illustrate this representation with a simple example, in order to clarify the operations performed in the main text. 
Let's start with the simple directed network depicted in Figure \ref{fig: example_network}:
\begin{figure}[H]
    \centering
    \begin{tikzpicture}[
        node distance=2.5cm,
        mynode/.style={circle, draw, fill=#1, text=white, minimum size=1cm, font=\bfseries} 
        % :)
    ]

    \node[mynode=teal!70!black] (1) {A};
    \node[mynode=teal!70!black] (2) [right of=1] {B};
    \node[mynode=teal!70!black] (3) [below right of=1] {C};
    \node[mynode=teal!70!black] (4) [below of=3] {D};
    \node[mynode=teal!70!black] (5) [below left of=1] {E};
    \node[mynode=teal!70!black] (6) [left of=5] {F};
    \node[mynode=teal!70!black] (7) [above of=6] {G};

    \draw[-{Stealth[scale=1.5]}] (1) -- (2);
    \draw[-{Stealth[scale=1.5]}] (2) -- (3);
    \draw[-{Stealth[scale=1.5]}] (3) -- (4);
    \draw[-{Stealth[scale=1.5]}] (4) -- (5);
    \draw[-{Stealth[scale=1.5]}] (3) -- (1);
    \draw[-{Stealth[scale=1.5]}] (6) -- (7);
    \draw[-{Stealth[scale=1.5]}] (1) -- (5);
    \draw[-{Stealth[scale=1.5]}] (5) -- (3);
    \end{tikzpicture}
    \caption{Example network.}
    \label{fig: example_network}
\end{figure}
The network has two disconnected components, composed respectively of 5 and 2 nodes. 
The matrix representation of the downstream adjacency of this network is given in figure \ref{fig: example_matrix_representation}:
\begin{figure}[H]
    \centering
    $\begin{array}{r|ccccccc}
      & \bf{A} & \bf{B} & \bf{C} & \bf{D} & \bf{E} & \bf{F} & \bf{G} \\
    \hline
    \bf{A} & 0 & 1 & 0 & 0 & 1 & 0 & 0 \\
    \bf{B} & 0 & 0 & 1 & 0 & 0 & 0 & 0 \\
    \bf{C} & 1 & 0 & 0 & 1 & 0 & 0 & 0 \\
    \bf{D} & 0 & 0 & 0 & 0 & 1 & 0 & 0 \\
    \bf{E} & 0 & 0 & 1 & 0 & 0 & 0 & 0 \\
    \bf{F} & 0 & 0 & 0 & 0 & 0 & 0 & 1 \\
    \bf{G} & 0 & 0 & 0 & 0 & 0 & 0 & 0 \\
    \end{array}$
    \caption{Matrix representation of the network. Nodes (firms) are indexed and reported both in rows and columns. The matrix has positive entries exclusively on pairs connected by a link.}
    \label{fig: example_matrix_representation}
\end{figure}
% \begin{figure}[H]
%     \centering
%     $\begin{array}{r|ccccccc}
%       & \bf{A} & \bf{B} & \bf{C} & \bf{D} & \bf{E} & \bf{F} & \bf{G} \\
%     \hline
%     \bf{A} & 0 & 0 & 1 & 0 & 0 & 0 & 0 \\
%     \bf{B} & 1 & 0 & 0 & 0 & 0 & 0 & 0 \\
%     \bf{C} & 0 & 1 & 0 & 0 & 1 & 0 & 0 \\
%     \bf{D} & 0 & 0 & 1 & 0 & 0 & 0 & 0 \\
%     \bf{E} & 1 & 0 & 0 & 1 & 0 & 0 & 0 \\
%     \bf{F} & 0 & 0 & 0 & 0 & 0 & 0 & 0 \\
%     \bf{G} & 0 & 0 & 0 & 0 & 0 & 1 & 0 \\
%     \end{array}$
%     \caption{Matrix representation of the network. Nodes (firms) are indexed and reported both in rows and columns. The matrix has positive entries exclusively on pairs connected by a link.}
%     \label{fig: example_matrix_representation}
% \end{figure}
This adjacency matrix is formed by filling the matrix entries corresponding to the presence of a link between two nodes of the network. 
The upstream flow of the network is easily described considering the transposed matrix, and the undirected network could be described by the symmetrized matrix.
If we wanted to consider the weighted network, instead of filling the entries with ones we would have inserted the corresponding weights. 
The existence of two disconnected components translates to the possibility of expressing the related matrix as a combination of diagonal blocks, as can be seen here in the case of the sub-matrices with indices $\bf{A}$ to $\bf{E}$ and $\bf{F}$-$\bf{G}$. 
In every matrix product these two sub-matrices can be treated independently, as can the sub-network of nodes $\bf{F}$ and $\bf{G}$. 
In the network depicted in figure \ref{fig: example_network} there is one sink node ($\bf{G}$). 
It corresponds, in the matrix representation, to an empty row, indicating the absence of links going out of the node. 
Similarly, we have a source node ($\bf{F}$) which corresponds to the presence of an empty column in the matrix.
More generally, the out-degree of a node corresponds to the sum of the entries in the associated row, and the in-degree is the sum of the associated column. 

\subsection{Centrality Measures and Shock Propagation}
Centrality measures associated to the matrix in Figure \ref{fig: example_matrix_representation} can be directly used to give an estimation of the effects on the whole network of localized shocks.
In particular the eigenvector centrality value of a node corresponds to a quantification of the global effects of a shock applied to it. 
Taking as an example the network represented in Figure \ref{fig: example_network} and applying a shock to node $\bf{A}$, after a time step the number of nodes influenced by the shock is equal to the number of out-going links from $\bf{A}$ (in this case nodes $\bf{B}$ and $\bf{E}$).
Written in terms of the adjacency matrix, this corresponds to the definition of (downstream) degree centrality:
\begin{equation}
    e_{i,1} = \sum_{j=1}^N A_{ij}
\end{equation}
where in this case $i=\bf{A}$, $N$ is equal to the number of nodes in the network, and $A_{ij}$ are adjacency matrix elements. 
After a second time step the shock propagates to nodes at distance two from $\bf{A}$ (in this case only node $\bf{C}$), and the sum becomes:
\begin{equation}
    e_{i,2} = \sum_{j=1}^N \sum_{k=1}^N A_{ik} A_{kj} = \sum_{j=1}^N  \left(A^2 \right)_{ij} 
\end{equation}
After a sufficient number of time steps this quantity approximates \eqref{eq: eigenvector_centrality_matrix}, the eigenvector centrality associated to node $i=\bf{A}$ (Katz and PageRank centrality being regularized versions of this measure).

Eigenvector centrality can also be used to evaluate the effects on a given node of a diffused shock on the network. 
The evolution of a distribution on the network over time is governed by the transpose of the adjacency matrix \ref{fig: example_matrix_representation}. 
The necessity to transpose the adjacency matrix can be understood in terms of the definition of matrix product: to evaluate the downstream eigenvector centrality of a node $i$ we sum the centrality scores of all the nodes reached by links coming from $i$. 
This implies summing over the rows of the adjacency matrix, in accordance with the definition of matrix multiplication on the right. 
On the other hand, to propagate a shock over the network, the effect on node $i$ is given by the sum of all the contributions coming from nodes with links pointing to $i$, which corresponds to summing over the columns of the adjacency matrix.
% In addition, to consider a meaningful propagation the columns of the transition matrix must be normalized. 
% Normalizing the columns we ensure that nodes cannot have an amplifying effect on shocks, and probability distributions are transformed to vectors summing to values lower than or equal to 1 when following the transitions in the network.
 % (as an example of the attenuating effect of the network, node $\bf{F}$ has no incoming connections, hence a shock applied to it does not propagate)
% The column-normalized (transposed) downstream matrix describes how a given distribution of goods is passed to the buyers, and the column-normalized upstream transition matrix describes how a given distribution of money is passed to the sellers.
The transition matrix associated to the adjacency structure of Figure \ref{fig: example_network} is thus given by:
% \begin{figure}[H]
%     \centering
%     $\begin{array}{r|ccccccc}
%       & \bf{A} & \bf{B} & \bf{C} & \bf{D} & \bf{E} & \bf{F} & \bf{G} \\
%     \hline
%     \bf{A} & 0 & 1 & 0 & 0 & 1 & 0 & 0 \\
%     \bf{B} & 0 & 0 & 1 & 0 & 0 & 0 & 0 \\
%     \bf{C} & 1 & 0 & 0 & 1 & 0 & 0 & 0 \\
%     \bf{D} & 0 & 0 & 0 & 0 & 1 & 0 & 0 \\
%     \bf{E} & 0 & 0 & 0 & 0 & 0 & 0 & 0 \\
%     \bf{F} & 0 & 0 & 0 & 0 & 0 & 0 & 1 \\
%     \bf{G} & 0 & 0 & 0 & 0 & 0 & 0 & 0 \\
%     \end{array}$
%     \caption{Transition matrix associated to the example network of Figure \ref{fig: example_network}.
%     The matrix is column-normalized, hence it transforms probability distributions to vectors summing to 1 or less.}
%     \label{fig: example_transition_matrix}
% \end{figure}
\begin{figure}[H]
    \centering
    $\begin{array}{r|ccccccc}
      & \bf{A} & \bf{B} & \bf{C} & \bf{D} & \bf{E} & \bf{F} & \bf{G} \\
    \hline
    \bf{A} & 0 & 0 & 1 & 0 & 0 & 0 & 0 \\
    \bf{B} & 1 & 0 & 0 & 0 & 0 & 0 & 0 \\
    \bf{C} & 0 & 1 & 0 & 0 & 1 & 0 & 0 \\
    \bf{D} & 0 & 0 & 1 & 0 & 0 & 0 & 0 \\
    \bf{E} & 1 & 0 & 0 & 1 & 0 & 0 & 0 \\
    \bf{F} & 0 & 0 & 0 & 0 & 0 & 0 & 0 \\
    \bf{G} & 0 & 0 & 0 & 0 & 0 & 1 & 0 \\
    \end{array}$
    \caption{Transition matrix associated to the example network of Figure \ref{fig: example_network}.}
    \label{fig: example_transition_matrix}
\end{figure}
This matrix can be used to describe (downstream) shock propagation through the network: for example starting with a small decrease in the production of a firm, we can use the transition matrix to propagate this change over the network, and study its long term effects. 
The final distribution of the effects is proportional to the eigenvector centrality of this transition matrix, or equivalently the upstream eigenvector centrality of the original matrix \ref{fig: example_matrix_representation}.

\subsection{Spectral properties of the matrix representation}
Once a matrix is associated to the network, a number of tools of matrix analysis can be applied to study its properties. 
In particular a singular value decomposition (SVD) can be applied to extract the spectrum of the matrix, and to investigate the presence of relevant lower-dimensional subspaces.
Given the large dimension of the network matrix, only the initial, largest values in the spectrum can be evaluated, but also for these quantities we find a power-law behavior.
Assuming this behavior is not limited to the initial portion of the spectrum we can estimate the relevant exponents and infer the decay properties of singular values deeper in the spectrum (the validity of a power-law behavior for the singular values associated to many real-world networks is confirmed in \cite{eikmeier2017revisiting}). 
More precisely, we analyze four matrix representations of the network: the adjacency matrix, the weighted matrix described in Section \ref{sec:weighted_links}, and the row- and column-normalized weighted matrices used in the evaluation of upstream and downstream PageRank centrality. 
For each matrix we evaluate the first 1,000 singular values, and fit the cumulative sum of these values with a power law to extend the sum further in the spectrum.\footnote{While the uncertainties on the exponents are negligible we expect significant biases, given that we are estimating the asymptotic behavior based only on the first few values of the spectrum.} 
The resulting exponents and intercepts are listed in Table \ref{tab:svd_spectra}:
\begin{table}[H]
    \centering
    \small
    \begin{tabular}{lcc}
    %\toprule 
    {} & \multicolumn{1}{c}{Exponent} &\multicolumn{1}{c}{Intercept} \\
    \midrule
    % Adjacency                   & 0.59 & 1,750.96 \\
    % Weighted                    & 0.19 & 57.92$\times 10^9$ \\
    % Weighted, row-normalized    & 0.67 & 69.24 \\
    % Weighted, column-normalized & 0.52 & 326.53 \\
    Adjacency                   & 0.59 & 1,751 \\
    Weighted                    & 0.19 & 57.92$\times 10^9$ \\
    Weighted, row-normalized    & 0.67 & 69 \\
    Weighted, column-normalized & 0.52 & 327 \\
    \bottomrule
    \end{tabular}
    \caption{Estimated intercepts and exponents for the cumulative sums of the singular values of adjacency matrix, weighted matrix, row-normalized and column-normalized weighted matrix. Standard deviations are not reported, given their values are negligible with respect to the expected bias.}
    \label{tab:svd_spectra}
\end{table}
In addition we show the growth of the cumulative sums of the various spectra in Figure \ref{fig:svd_spectra}:
\begin{figure}[H]
\centering
  \includegraphics[width=0.9\linewidth]{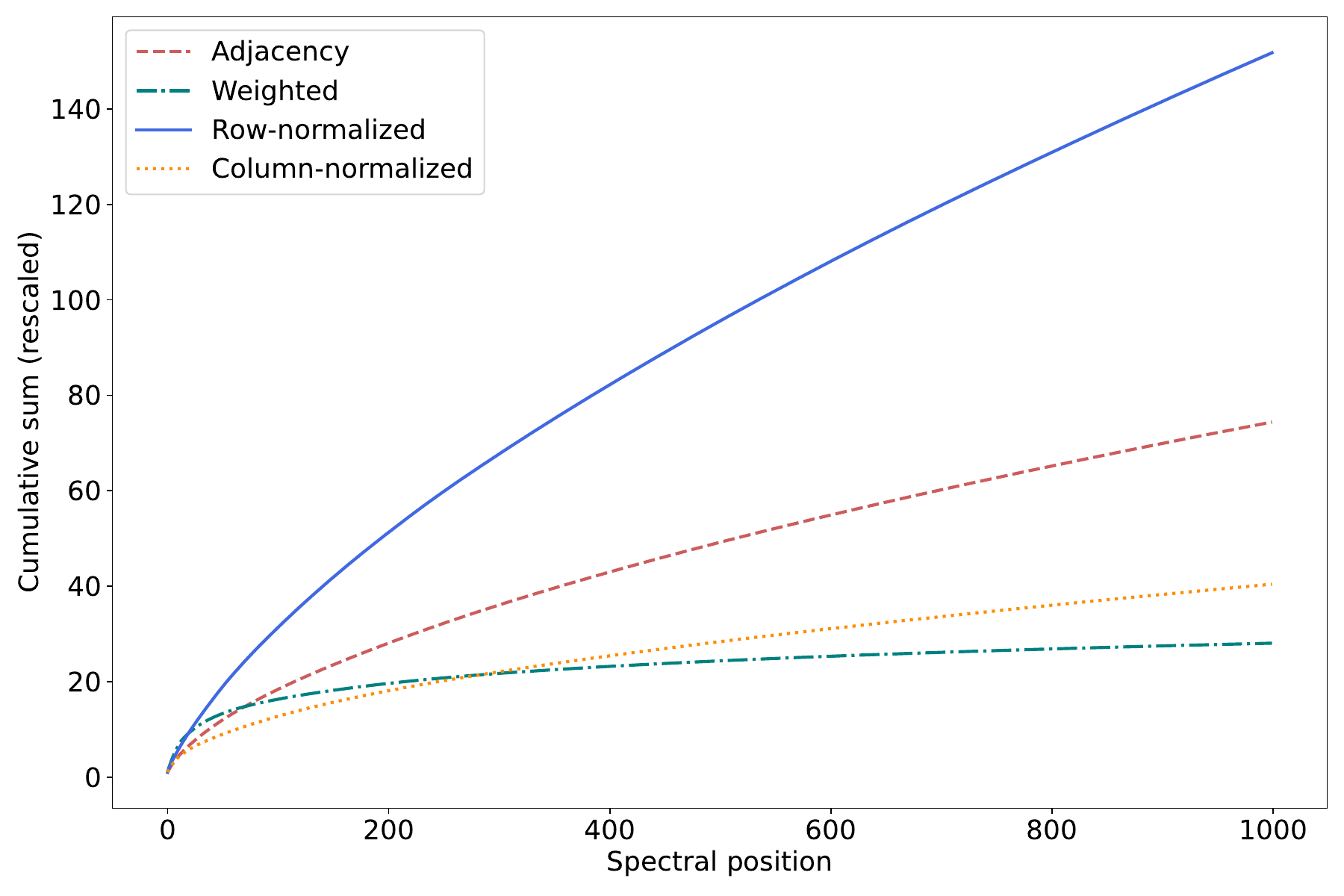}
  \caption{Rescaled cumulative sums of the first 1000 singular values for the adjacency, weighted, row-normalized and column-normalized matrix representations of the network. All the sums are normalized by the value of the first singular value, in order to have a common starting point.}
  \label{fig:svd_spectra}
\end{figure} 
As clearly visible from Figure \ref{fig:svd_spectra} and as confirmed by the values of the exponents in Table \ref{tab:svd_spectra}, the weighted matrix has the spectrum with the slowest-growing cumulative sum. 
This could be expected by the fact that the weighted matrix is effectively more sparse than the others, as explained in section \ref{sec:weighted_links}.
The values of the exponents listed in Table \ref{tab:svd_spectra} - with the assumption that these values remain significantly larger than zero also deeper in the spectrum - imply that a low-dimensional approximation for the listed matrices would not be useful. 
To capture a meaningful fraction of the total variation expressed in the full spectrum an impractically large number of singular values have to be kept in the approximation: in the case of the weighted matrix - the one with the fastest-decaying spectrum - to represent at least 80\% of the total variation more than 500,000 singular values are needed.
For the other three matrices the effective dimension needed to capture the same fraction of the total variation is in excess of one million, making such an approximation quite impractical.
As already stated this conclusion is strongly dependent on the power-law nature of the spectrum of the matrix representations of the network. 
While this assumption could prove wrong when analyzing deeper portions of the spectrum, the power-law behavior is confirmed in many real-world networks for which the full spectrum can be evaluated (see \cite{eikmeier2017revisiting}), and even in the presence of an upward bias in the exponents estimated here the qualitative conclusions are unlikely to be altered.

\clearpage
\section{Reporting Threshold}
\label{sec:appendix_rep_thres}
\setcounter{figure}{0}  
\setcounter{table}{0} 

To ensure that the quantities computed in Sections \ref{sec:density_and_mean_degree} and \ref{sec:density_and_mean_degree_breakdown} are comparable to those reported in the literature, it is important to consider that some of the datasets collecting information on transactions between firms provide specific constraints as part of their data collection methodology.\footnote{We refer the reader to \cite{bacilieri2026firm} for a comprehensive review of the datasets currently available worldwide.} The most common of these is a minimum threshold for recording a transaction. The following plots illustrate the behavior of our network under a scenario that synthetically replicates the presence of such a threshold in the data collection, considering only transactions with average invoice amount above a given value.\footnote{We performed the same exercise considering the total amount of the invoices between two firms. The results are consistent with what is presented in this section, and are reported in Appendix \ref{subsec:appendix_plot_threshold_total_gross}.}
We begin by illustrating how the number of firms and the number of links vary with respect to the magnitude of the threshold.
\begin{figure}[H]
\centering
  \includegraphics[width=1\linewidth]{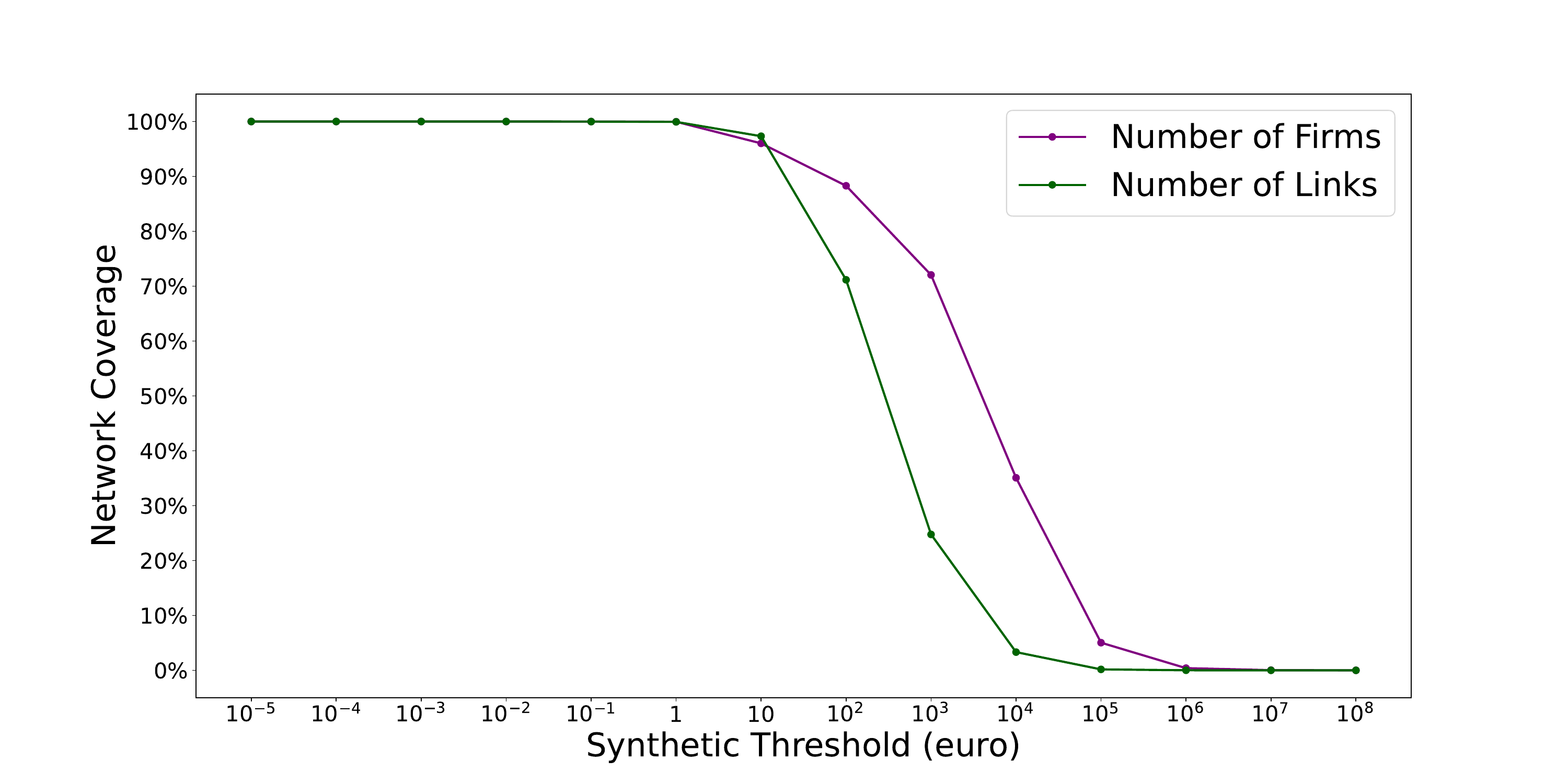}
  \caption{Variation of the number of nodes and links of the Network with respect to the magnitude of a synthetic threshold on the average invoice amount.}
  \label{fig:firms_links_cutoff_mean_link}
\end{figure} 
Notably, the number of links is reduced by approximately half when the threshold lies between $10^2$ euro and $10^3$ euro (as in Belgium, for instance, where the threshold is 250 euro). The plot also shows that for a comparable reduction of the number of nodes (a node is excluded from the network when it has no remaining links) the threshold has to be of order $10^4$. %This discrepancy suggests that at least half of the firms have an average transaction value on the order of $10^4$, whereas the overall mean of firms’ average invoice values is 12.
Finally, the structure of the network remains basically unaffected when the threshold is up to an order of $10$ euro, as in Hungary after 2021, where the threshold was lowered to 2.5 euro (1,000 HUF).

\begin{figure}[H]
\centering
  \includegraphics[width=1\linewidth]{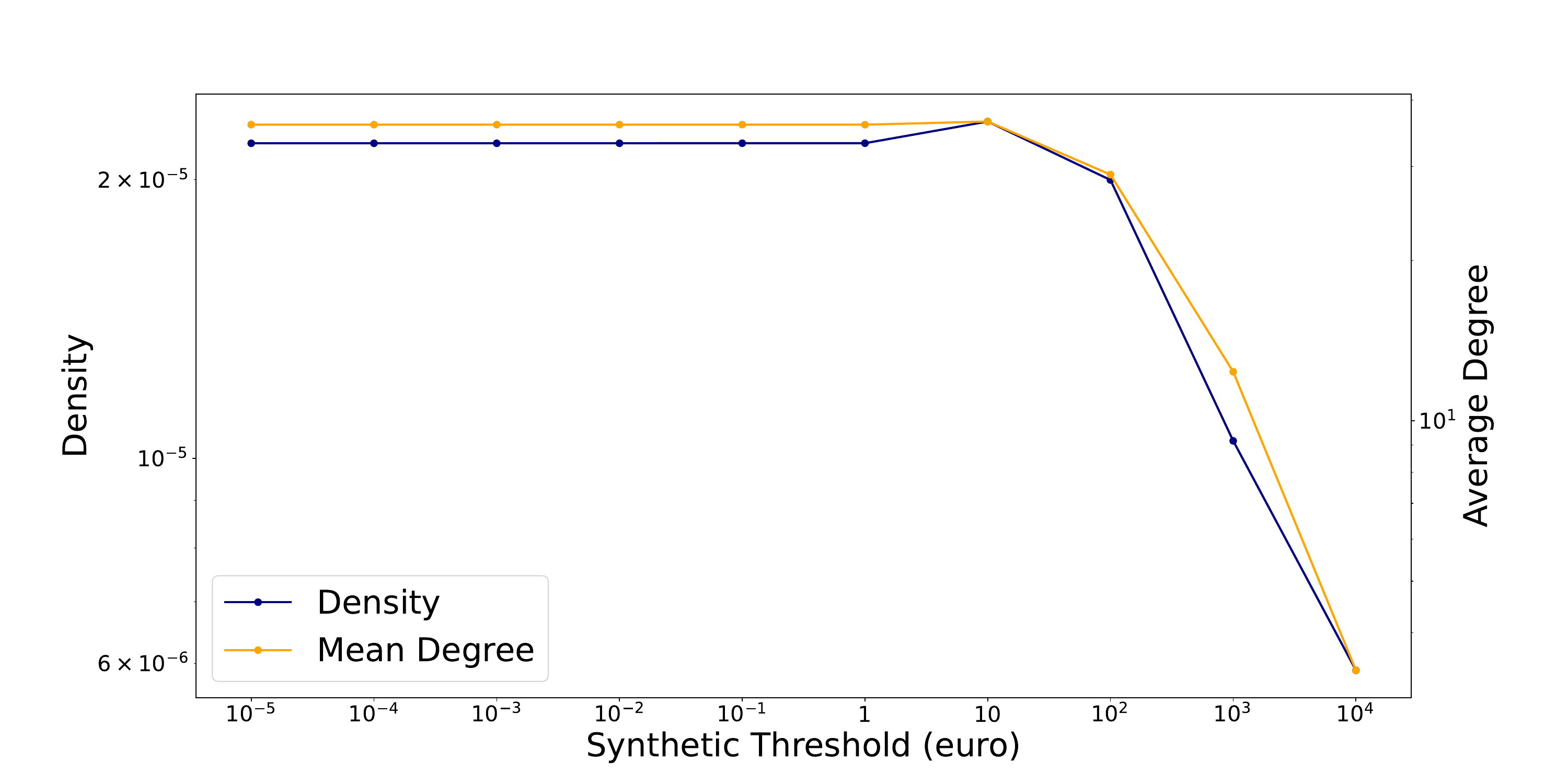}
  \caption{Variation of the Density and Mean Degree in the Network with respect to the magnitude of the synthetic threshold on the average invoice amount.}
  \label{fig:density_mean_deg_cutoff_mean_link}
\end{figure} 

On the other hand, it is noteworthy that when looking at the density and mean degree, beyond a threshold of about 10 euro they both decrease as the threshold order increases. We report the variation of these quantities up to a maximum threshold on the order $10^4$ euro, as it is the highest threshold reported in the literature (also, with higher thresholds, the reduced number of nodes makes the results less meaningful). This observation may appear to contradict the findings presented in Figure \ref{fig:density_mean_degree_wrt_Ncategory}. However, it is important to note that, in this case, we are selecting links based on the threshold, whereas the analysis in that figure was based on a nodes selection.

%On the other hand, it is interesting how both the density and the mean degree decrease as the threshold size increases. We report the variation for these quantities for a maximum threshold of order 10 4 of order 10,000 because this is the maximum order of threshold reported in the literature. This seems to contradict what is reported in Figure X in the REF section. However, it should be remembered that we are only selecting links here, whereas there we were selecting nodes.

\subsection{Plot of Reporting Threshold by Total Gross Amount of Invoice}
\label{subsec:appendix_plot_threshold_total_gross}

\begin{figure}[H]
\centering
  \includegraphics[width=1\linewidth]{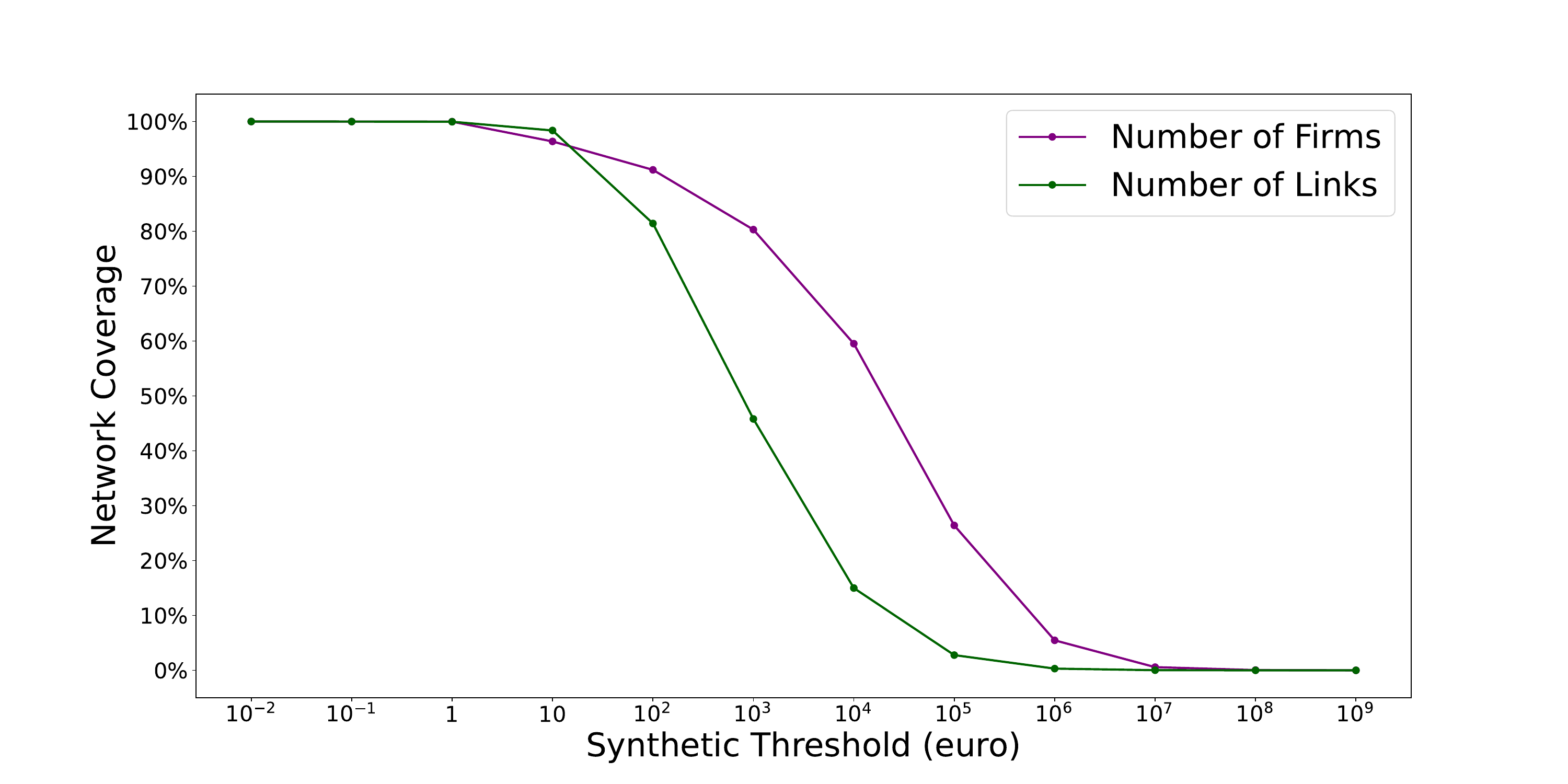}
  \caption{Variation of the number of nodes and links of the Network with respect to the magnitude of a synthetic threshold on the total invoice amount.}
  \label{fig:firms_links_cutoff}
\end{figure} 

\begin{figure}[H]
\centering
  \includegraphics[width=1\linewidth]{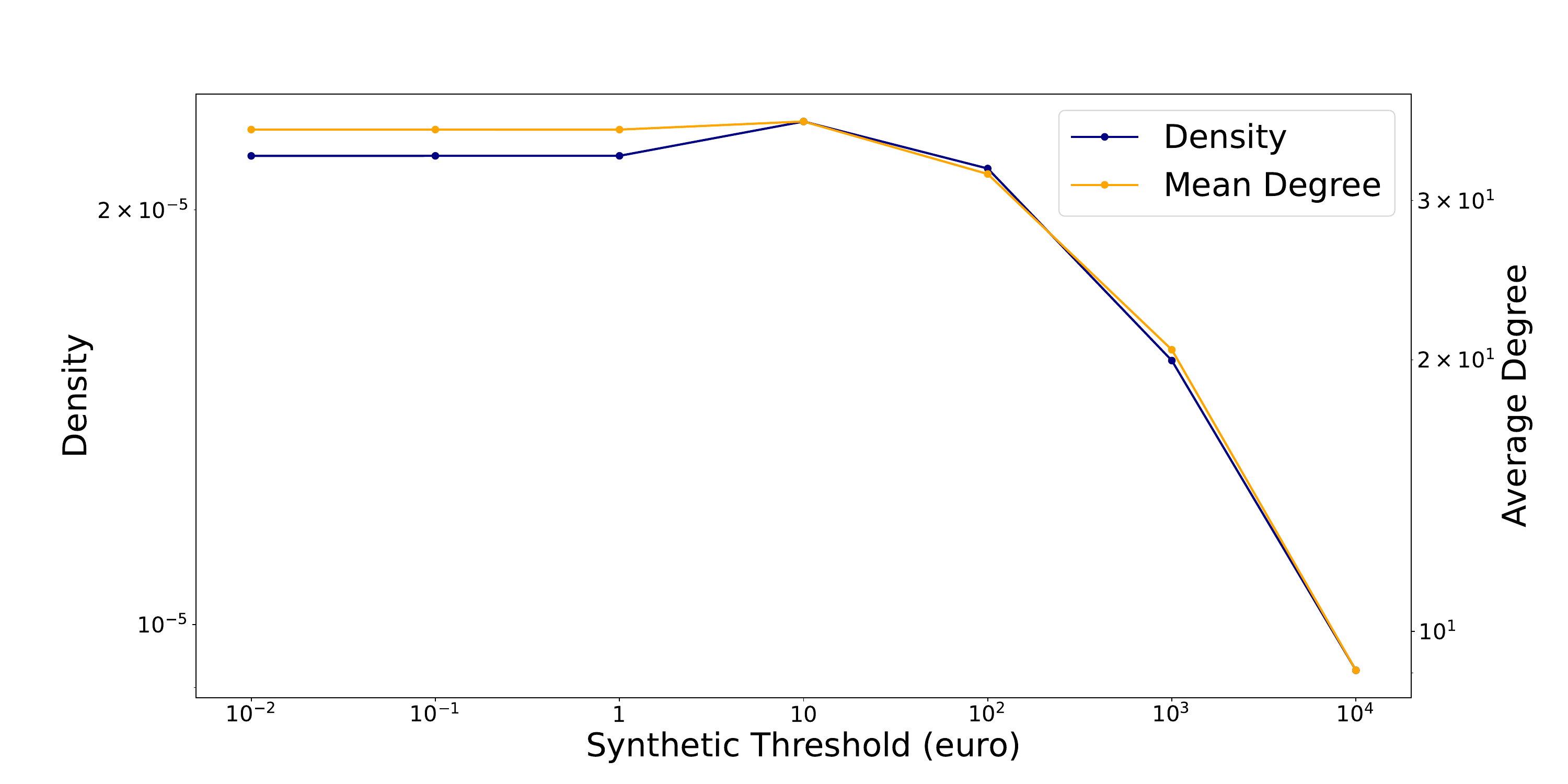}
  \caption{Variation of the Density and Mean Degree in the Network with respect to the magnitude of a synthetic threshold on the total invoice amount.}
  \label{fig:density_mean_deg_cutoff}
\end{figure} 

\clearpage

\section{Buyer-only and Seller-only Firms}
\label{subsection:SinkSources}
\setcounter{figure}{0}  
\setcounter{table}{0} 

Nodes characterized by one type of link only (i.e. either exclusively buyers or sellers) are not uncommon in business relationship networks. In our specific case, however, some links could have been removed during the data cleaning process.\footnote{See Section \ref{sec:FromDataToNetwork}, where we considered only relations between firms of type B.}
When looking at the whole dataset, considering entities of all types, $71\%$ of the seller-only firms and $95\%$ of the buyer-only firms show the same structure, thereby indicating that the lack of connections is not dependent on data pre-processing.  
The underlying reasons for this particular behavior may vary, potentially arising from trade-related factors (e.g., firms engaged solely in importing or exporting activities with foreign markets) or commercial considerations (e.g., firms established for specific, one-time business transactions).
\begin{table}[H]
    \centering
    \small
    \begin{tabular}{llrr}
    \multicolumn{1}{c}{Firm Type} &
    \multicolumn{1}{c}{Network} &
    \multicolumn{1}{c}{Share}&
    \multicolumn{1}{c}{Count}\\
   \midrule
    Seller-only & B $\rightarrow$ B &   0.67 \% &  10,944 \\
                & Complete          &   0.44 \% &   7,734 \\
    Buyer-only  & B $\rightarrow$ B &  38.02 \% & 624,772 \\
                & Complete          &  34.10 \% & 594,770 \\
   \bottomrule
    \end{tabular}
    \caption{Statistics on sinks and sources, nodes that have only a single type of link: sources only have out-going links and sinks only have in-going links. In production networks they represents firms only selling (sources) or only buying (sinks).}
    \label{tab:SinkSourcesStats}
\end{table}

\begin{figure}[H]
\centering
  \includegraphics[width=1\linewidth]{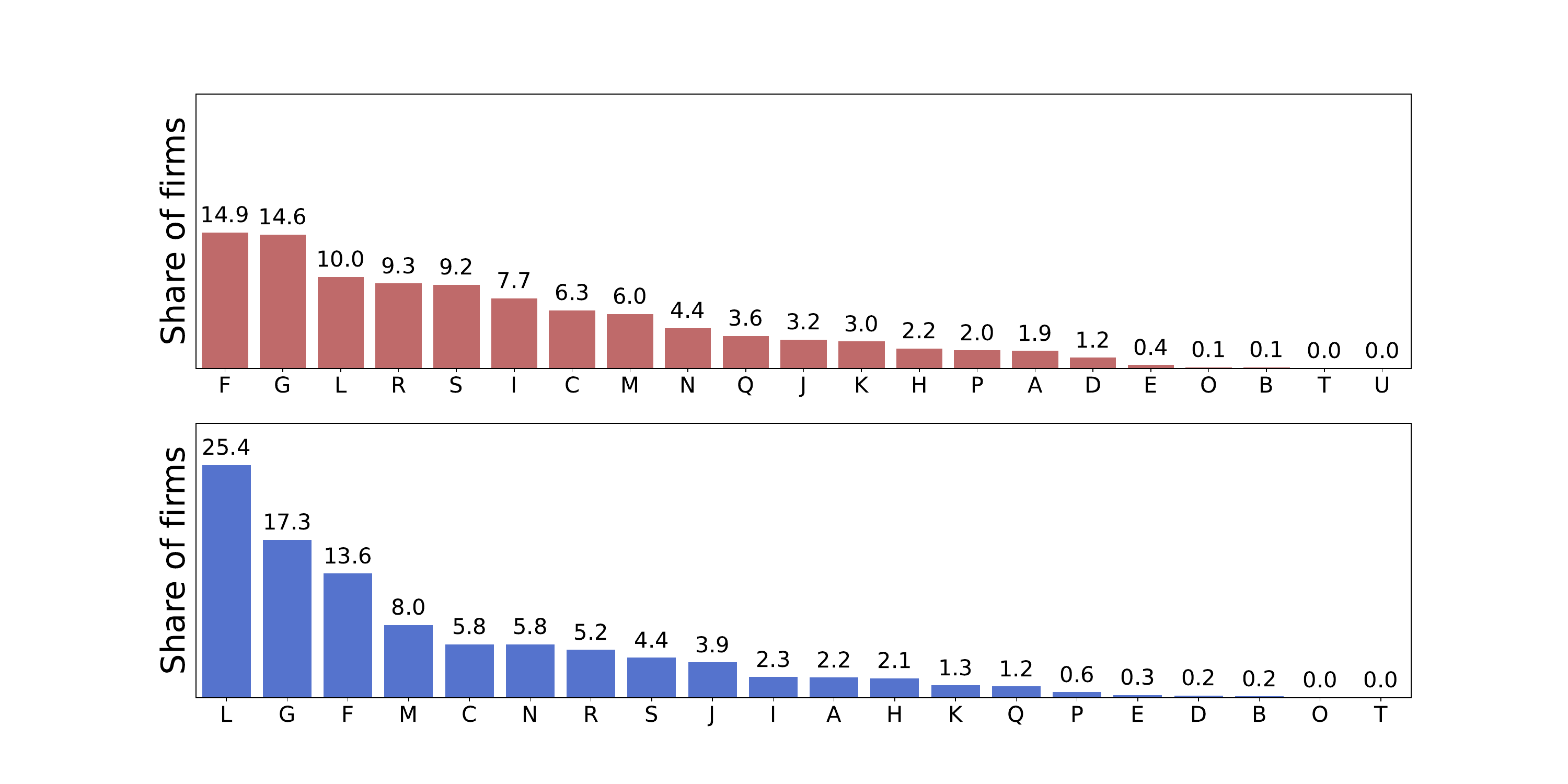}
  \caption{Share of buyer-only (top) and seller-only firms (bottom) for each sector.}
  \label{fig:seller_buyer_only_cat_breakdown}
\end{figure}

\begin{figure}[H]
\centering
  \includegraphics[width=1\linewidth]{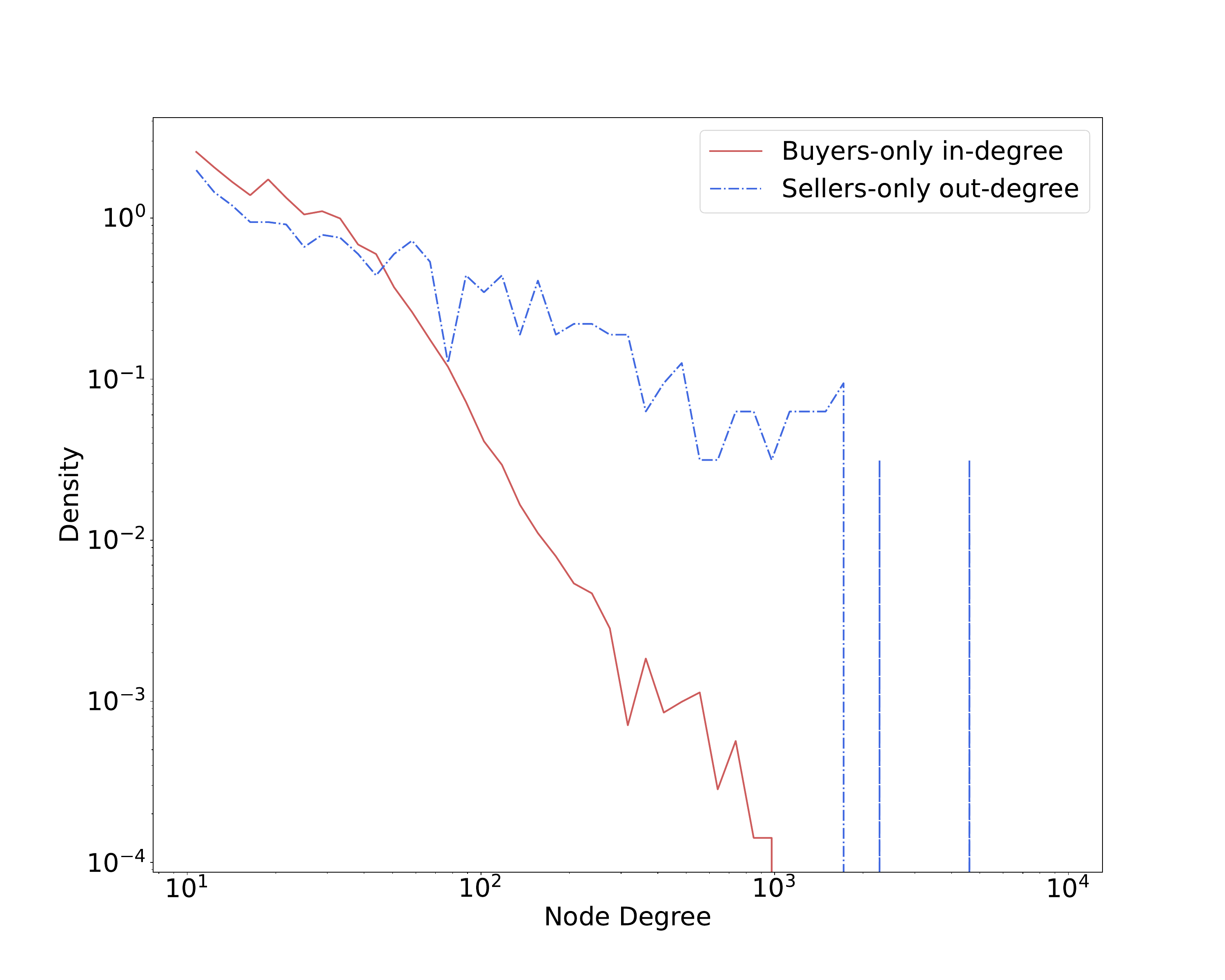}
  \caption{Density histograms of out-degree of seller-only firms and in-degree of buyer-only firms. The x-axis is partitioned into 100 logarithmically spaced bins and the y-axis reports the density of firms falling within each bin.}
  \label{fig:sellers_buyers_only_degree_log_scale}
\end{figure}

As seen in Table \ref{tab:buyer_seller_only_confronto} there is no clear trend in the proportion of this type of nodes over time. 
Nonetheless, the share of buyer-only firms observed for Italy appears higher than in other countries, suggesting the possibility of further investigation into the underlying factors contributing to such a high presence.
\begin{table}[H]
    \centering
    \small
    \begin{tabular}{lcrrr}
    \multicolumn{1}{c}{Country} &
    Year &
    \multicolumn{1}{c}{\makecell{Transaction \\  threshold (euro)}}& \multicolumn{1}{c}{\makecell{Share of \\  seller-only}} & \multicolumn{1}{c}{\makecell{Share of \\ Buyer-only  }}\\
   \midrule
   \textbf{Italy} & \textbf{2019} & \textbf{0.00} & \textbf{0.7\%} &\textbf{38.0\%} \\
   Ecuador & 2015 & 0.00   & 15.3\% & 20.0\%\\
   Hungary & 2021 & 2.50   &  19.4\% & 13.7\%\\
   Belgium & 2014 & 250.00 & 0.1\% & 15.4\%\\
   \bottomrule
    \end{tabular}
    \caption{Comparison of share of buyer-only and seller-only firms between Italy and other similar datasets.}
    \label{tab:buyer_seller_only_confronto}
\end{table}
Figure \ref{fig:seller_buyer_only_cat_breakdown} illustrates the distribution of these nodes by sector of economic activity. Notably, a substantial share corresponds to sector L - real estate activities. 
Understanding how link distributions vary when looking at these types of firms is also of interest.
Figure \ref{fig:sellers_buyers_only_degree_log_scale} illustrates that both of these distributions present power-law tails.

\begin{figure}[H]
\centering
  \begin{subfigure}[b]{0.49\textwidth}
    \centering
    \includegraphics[width=1\linewidth]{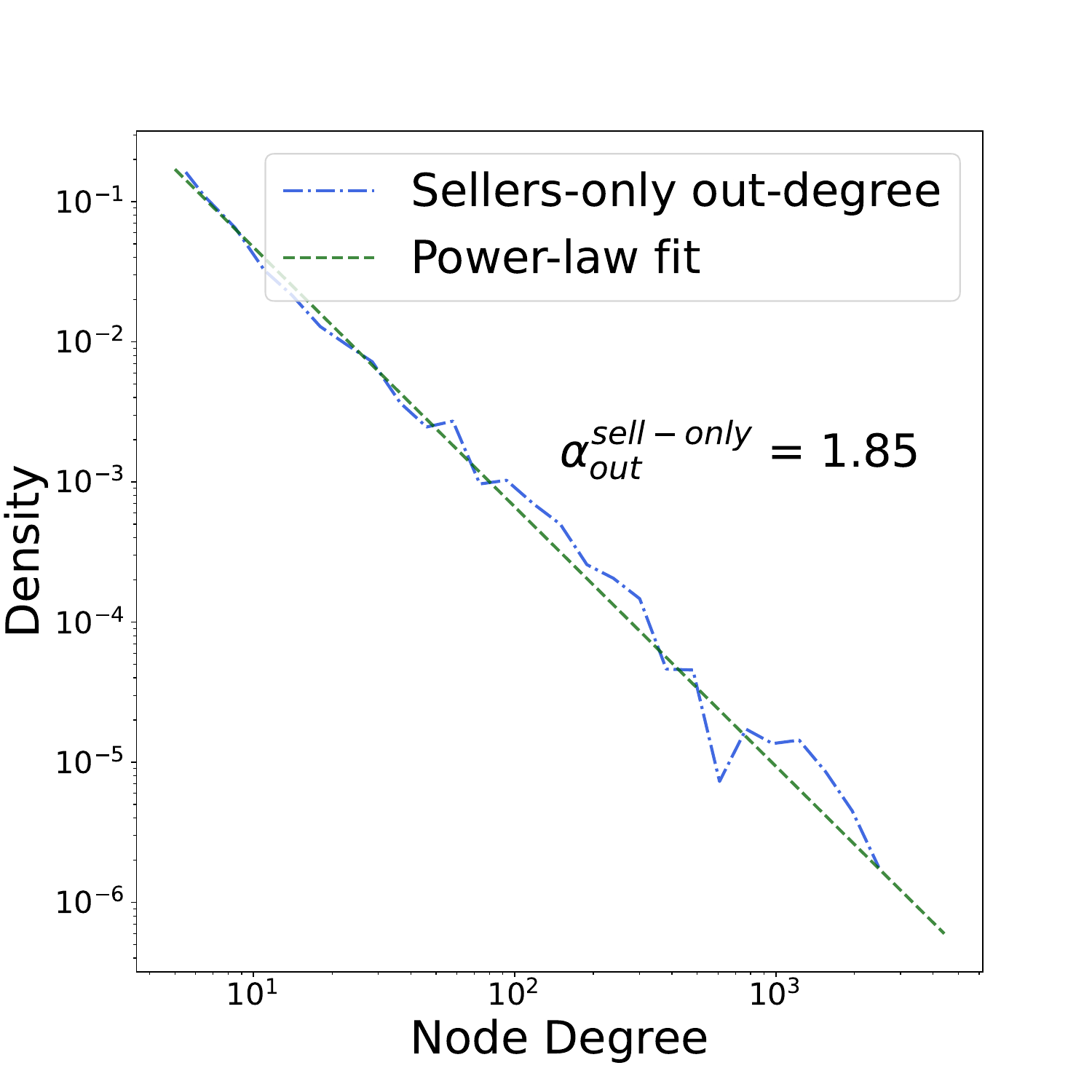}
    \caption{Seller-only out-degree}
    \label{fig:buyers_only_in_degree_log_scale_tail_fit}
    %\vspace{4ex}
  \end{subfigure}%% 
  \begin{subfigure}[b]{0.49\textwidth}
    \centering
    \includegraphics[width=1\linewidth]{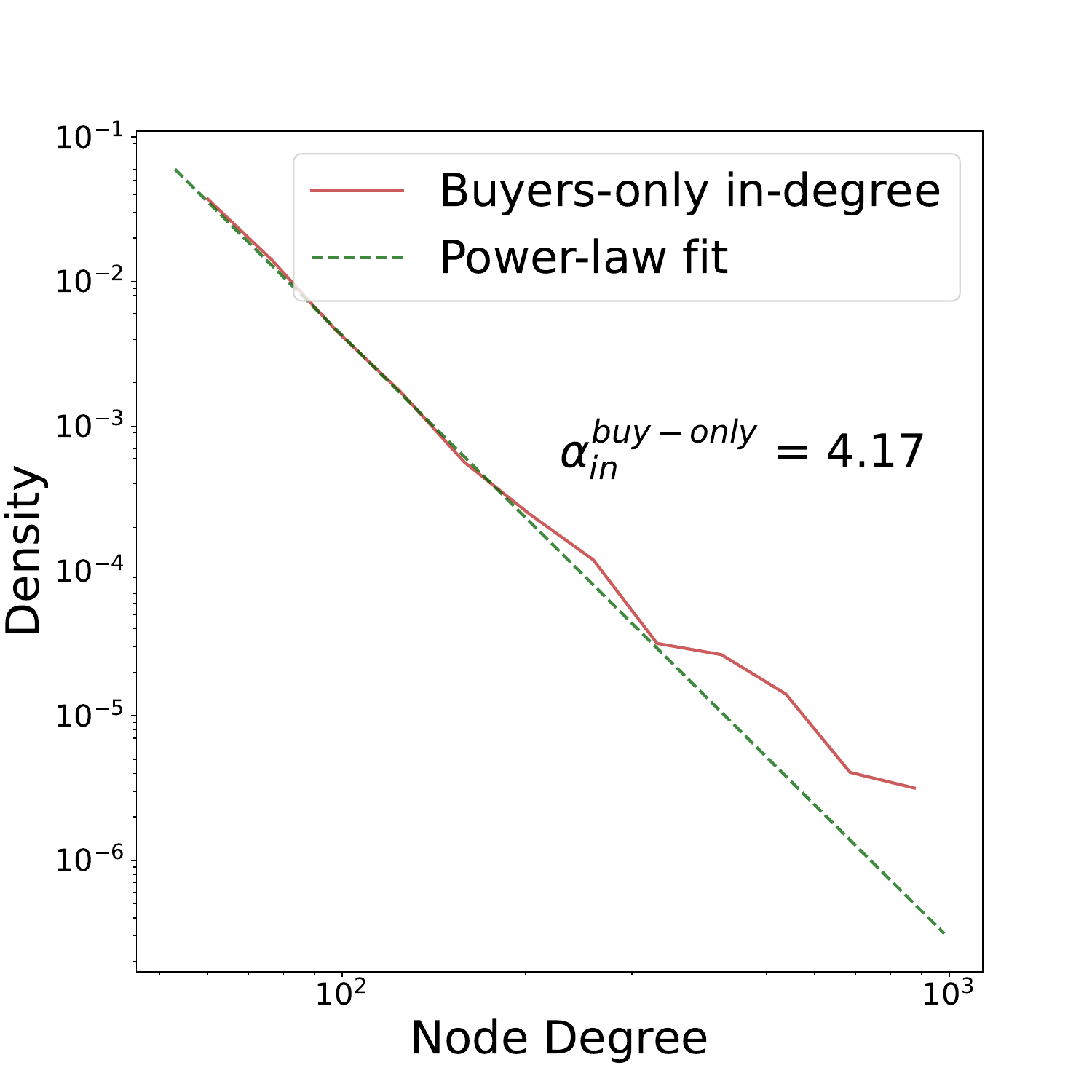}
    \caption{Buyer-only in-degree} 
    \label{fig:sellers_only_out_degree_log_scale_tail_fit} 
    %\vspace{4ex}
  \end{subfigure} 
  \caption{Tail exponents fit for the out-degree of seller-only firms (left hand side) and in-degree of buyer-only firms (right hand side) distributions. 
  The buyer-only in-degree distributions has the same properties of the in-degree distributions for the whole network. 
  The seller-only out-degree distributions, instead, shows greater variability than the out-degree distributions for the complete network. 
  The exponent is smaller than 2, implying an infinite mean for the distribution of degrees.}
  \label{fig:buyer_sellers_only_tail_fit} 
\end{figure}

When estimating the tail exponents (Figure \ref{fig:buyer_sellers_only_tail_fit}) we find that the in-degree distribution of buyer-only firms is not significantly different from that of the whole network. 
Both distributions have finite second moment (this one also has finite third moment, since the exponent is greater than 4 in absolute value), implying a moderate variability of the number of links. 
In contrast, the out-degree distribution of the seller-only firms shows an exponent that is not only lower than the one observed for nodes in the complete network but also falls below the value of 2. 
This implies an even greater variability in the number of links associated with this type of node, since a power-law distribution with exponent smaller than 2 has infinite mean.
This phenomenon is noteworthy and could deserve further investigation in future works, both in terms of its economic implications and underlying factors contributing to this distinctive structural property.

\clearpage
\section{Centrality Plots}
\label{subsec:appendix_plot_global_centrality}
\setcounter{figure}{0}  
\setcounter{table}{0}

\begin{figure}[H]
\centering
  \begin{subfigure}[b]{0.49\textwidth}
    \centering
    \includegraphics[width=1\linewidth]{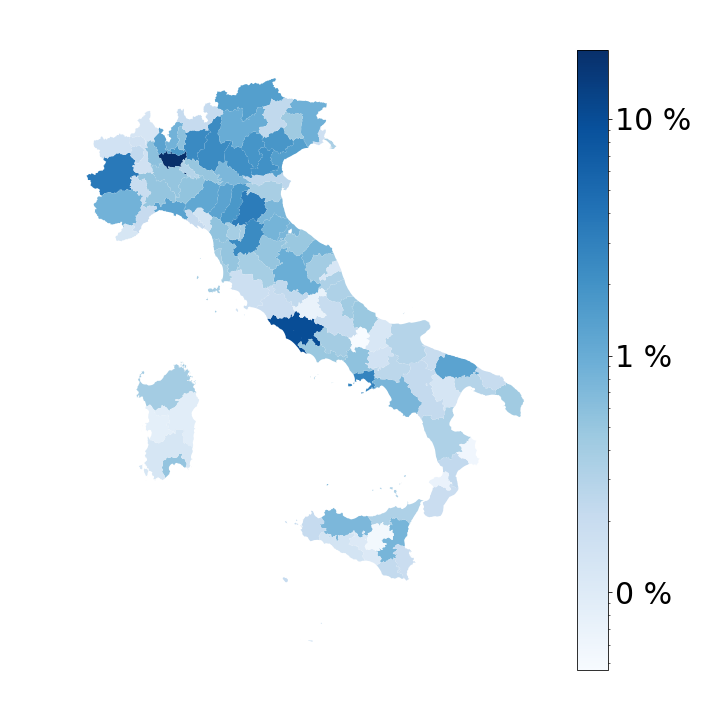}
    \caption{Eigenvector Centrality}
    \label{fig:eigenvector_downstream_provincie}
    %\vspace{4ex}
  \end{subfigure}%% 
  \begin{subfigure}[b]{0.49\textwidth}
    \centering
    \includegraphics[width=1\linewidth]{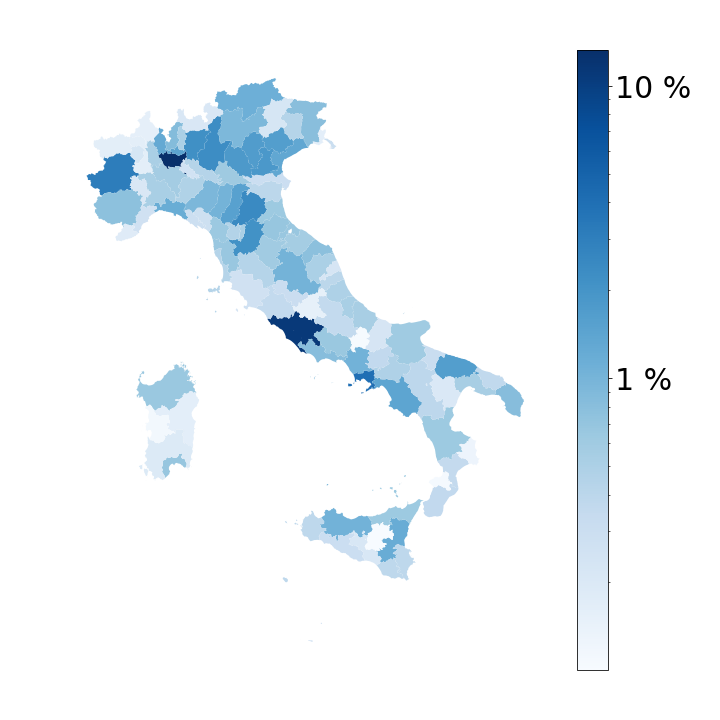}
    \caption{Katz Centrality}
    \label{fig:katz_downstream_provincie} 
    %\vspace{4ex}
  \end{subfigure} 
  \begin{subfigure}[b]{0.49\textwidth}
    \centering
    \includegraphics[width=1\linewidth]{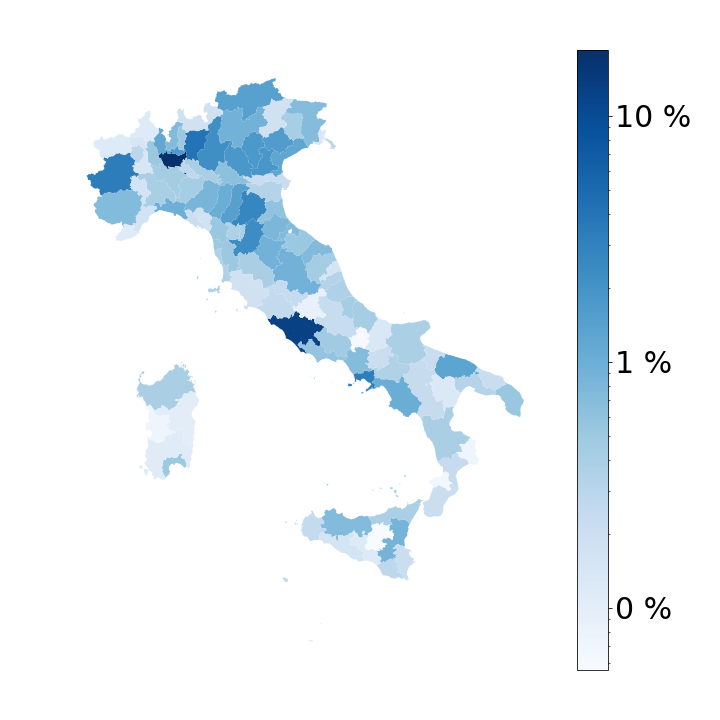}
    \caption{PageRank Centrality}
    \label{fig:pagerank_downstream_provincie} 
  \end{subfigure}%%
  \begin{subfigure}[b]{0.49\textwidth}
    \centering
    \includegraphics[width=1\linewidth]{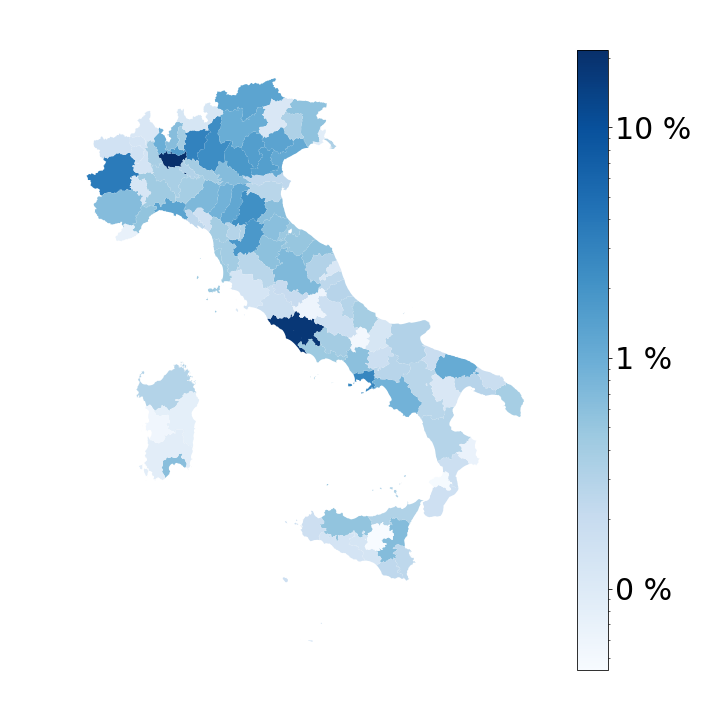}
    \caption{Weighted PageRank Centrality}
    \label{fig:pagerank_importi_downstream_provincie} 
  \end{subfigure} 
  \caption{Plots of the downstream global centrality by firms in the provinces. %{correlazione tra share di firms nelle provincie e share di centralità per provinicia così facciamo vedere che sono correlati? è una prova, potrebbe anche non portare a niente}
  }
  \label{fig:centrality_dowstream_provincie} 
\end{figure}

\begin{figure}[H]
\centering
  \begin{subfigure}[b]{0.49\textwidth}
    \centering
    \includegraphics[width=1\linewidth]{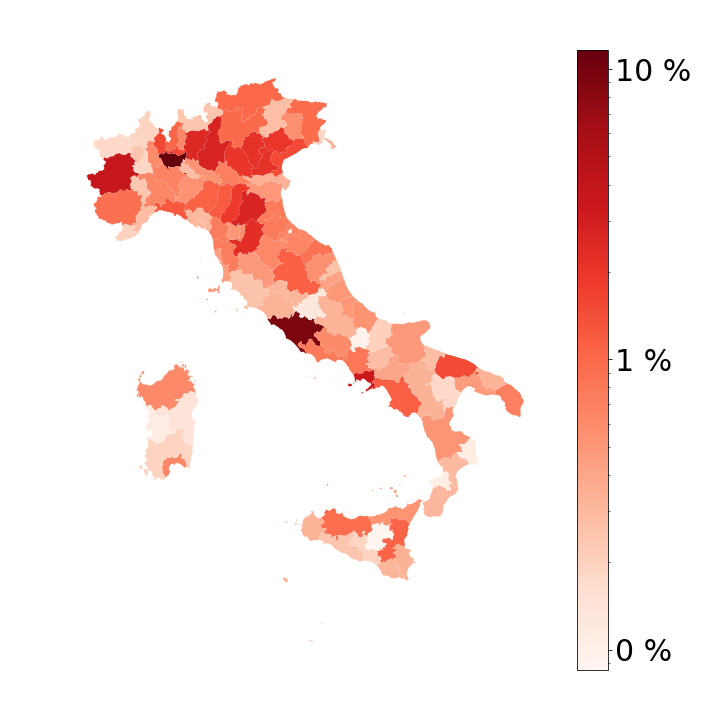}
    \caption{Eigenvector Centrality}
    \label{fig:eigenvector_upstream_provincie}
    %\vspace{4ex}
  \end{subfigure}%% 
  \begin{subfigure}[b]{0.49\textwidth}
    \centering
    \includegraphics[width=1\linewidth]{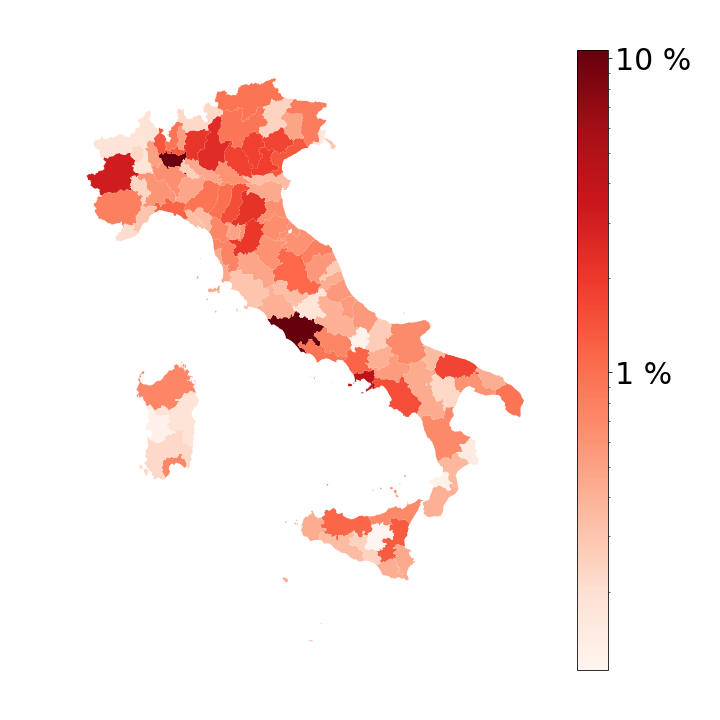}
    \caption{Katz Centrality}
    \label{fig:katz_upstream_provincie} 
    %\vspace{4ex}
  \end{subfigure} 
  \begin{subfigure}[b]{0.49\textwidth}
    \centering
    \includegraphics[width=1\linewidth]{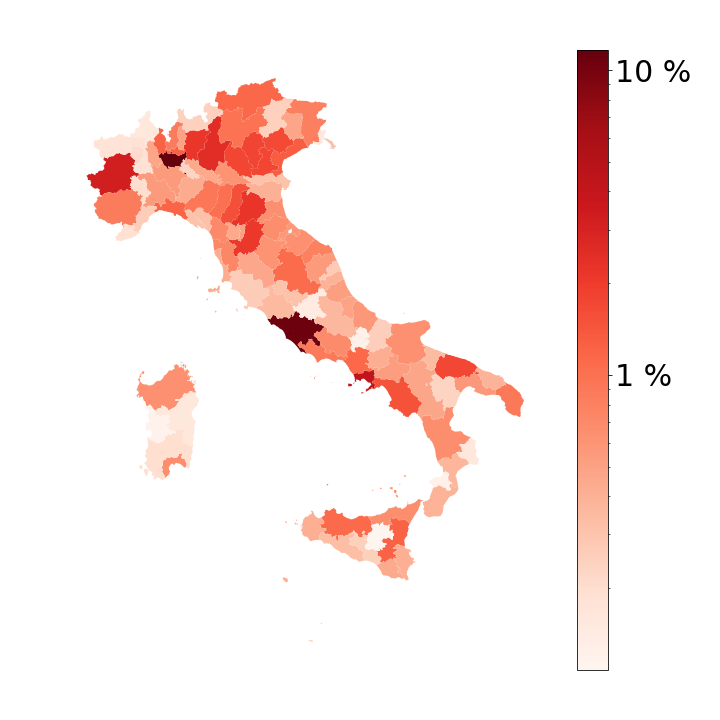}
    \caption{PageRank Centrality}
    \label{fig:pagerank_upstream_provincie} 
  \end{subfigure}%%
  \begin{subfigure}[b]{0.49\textwidth}
    \centering
    \includegraphics[width=1\linewidth]{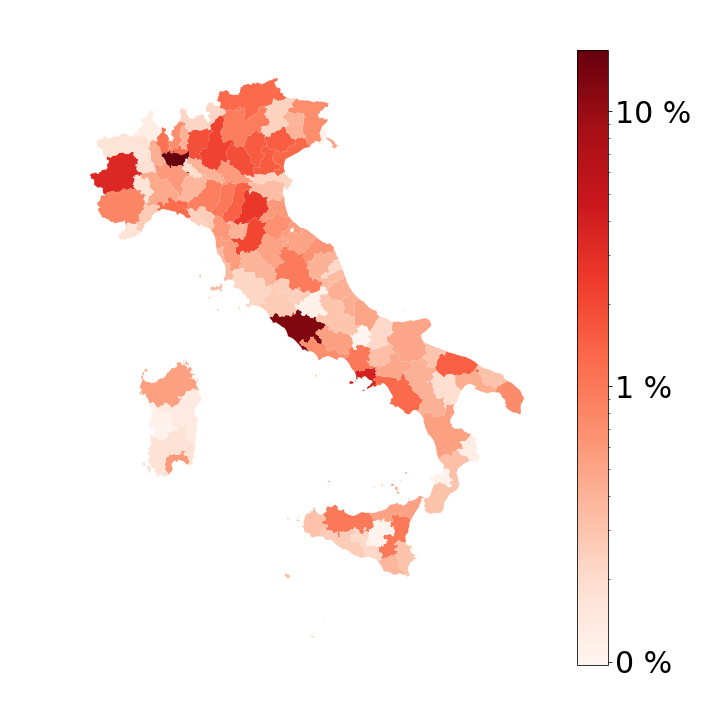}
    \caption{Weighted PageRank Centrality}
    \label{fig:pagerank_importi_upstream_provincie} 
  \end{subfigure} 
  \caption{Plots of the upstream global centrality by firms in the provinces.}
  \label{fig:centrality_upstream_provincie} 
\end{figure}

\begin{figure}[H]
\centering
  \begin{subfigure}[b]{0.49\textwidth}
    \centering
    \includegraphics[width=1\linewidth]{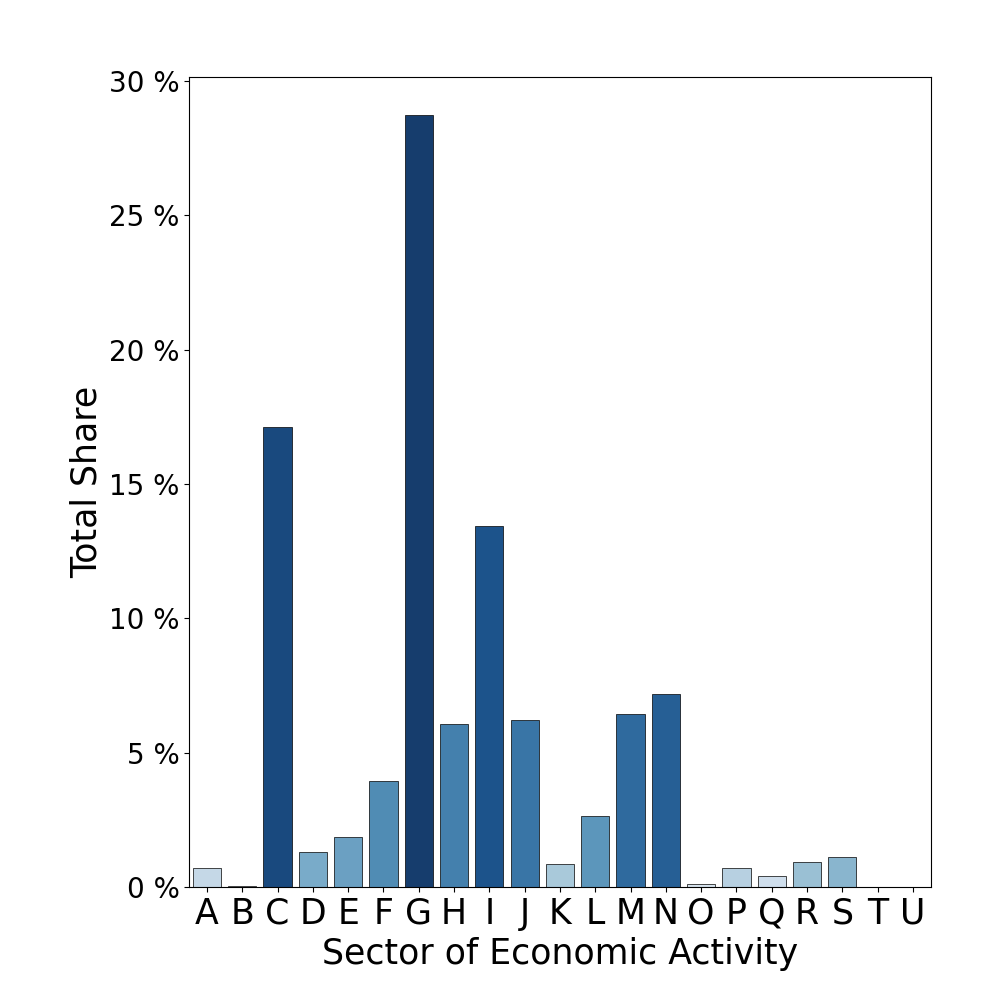}
    \caption{Eigenvector Centrality}
    \label{fig:eigenvector_downstream_category}
    %\vspace{4ex}
  \end{subfigure}%% 
  \begin{subfigure}[b]{0.49\textwidth}
    \centering
    \includegraphics[width=1\linewidth]{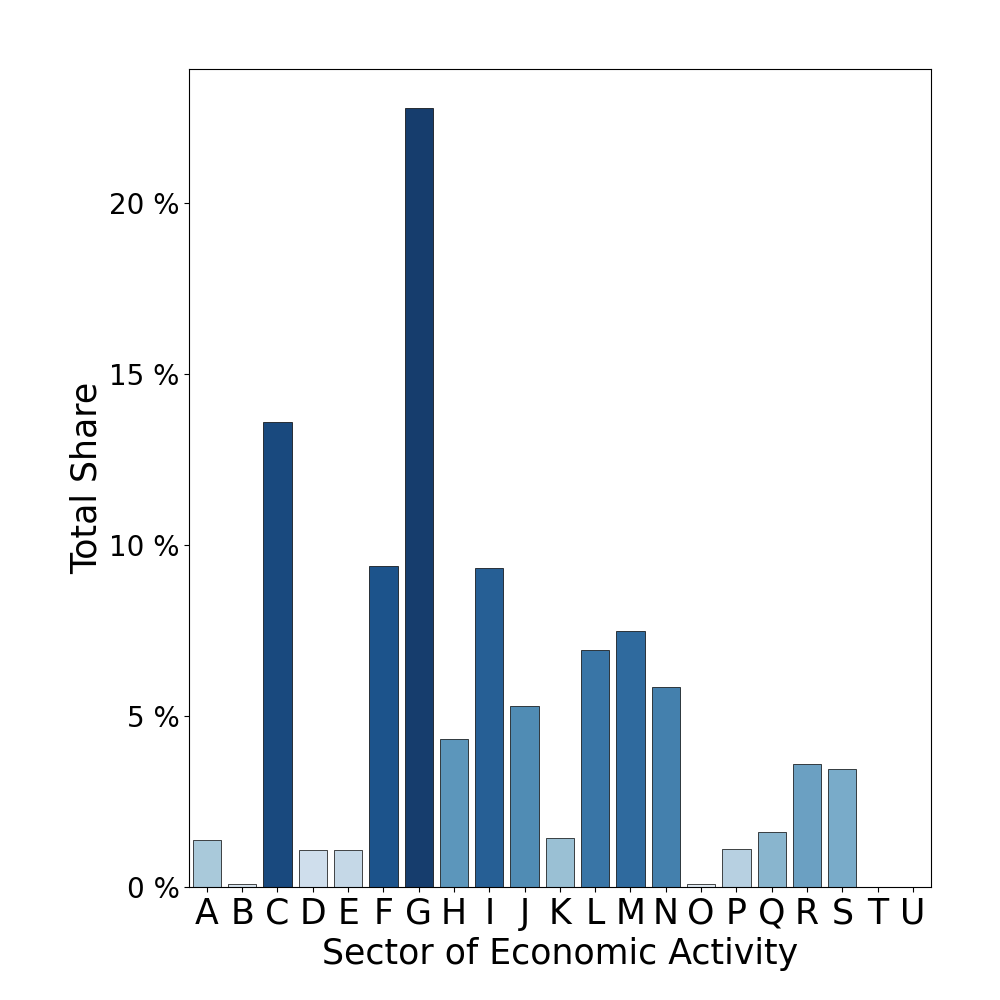}
    \caption{Katz Centrality}
    \label{fig:katz_downstream_category} 
    %\vspace{4ex}
  \end{subfigure} 
  \begin{subfigure}[b]{0.49\textwidth}
    \centering
    \includegraphics[width=1\linewidth]{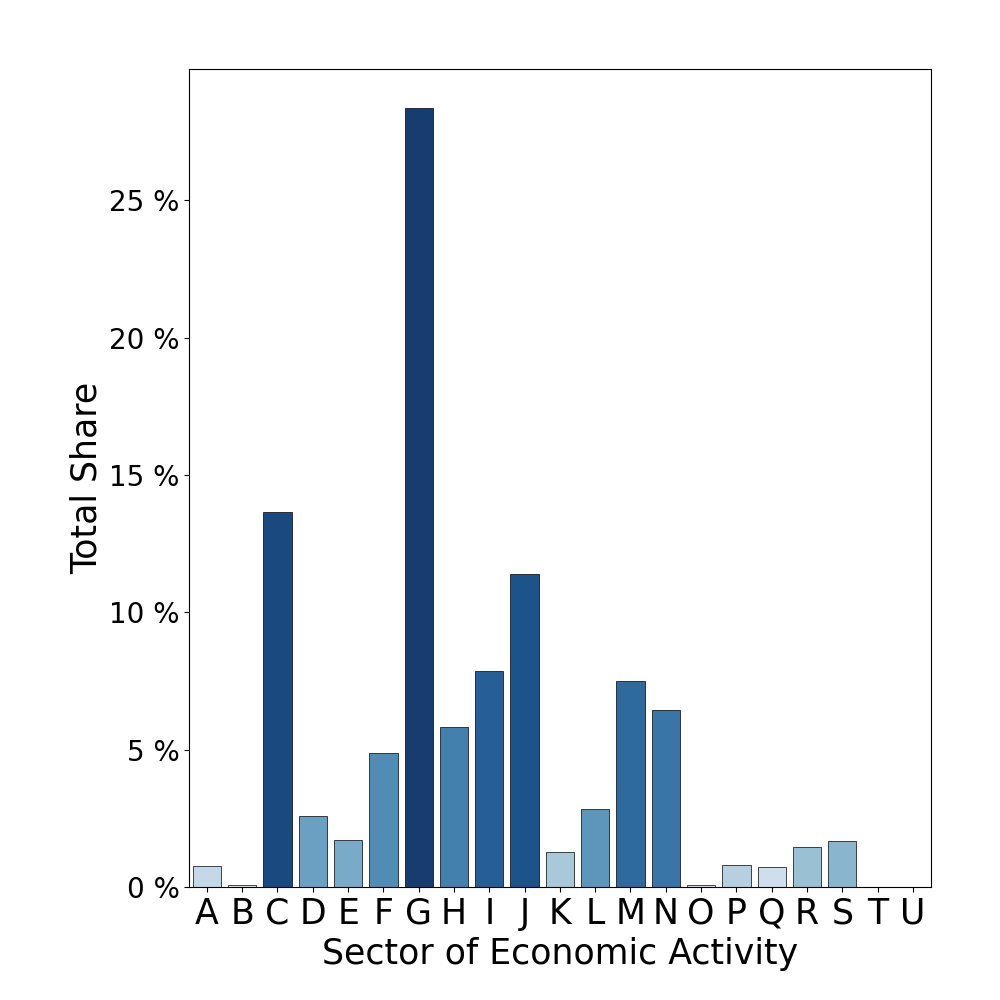}
    \caption{PageRank Centrality}
    \label{fig:pagerank_downstream_category} 
  \end{subfigure}%%
  \begin{subfigure}[b]{0.49\textwidth}
    \centering
    \includegraphics[width=1\linewidth]{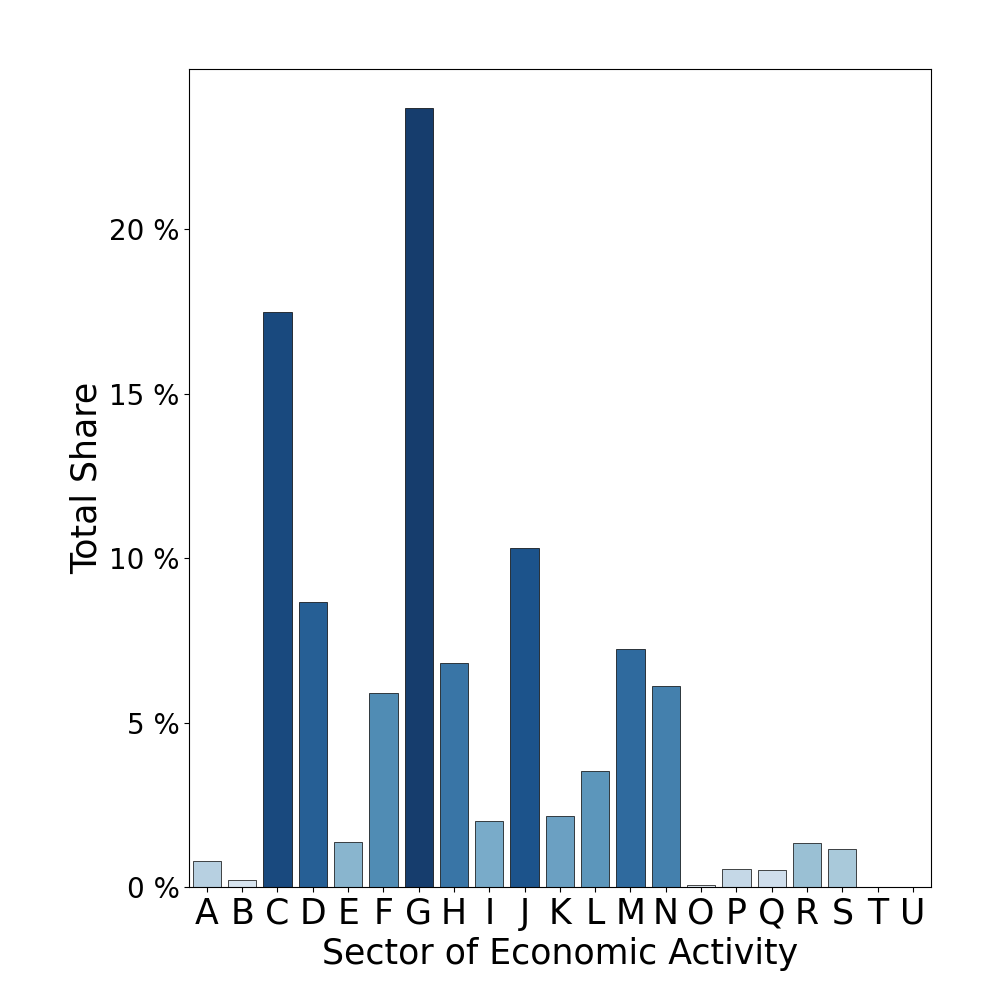}
    \caption{Weighted PageRank Centrality}
    \label{fig:pagerank_importi_downstream_category} 
  \end{subfigure} 
  \caption{Plots of the downstream global centrality by firms in category of economic activity.}
  \label{fig:centrality_downstream_provincie} 
\end{figure}

\begin{figure}[H]
\centering
  \begin{subfigure}[b]{0.49\textwidth}
    \centering
    \includegraphics[width=1\linewidth]{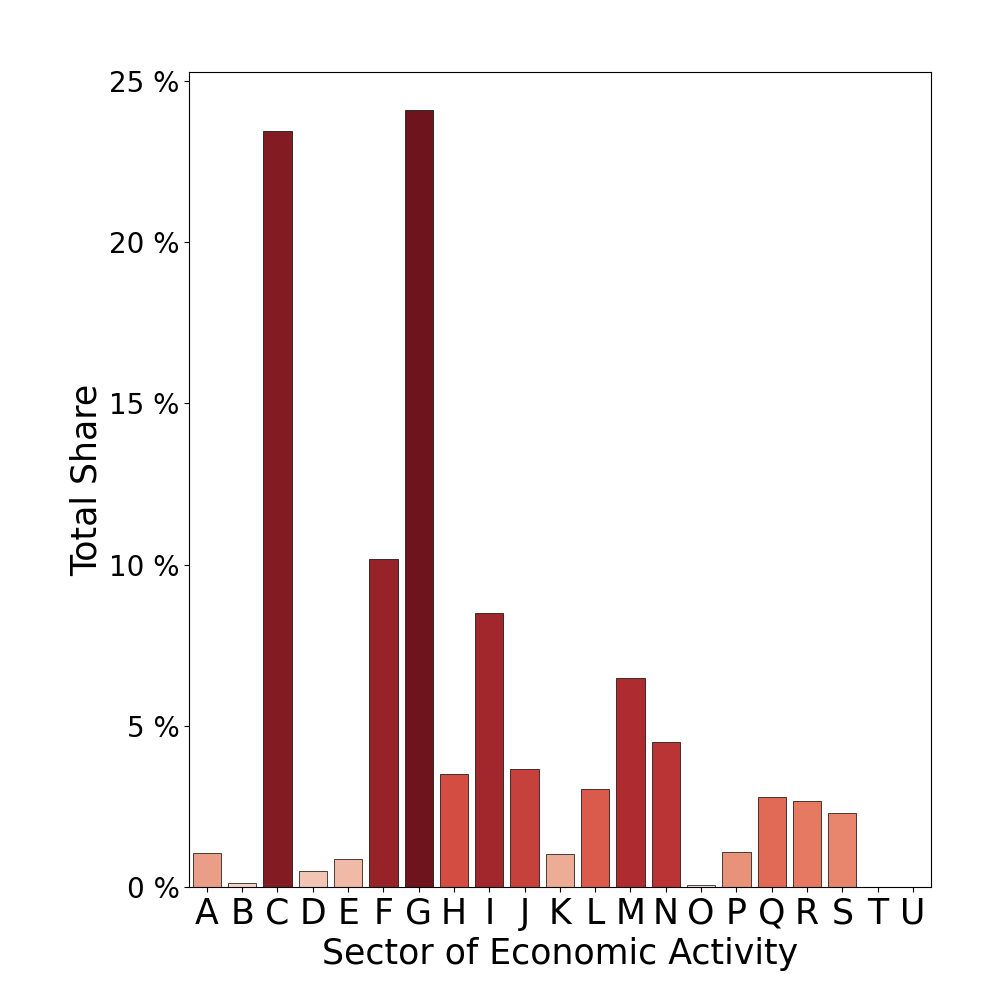}
    \caption{Eigenvector Centrality}
    \label{fig:eigenvector_upstream_category}
    %\vspace{4ex}
  \end{subfigure}%% 
  \begin{subfigure}[b]{0.49\textwidth}
    \centering
    \includegraphics[width=1\linewidth]{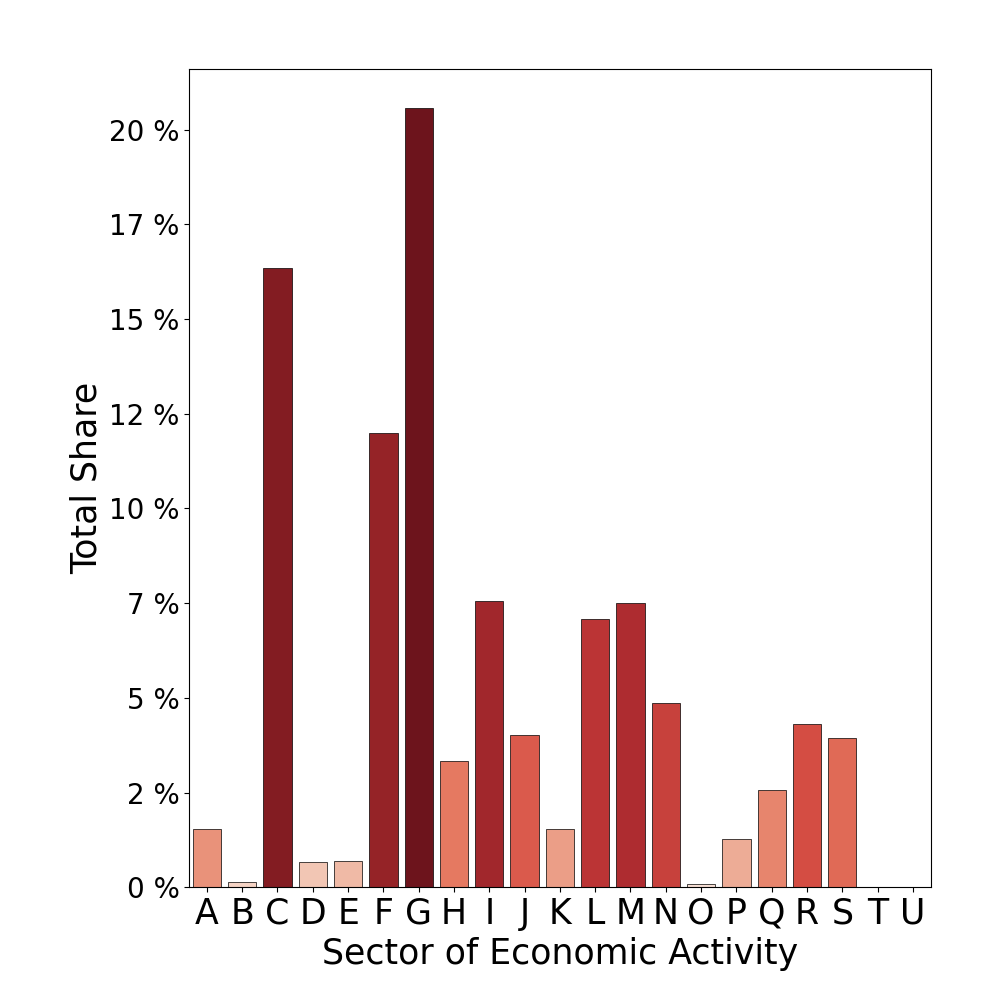}
    \caption{Katz Centrality}
    \label{fig:katz_upstream_category} 
    %\vspace{4ex}
  \end{subfigure} 
  \begin{subfigure}[b]{0.49\textwidth}
    \centering
    \includegraphics[width=1\linewidth]{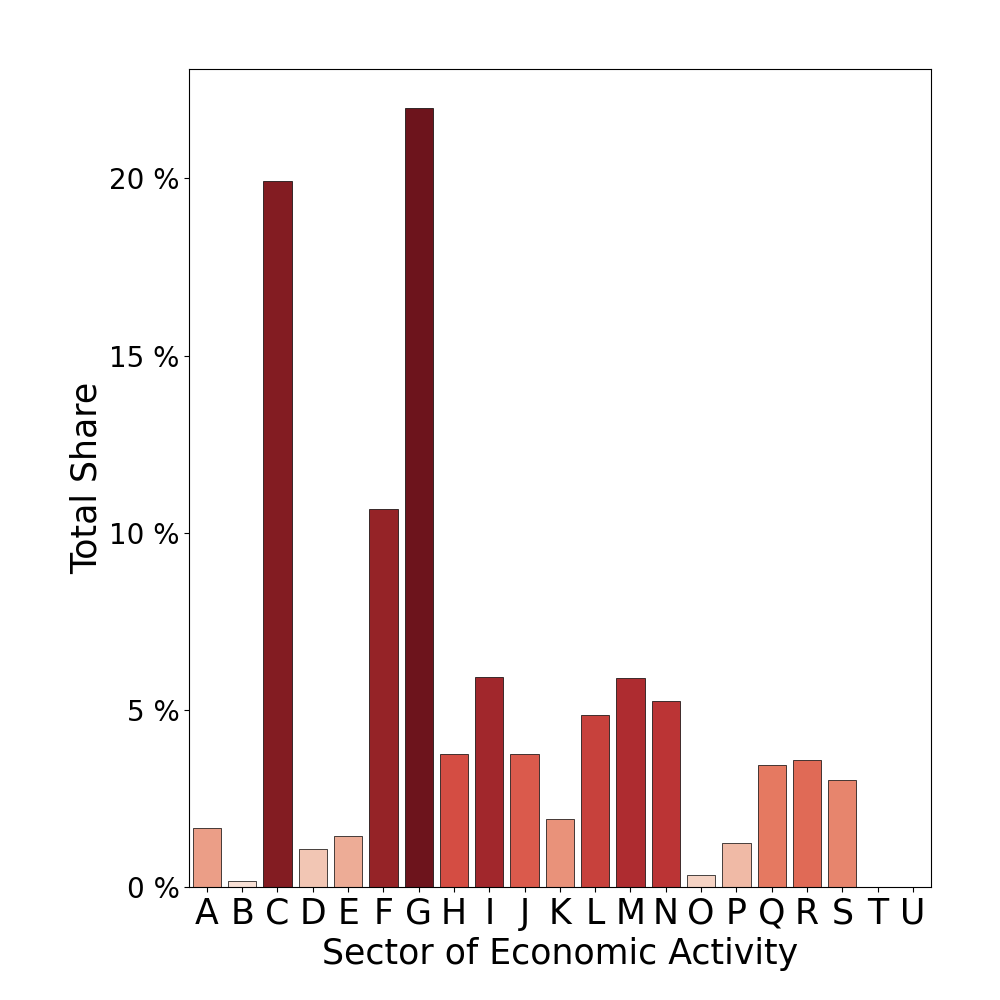}
    \caption{PageRank Centrality}
    \label{fig:pagerank_upstream_category} 
  \end{subfigure}%%
  \begin{subfigure}[b]{0.49\textwidth}
    \centering
    \includegraphics[width=1\linewidth]{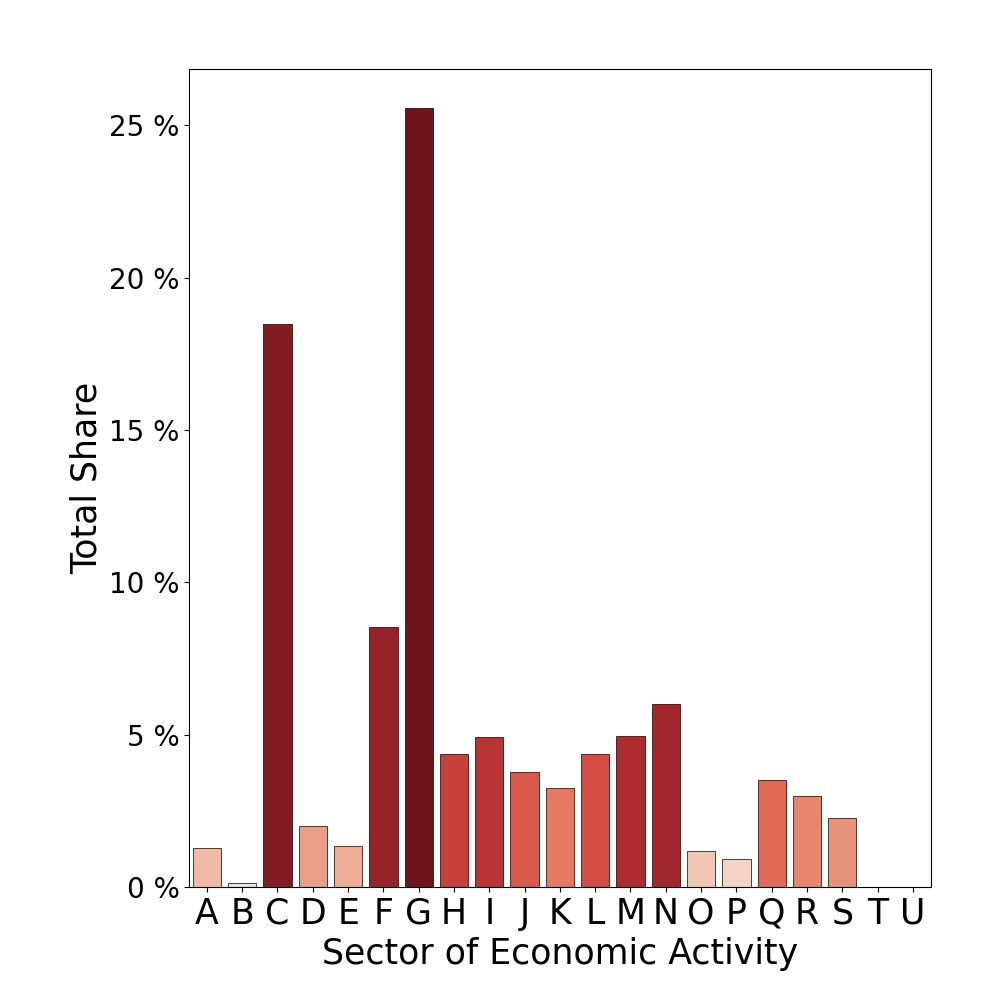}
    \caption{Weighted PageRank Centrality}
    \label{fig:pagerank_importi_upstream_category} 
  \end{subfigure}   
  \caption{Plots of the upstream global centrality by firms in sector of economic activity.}
  \label{fig:centrality_upstream_categorie} 
\end{figure}

\subsection{Global Centrality Plots by 2-digit Sector of Economic Activity}
\label{subsec:appendix_plot_centrality}

\begin{figure}[H]
\centering
  \begin{subfigure}[t]{\textwidth}
    \centering
    \includegraphics[width=0.99\linewidth]{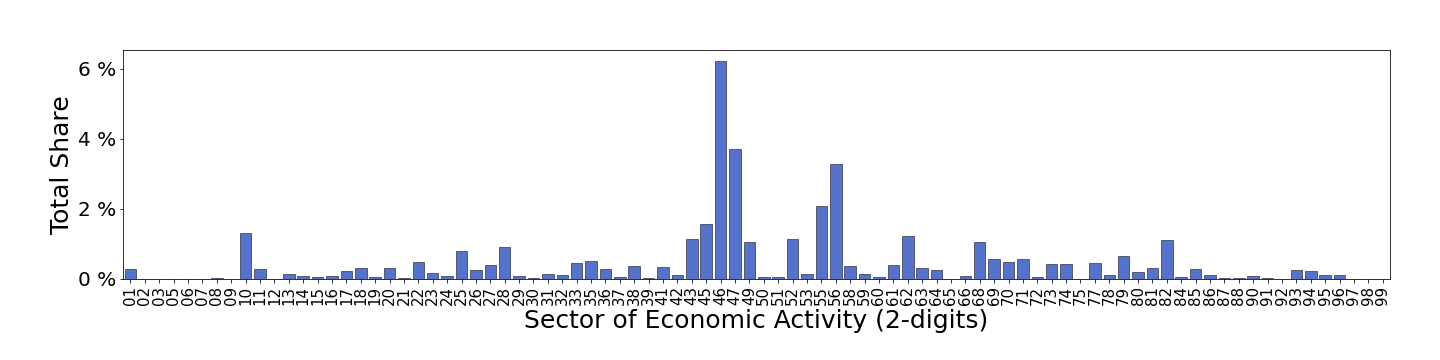}
    \caption{Eigenvector Centrality} 
    \label{fig:eigenvector_downstream_ateco2}
    %\vspace{4ex}
  \end{subfigure}
  \begin{subfigure}[t]{\textwidth}
    \centering
    \includegraphics[width=0.99\linewidth]{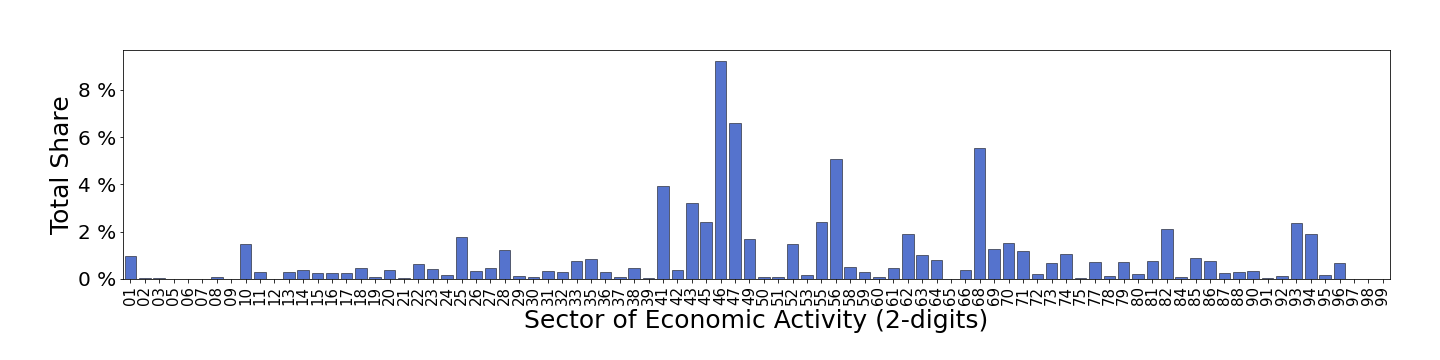}
    \caption{Katz Centrality} 
    \label{fig:katz_downstream_ateco2} 
    %\vspace{4ex}
  \end{subfigure}
  \begin{subfigure}[t]{\textwidth}
    \centering
    \includegraphics[width=0.99\linewidth]{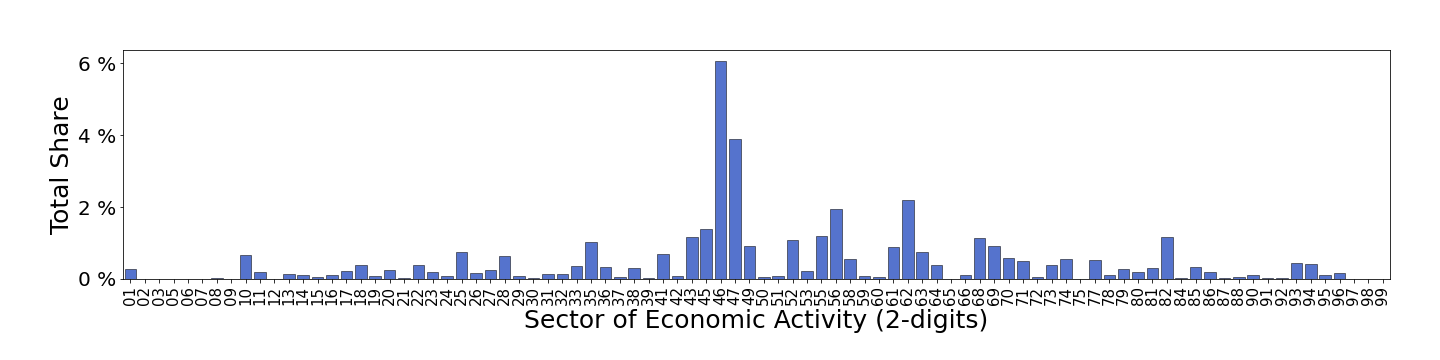}
    \caption{PageRank Centrality} 
    \label{fig:pagerank_downstream_ateco2} 
  \end{subfigure}
  \begin{subfigure}[t]{\textwidth}
    \centering
    \includegraphics[width=0.99\linewidth]{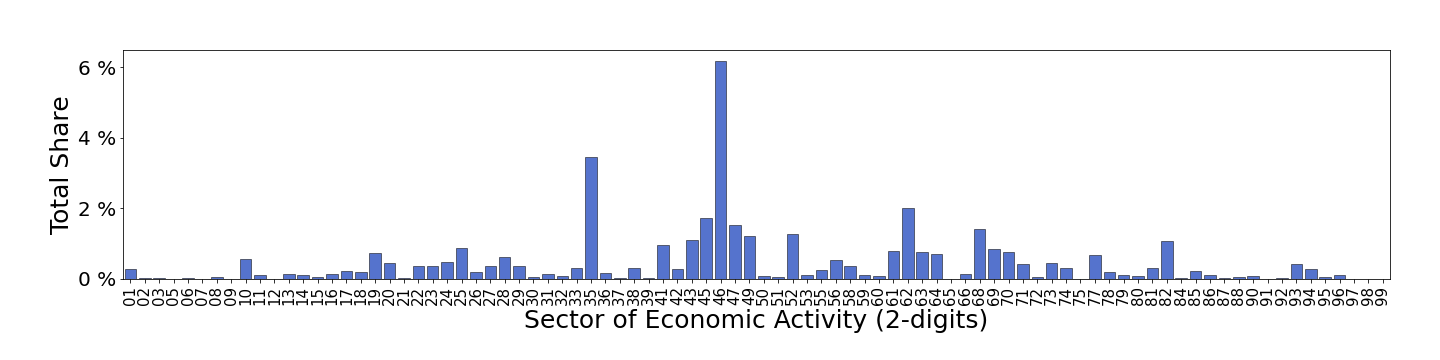}
    \caption{Weighted PageRank Centrality}
    \label{fig:pagerank_importi_downstream_ateco2} 
  \end{subfigure}
  \caption{Plots of the downstream global centrality aggregated by in 2-digit sector of economic activity.}
  \label{fig:centrality_downstream_ateco2} 
\end{figure}

\begin{figure}[H]
\centering
  \begin{subfigure}[t]{\textwidth}
    \centering
    \includegraphics[width=0.99\linewidth]{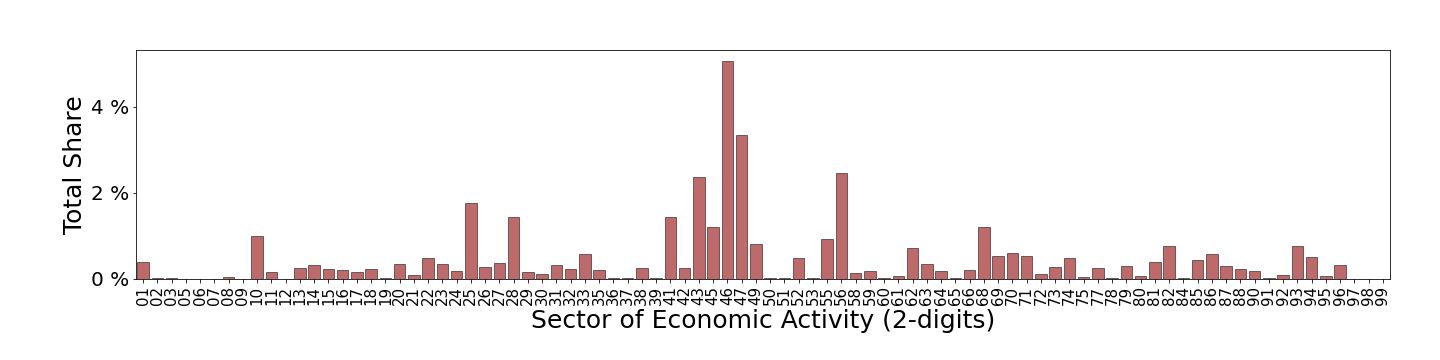}
    \caption{Eigenvector Centrality} 
    \label{fig:eigenvector_upstream_ateco2}
    %\vspace{4ex}
  \end{subfigure}
  \begin{subfigure}[t]{\textwidth}
    \centering
    \includegraphics[width=0.99\linewidth]{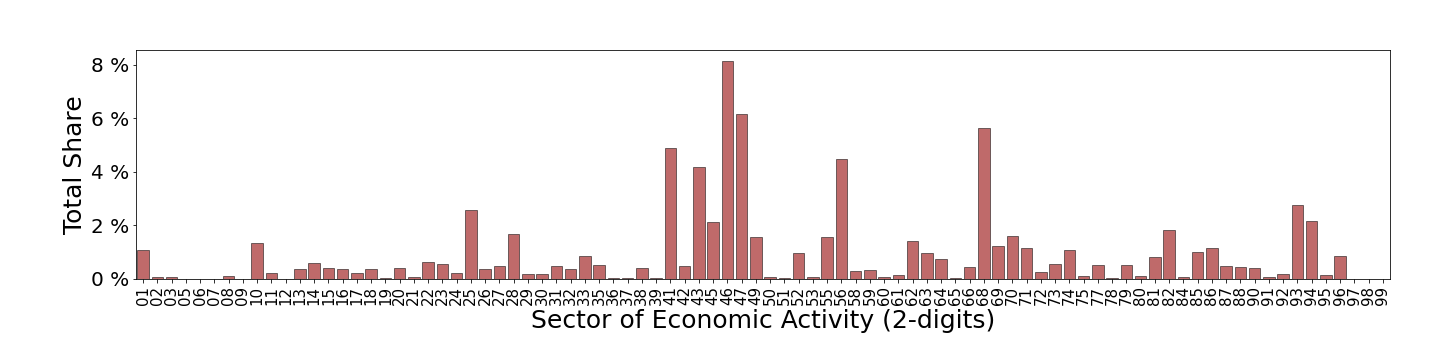}
    \caption{Katz Centrality} 
    \label{fig:katz_upstream_ateco2} 
    %\vspace{4ex}
  \end{subfigure}
  \begin{subfigure}[t]{\textwidth}
    \centering
    \includegraphics[width=0.99\linewidth]{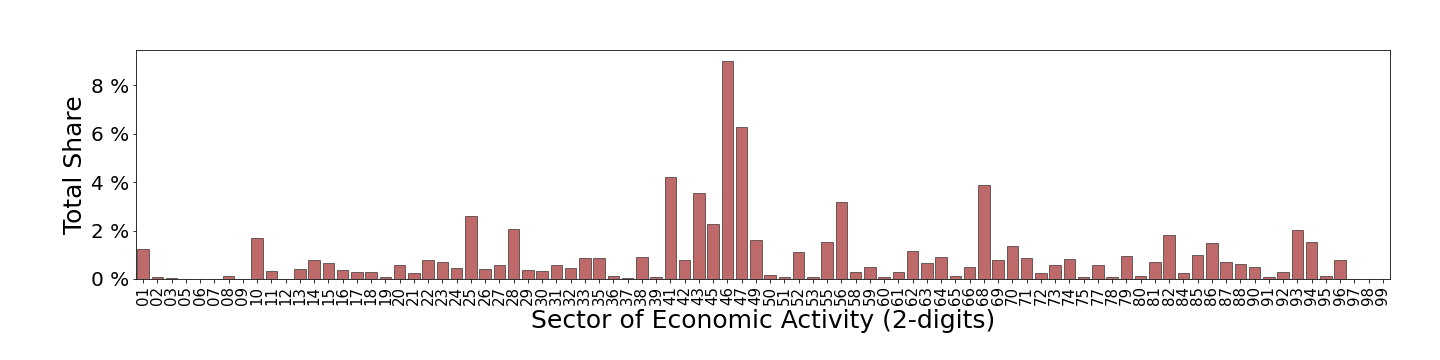}
    \caption{PageRank Centrality} 
    \label{fig:pagerank_upstream_ateco2} 
  \end{subfigure}
  \begin{subfigure}[t]{\textwidth}
    \centering
    \includegraphics[width=0.99\linewidth]{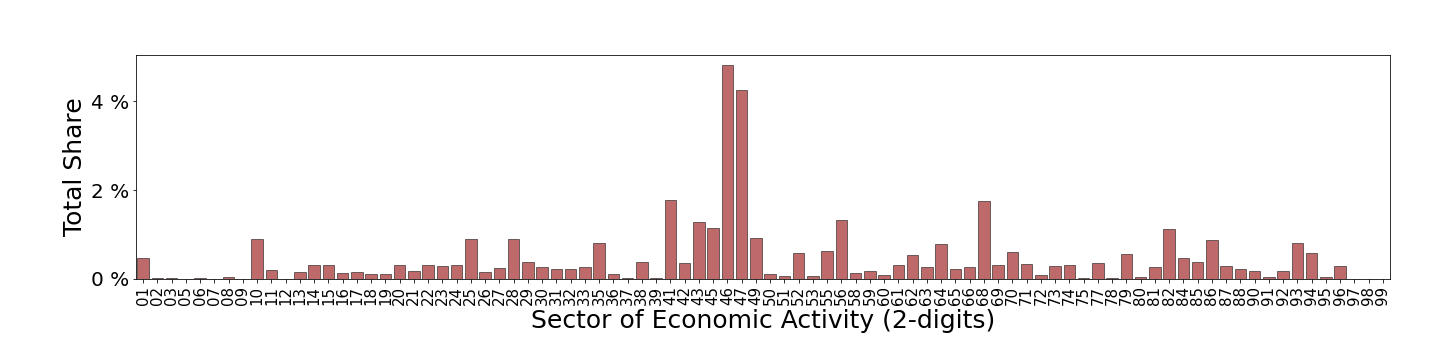}
    \caption{Weighted PageRank Centrality}
    \label{fig:pagerank_importi_upstream_ateco2} 
  \end{subfigure}
  \caption{Plots of the upstream global centrality aggregated by 2-digit sector of economic activity.}
  \label{fig:centrality_upstream_ateco} 
\end{figure}

\subsubsection{Degree Centrality Plots by 2-digit Sector of Economic Activity}

\begin{figure}[H]
\centering
  \begin{subfigure}[t]{\textwidth}
    \centering
    \includegraphics[width=0.99\linewidth]{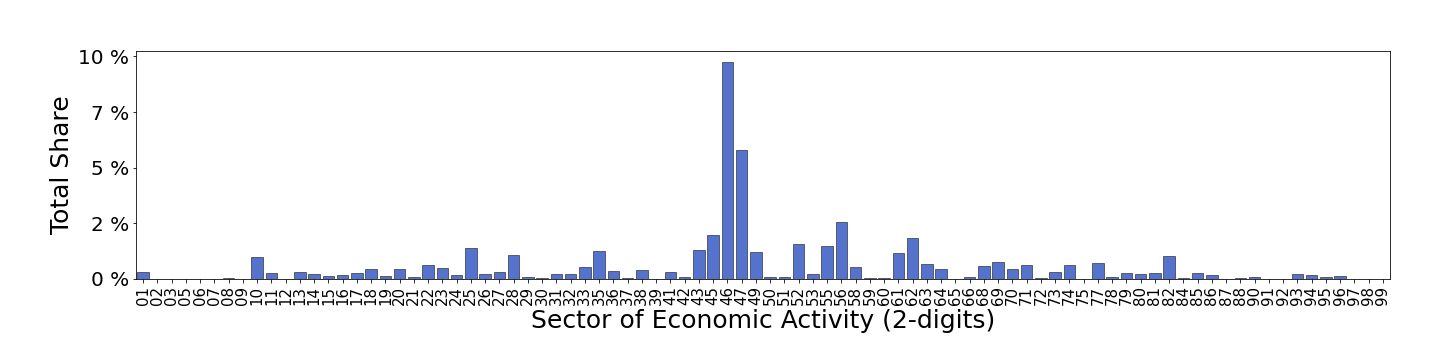}
    \caption{Downstream Local Centrality (out-Degree)} 
    \label{fig:grado_out_downstream_ateco2}
    %\vspace{4ex}
  \end{subfigure}
  \begin{subfigure}[t]{\textwidth}
    \centering
    \includegraphics[width=0.99\linewidth]{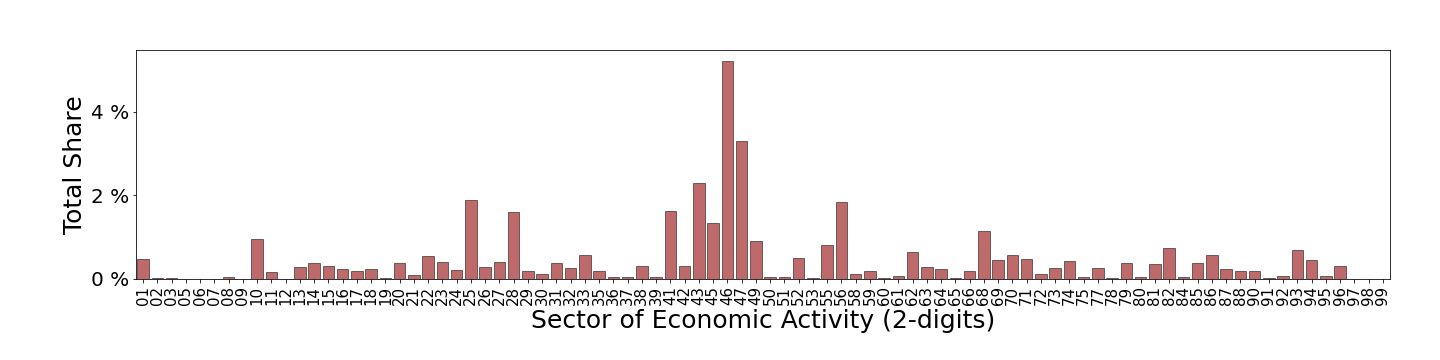}
    \caption{Upstream Local Centrality (in-Degree)}  
    \label{fig:grado_in_upstream_ateco2} 
    %\vspace{4ex}
  \end{subfigure}
  \caption{Degree centrality by 2-digit sector of economic activity.}
  \label{fig:degree_centrality_ateco2} 
\end{figure}

% \subsection{Proofs}
% \label{subsec:appendix_proofs}
% The average degree with link threshold $f$ is:
% \begin{equation}
%     \rho(f) = \frac{1}{N(f)} \sum_{i=1}^{N(f)} k_i(f)
% \end{equation}
% where $N(f)$ is the number of nodes containing at least one link above the threshold, and $k_i(f)$ is the number of links above threshold for node $i$.
% Now suppose we decrease the threshold to $f - \delta f$, with $\delta f$ small enough that only $1$ new node appears.
% The average degree becomes:
% \begin{align}
%     \rho(f-\delta f) &= \frac{1}{N(f)+1} \sum_{i=1}^{N(f)+1} k_i(f - \delta f) \\
%     &= \frac{1}{N(f)+1} \sum_{i=1}^{N(f)} \left[ k_i(f) + \delta k_i(f) \right] + \frac{k_{N(f)+1}(f-\delta f)}{N(f)+1} \\
%     &= \left(\frac{N(f)}{N(f)+1} \right)\rho(f) + \frac{1}{N(f)+1} \sum_{i=1}^{N(f)} \delta k_i(f) + \frac{k_{N(f)+1}(f-\delta f)}{N(f)+1} \\
% \end{align}
% So the difference in average degree, to first order in $\frac{1}{N(f)}$, is:
% \begin{equation}
%     \delta \rho(f) = \frac{1}{N(f)}\left[ - \rho(f) + \sum_{i=1}^{N(f)} \delta k_i(f) + k_{N(f)+1}(f-\delta f) \right]
% \end{equation}
% While the first term is of order $1$ (it does not grow with $N(f)$), the second term will typically be of order $N(f)$, hence the average degree typically increases when the threshold $f$ is lowered.
\clearpage
\section{Additional Figures and Tables}
\label{app:additional_figures}
% \setcounter{figure}{0}  
% \setcounter{table}{0}
% % \begin{table}[H]
% %     \centering
% %     \small
% %     \begin{tabular}{lccc}
% %     %\toprule
% %      & Complete Dataset & B $\rightarrow$ B & \% \\
% %     \midrule
% %     total gross amount (euro mld) &  2.705,61 & 1.689,36 &  62,44 \% \\
% %     number of entities (mln)      & 1,74      & 1,64     &  94,18 \% \\
% %     number of rows (mln)          & 163,15    & 59,07     &  36,21 \% \\
% %     \bottomrule
% %     \end{tabular}
% %     \caption{Comparison of the complete dataset with the business-to-business subset of transactions.}
% %     \label{tab:stats_dataset}
% % \end{table}
% \vspace{-0.7cm}
% \begin{figure}[H]
% \centering
% \begin{minipage}{.5\textwidth}
%   \centering
%   \begin{center}
%    \begin{tabular}{ccc}
%     \toprule 
%     Seller ID & Buyer ID & Amount\\
%     \midrule
%     A001 & B002 & 1,936.27\\
%     \bottomrule
%     \end{tabular} 
% \end{center}
% \end{minipage}%
% \begin{minipage}{.5\textwidth}
%   \centering
%   \includegraphics[width=.99\linewidth]{plots/from_data_2_network.pdf}
% \end{minipage}
% \caption{Record of the dataset (left hand side) as seen as part of the network (right hand side). The seller firm is the starting point of the link and the buyer firm is the ending point. The relation is not symmetrical, so the link is an arrow from the seller to the buyer.}
%   \label{fig:fromDataToNetwork}
% \end{figure}

 \begin{figure}[H]
    \centering
    \includegraphics[width=0.75\linewidth]{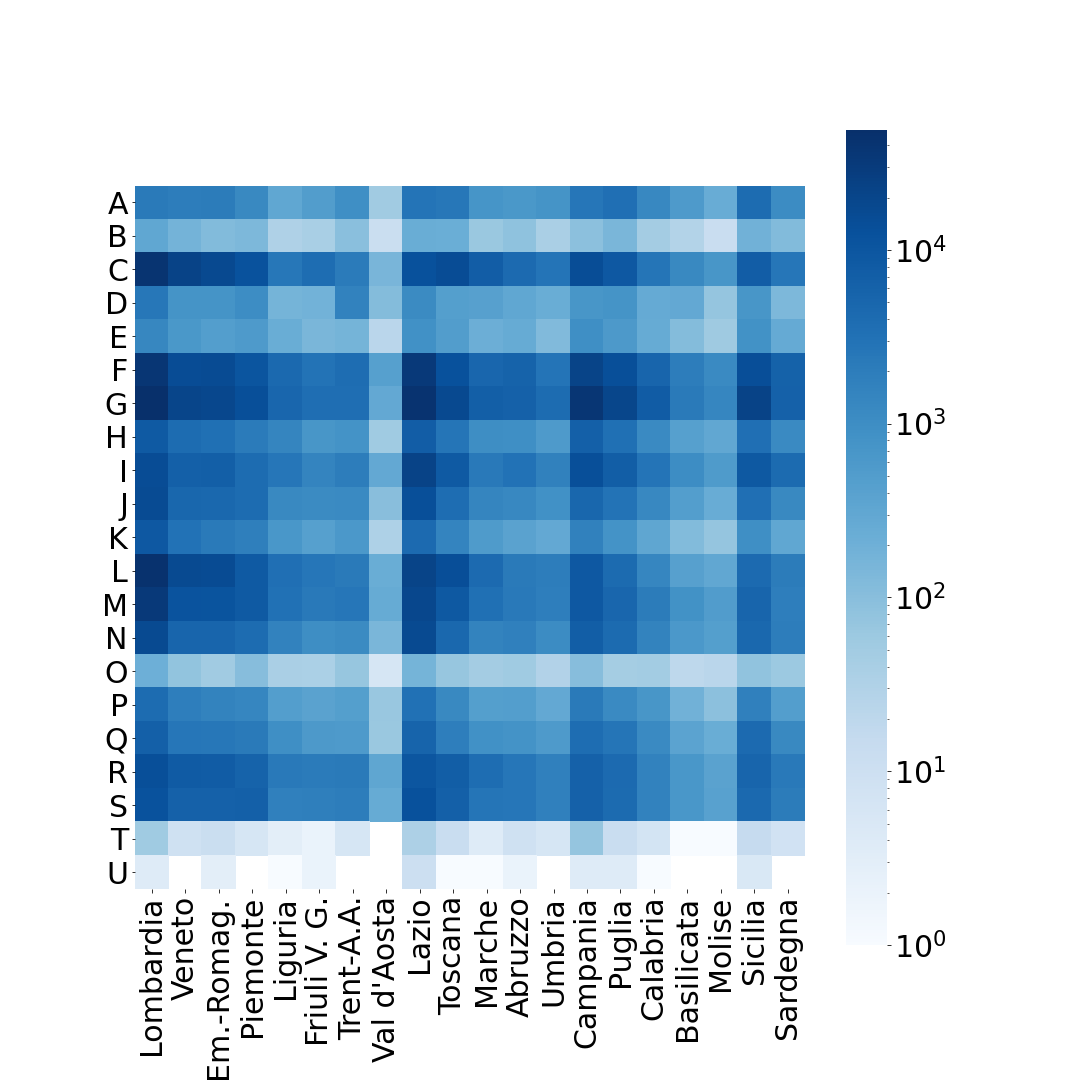}
    \caption{Distribution of Italian firms by intersection of regional division and sector of economic activity. Sectors G, F and C consistently remain the most represented also when considering the region where the registered office is located.}
    \label{fig:heatmap_category_regioni_power}
\end{figure}

\begin{figure}[H]
\centering
\begin{minipage}{.5\textwidth}
  \centering
  \includegraphics[width=1\linewidth]{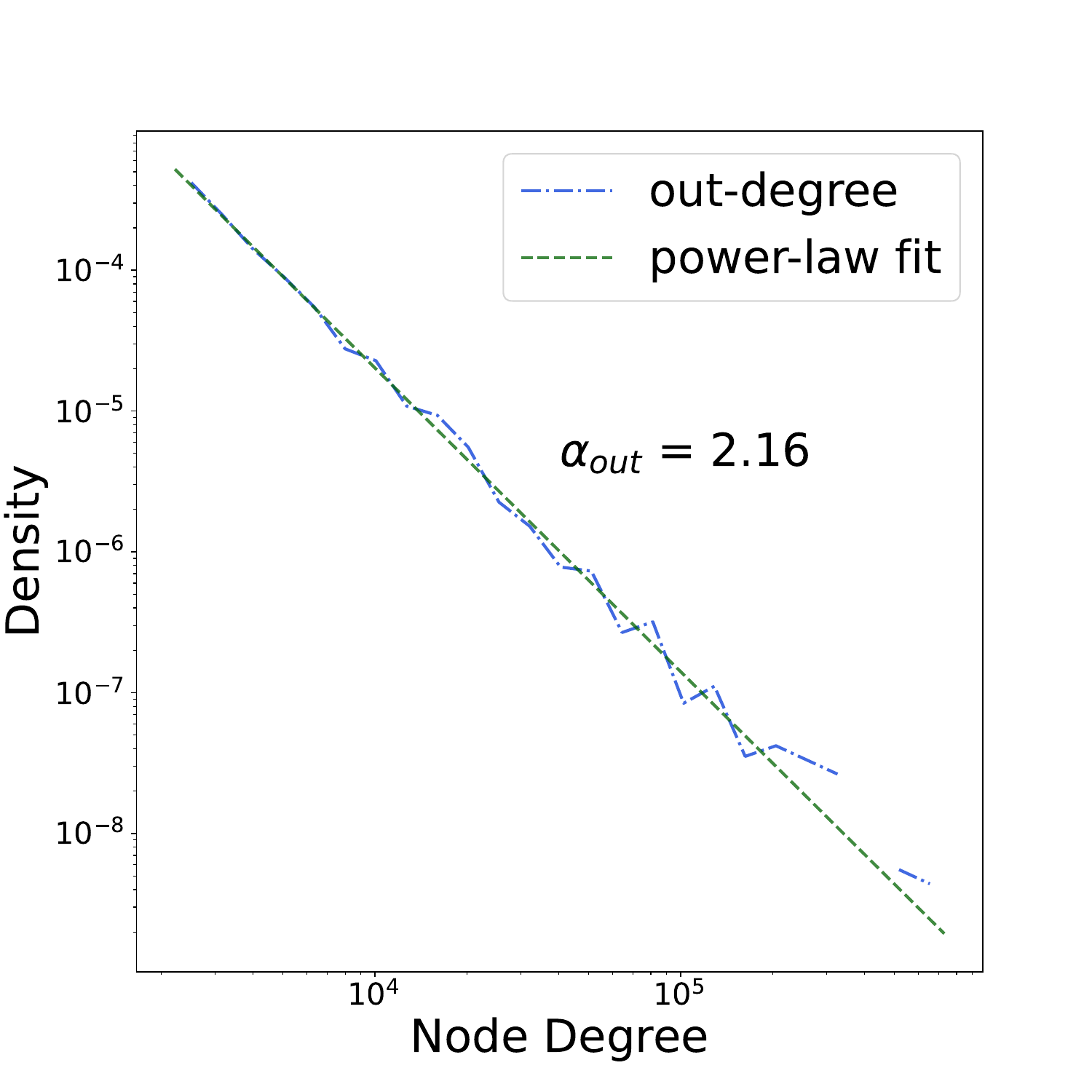}
\end{minipage}%
\begin{minipage}{.5\textwidth}
  \centering
  \includegraphics[width=1\linewidth]{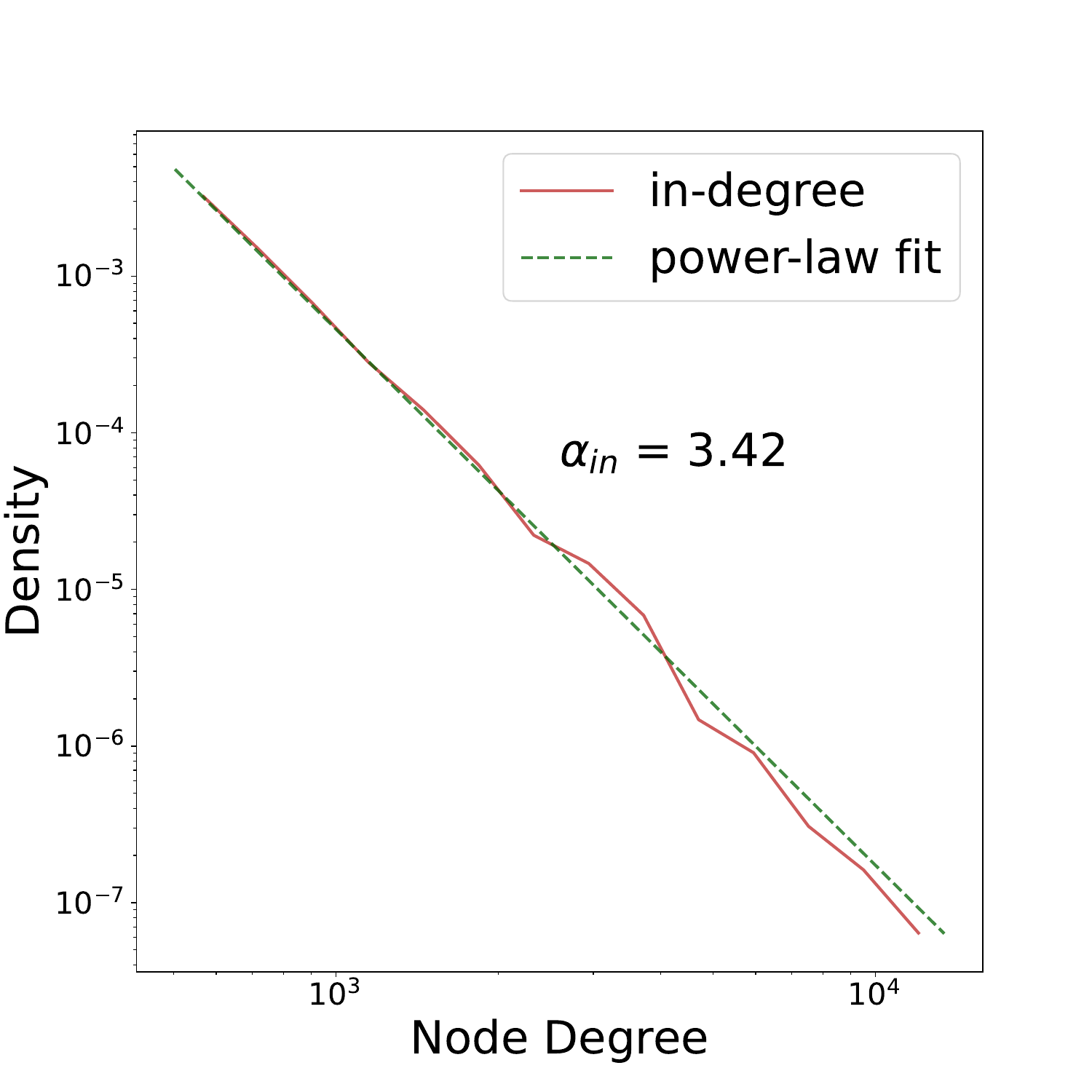}
\end{minipage}
\caption{Tail exponents fit for the out-degree (left hand side) and in-degree (right hand side) distributions. The out-degree distribution is characterized by a smaller tail exponent, which is indicative of a distribution with a greater dispersion.}%\mi{Appendice?}
  \label{fig:in_out_degree_log_scale_tail_fit}
\end{figure}

\begin{figure}[H]
\centering
  \includegraphics[width=0.9\linewidth]{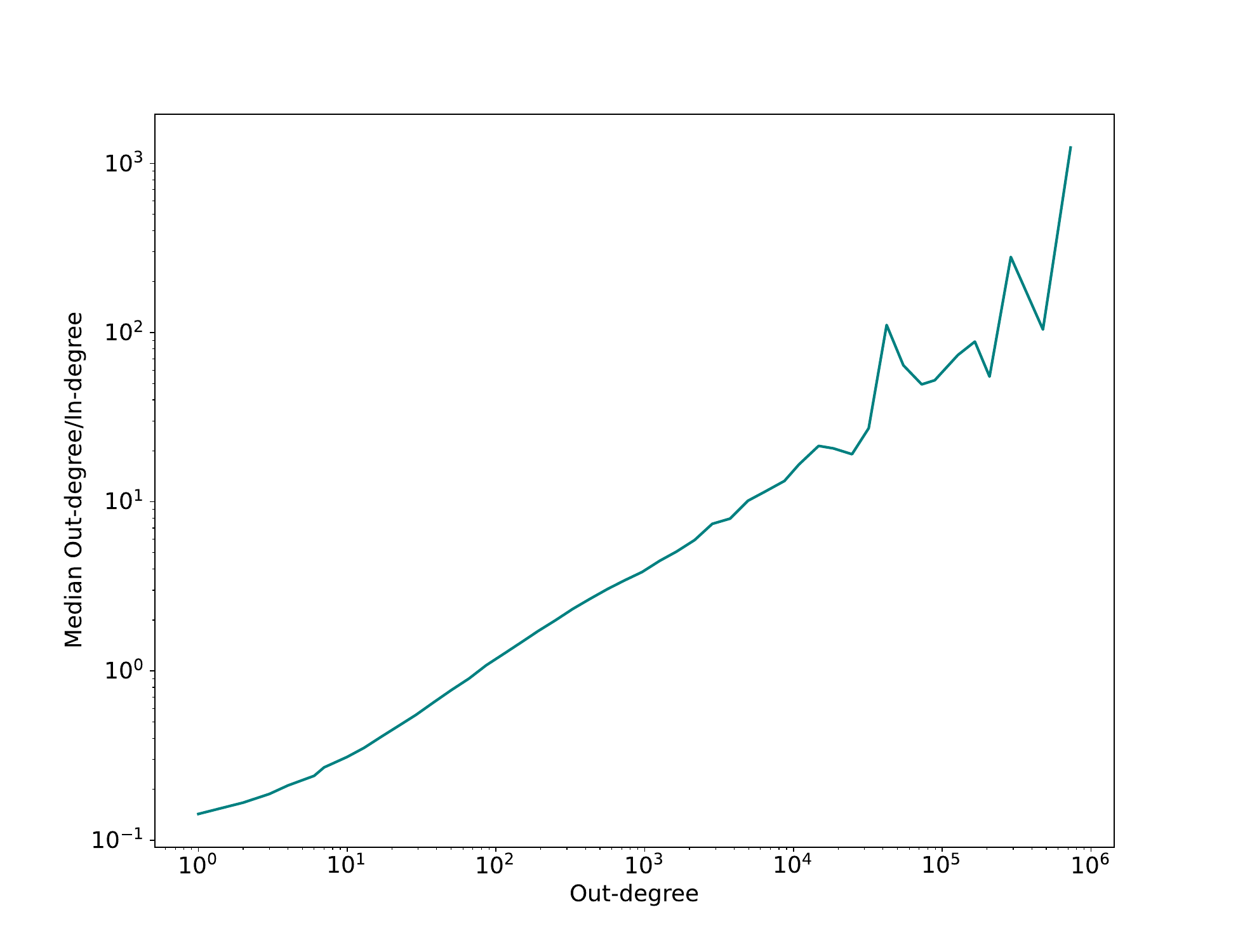}
  \caption{Median out-degree/in-degree ratio as a function of the out-degree of a node. Even though firms with a small number of out-going links have typically more suppliers than buyers, the ratio increases approximately as a power law, confirming the results of section \ref{sec:degree_distribution}.}
  \label{fig:degree_ratio}
\end{figure} 

\begin{figure}[H]
\centering
  \includegraphics[width=0.9\linewidth]{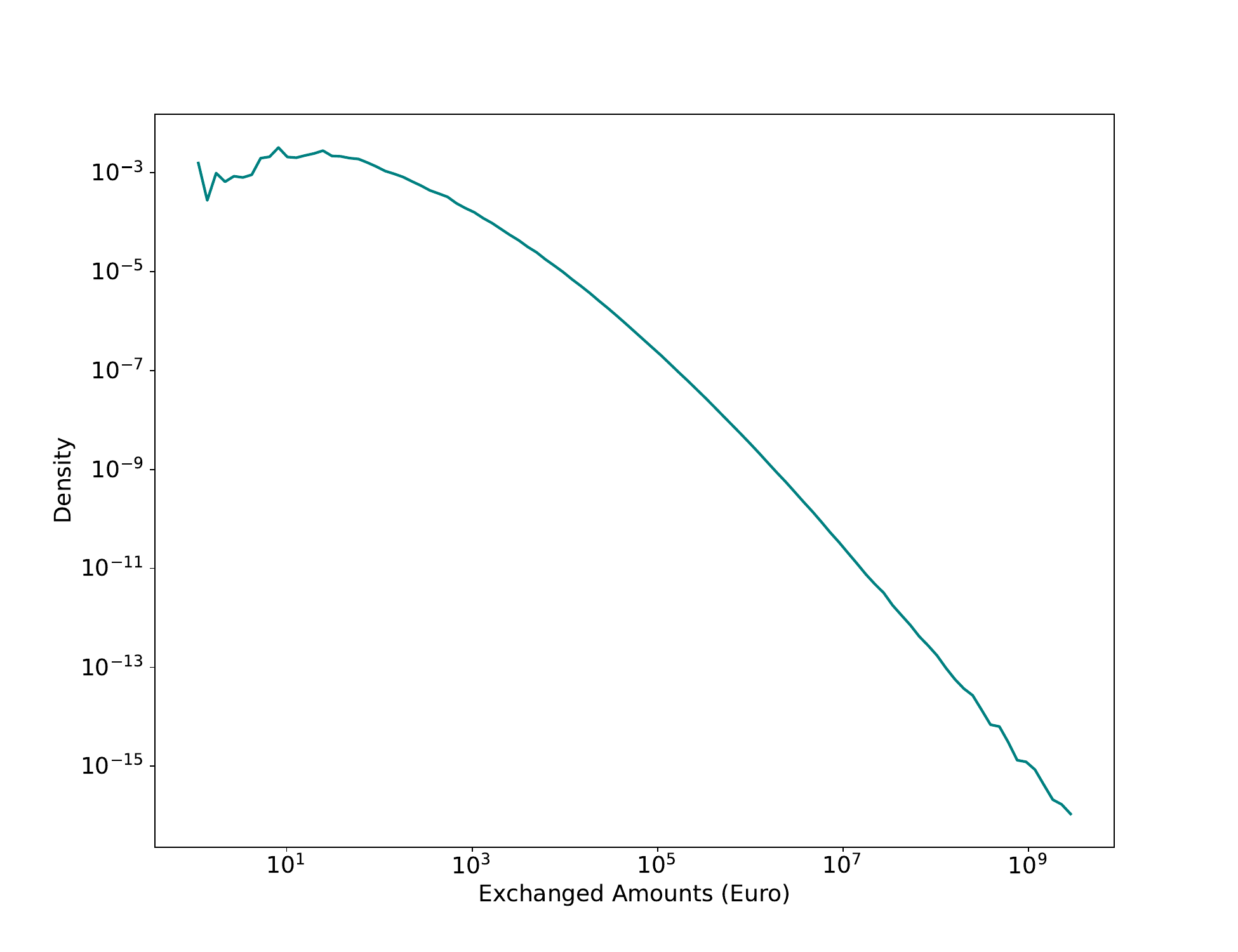}
  \caption{Density of invoice amounts. The x-axis is partitioned into 100 logarithmically spaced bins and the y-axis reports the density of invoices falling within each bin.}
  \label{fig:exchanged_amounts}
\end{figure}

\section{Provinces and Sector Codes}
\setcounter{figure}{0}  
\setcounter{table}{0}
\vspace{-5pt} 
\subsection{One-Letter Sector of Economic Activity}
\label{sec:app_tabella_settori}

%\begin{tabular}{p{12.5cm} p{5cm}}
\begin{table}[H]
    \centering
    \begin{tabular}{p{2cm}p{7cm}p{3cm}}
    \textbf{One-letter Sector} & \textbf{Description} & \textbf{2-digit Sectors} \\ [0.5ex] 
 \hline\hline
    A & agriculture, forestry and fishing & 01, 02, 03 \\ \hline
    B & mining and quarrying & 05, 06, 07, 08, 09 \\ \hline
    C & manufacturing & 10, 11, 12, 13, 14, 15, 16, 17, 18, 19, 20, 21, 22, 23, 24, 25, 26, 27, 28, 29, 30, 31, 32, 33 \\ \hline
    D & electricity, gas, steam, and air conditioning supply & 35 \\ \hline
    E & water supply, sewerage, waste management and sanitation activities & 36, 37, 38, 39 \\ \hline
    F & construction & 41, 42, 43 \\ \hline
    G & wholesale and retail trade; repair of motor vehicles and motorcycles & 45, 46, 47 \\ \hline
    H & transport and storage & 49, 50, 51, 52, 53 \\ \hline
    I & hotels and restaurants & 55, 56 \\ \hline
    J & information and communication services & 58, 59, 60, 61, 62, 63 \\ \hline
    K & financial and insurance activities & 64, 65, 66 \\ \hline
    L & real estate activities & 68 \\ \hline
    M & professional, scientific and technical activities & 69, 70, 71, 72, 73, 74, 75 \\ \hline
    N & renting, travel agencies, business support activities & 77, 78, 79, 80, 81, 82 \\ \hline
    O & public administration and defense; compulsory social security & 84 \\ \hline
    P & education & 85 \\ \hline
    Q & health and social work & 86, 87, 88 \\ \hline
    R & recreational, cultural and sporting activities & 90, 91, 92, 93 \\ \hline
    S & other service activities & 94, 95, 96 \\ \hline
    T & activities of households and collective households as employers of domestic personnel & 97, 98 \\ \hline
    U & extraterritorial organizations and entities & 99 \\ \hline
    \end{tabular}
    \caption{Correspondence between one-letter sector of economic activity, activity description and 2-digit sectors.}
    \label{tab:my_label}
\end{table}

\subsection{Maps}
\label{subsec:appendix_maps}

\subsubsection{Regions}
\label{sec:app_mappa_regione}
\begin{figure}[H]
\centering
  \includegraphics[width=1\linewidth]{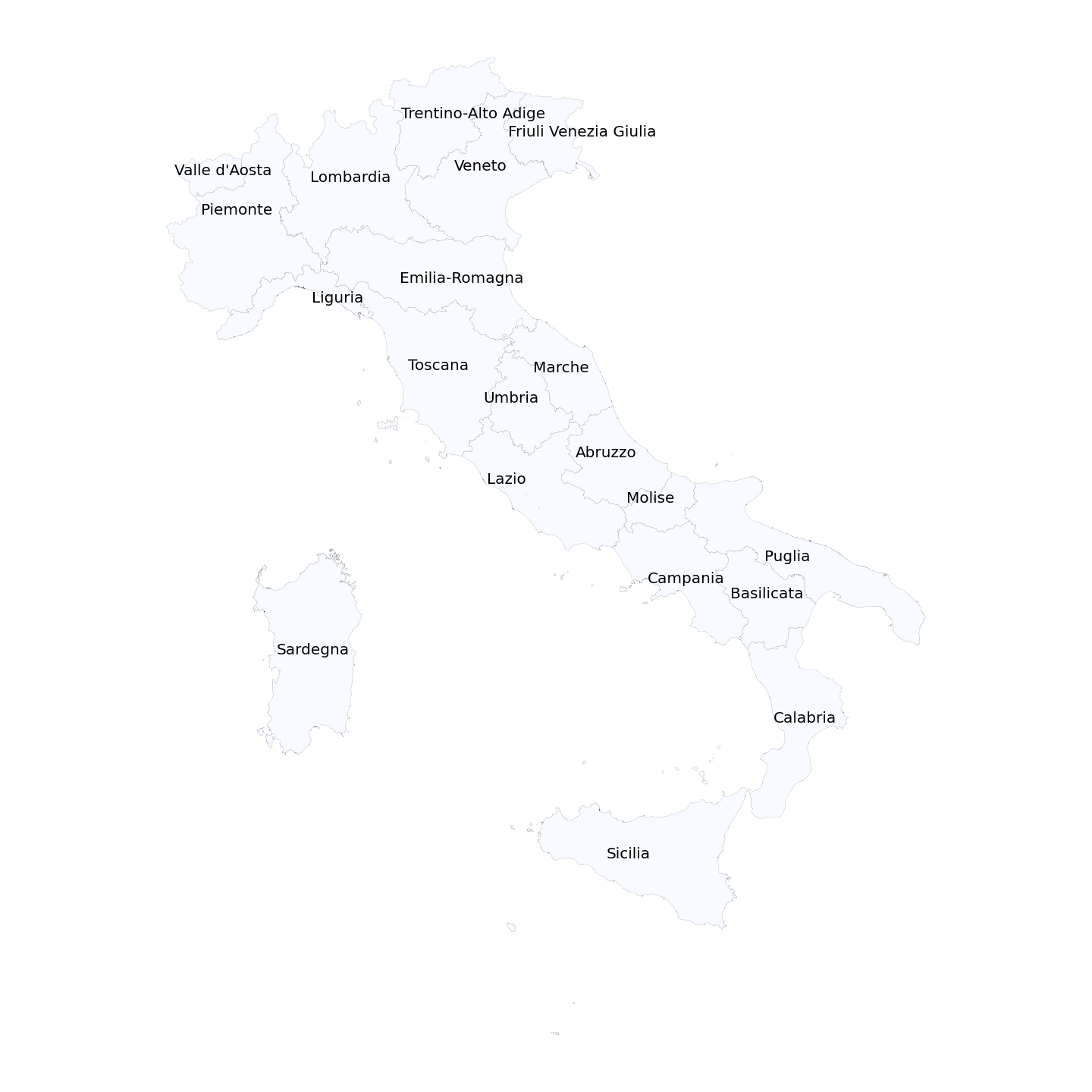}
  \caption{Map of Italian regions.}
  \label{fig:mappa_regioni}
\end{figure} 

\subsubsection{Provinces}
\label{sec:app_mappa_province}
\begin{figure}[H]
\centering
  \includegraphics[width=1\linewidth]{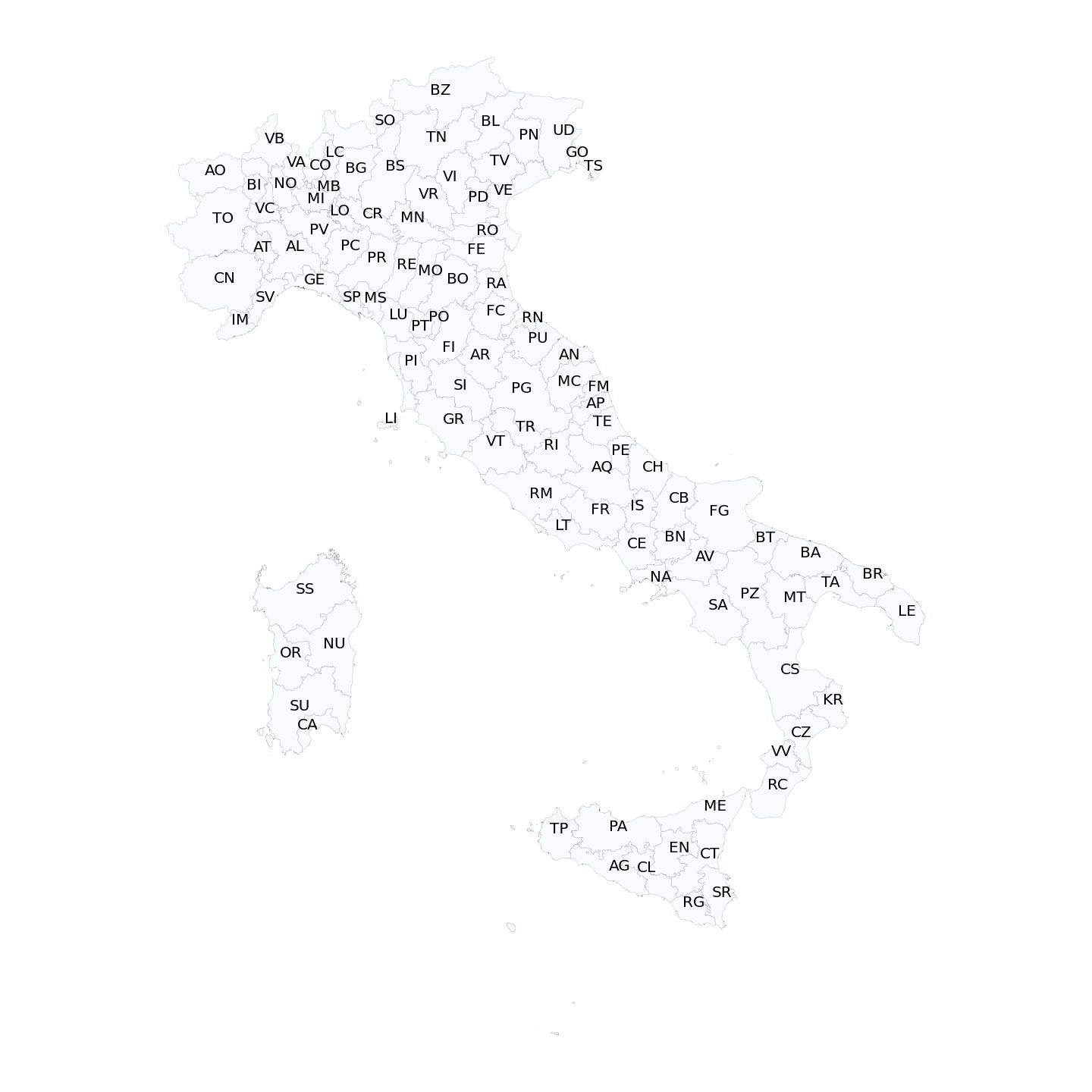}
  \caption{Map of Italian provinces.}
  \label{fig:mappa_provincie}
\end{figure} 

% \newpage
% \thispagestyle{empty}
% \listoffigures

% \begin{figure}[H]
% \centering
%  \includegraphics[width=1\linewidth]{plots/2019_cat_heatmap.pdf}
%  \caption{.}
%  \label{fig:2019_cat_heatmap}
% \end{figure} 
% 

\end{document}